\documentclass[12pt,a4paper]{article}
\usepackage{amssymb}
\usepackage{amsmath}
\usepackage{amsfonts}
\usepackage{geometry}
\usepackage{enumerate}
\usepackage{float}
\usepackage{rotating}
\usepackage{multirow}
\usepackage{comment}
\usepackage[round]{natbib}
\usepackage{url}

\restylefloat{table}
\begin{document}

\title{Identification and Estimation of Seller Risk Aversion in Ascending
Auctions\thanks{Earlier versions of this work had been presented under the title
\textquotedblleft \textit{Semiparametric Estimation of Ascending Auctions with Risk Averse Sellers.}\textquotedblright\ We thank Miguel Delgado, Emmanuel Guerre, Ming Li, Jingfeng Lu, and the seminar participants at City, University of London, Durham University, Nanyang Technological University, Universidad Carlos III de Madrid, and the Asian Meeting of the Econometric Society in Abu Dhabi for helpful comments and discussions. We also thank Raissa Arantes for research assistance. We gratefully acknowledge financial support from the Carlos Chagas Filho Foundation for Research Support of the State of Rio de Janeiro (2022JCN-281395) [Gimenes] and the British Academy Newton Advanced Fellowship (NAFR1180122) [Srisuma].}
\thanks{\textit{E-mail addresses}: \texttt{ngimenes@econ.puc-rio.br}, \texttt{qi.tonghui@nus.edu.sg}, \texttt{s.srisuma@nus.edu.sg}}}

\author{Nathalie Gimenes \\
PUC-Rio \and Tonghui Qi \\
National University of Singapore \and Sorawoot Srisuma \\
National University of Singapore}
\date{February 4, 2026}
\maketitle

\begin{abstract}
This paper shows how to identify and estimate the seller’s risk parameter in an ascending auction. We consider a semiparametric model where the seller has a parametric utility function (such as CARA or CRRA) and the distribution of bidder valuations is modeled flexibly. We provide primitive conditions under which the risk parameter is identified and show that it can be consistently estimated with an asymptotically normal limiting distribution under standard regularity conditions. A Monte Carlo study demonstrates good finite-sample performance of the proposed estimator. We apply our approach to foreclosure real estate auction data from S\~{a}o Paulo. We find evidence that sellers are risk-averse, which leads to a much better fit to the data than a model with risk-neutral sellers, which would substantially underpredict the reserve price relative to what is observed.


\end{abstract}

\section{Introduction}

Knowing the seller’s risk attitude is essential for auction design and policy analysis, as seller risk preferences affect supply-side welfare and equilibrium outcomes. The theoretical literature has shown that many canonical results derived under risk neutrality no longer hold when sellers are risk averse. For example, the Revenue Equivalence Theorem does not hold with a risk-averse seller, and while some optimal auction results in \cite{myerson1981optimal} extend to risk-averse sellers, they do so only in some cases (\cite{sundararajan2020robust}). More generally, seller risk aversion affects key design choices, including reserve prices and auction formats (\cite{waehrer1998auction}, \cite{moreno2017reserve}). These insights highlight why policymakers seeking to predict auction outcomes and design effective auction mechanisms would want to learn about seller risk attitudes.

Prior to our study, to the best of our knowledge, there are no results on the identification of seller risk
aversion or on how to estimate it in any auction format. Our paper addresses
this gap in the literature. We study a semiparametric model of an ascending auction in which
the seller has a parametric utility function (such as CARA or CRRA), while
bidder valuations and independent and satisfy a linear quantile specification. We show that the risk-aversion parameter is identified under primitive conditions. We then propose an estimator for the
risk-aversion parameter and show that our estimator is consistent and
asymptotically normal under standard regularity conditions.

Using data on winning bids, reserve price, and value of the auctioned object, we can identify
components of the first-order condition (FOC) that characterizes the
seller's optimal reserve price chosen to maximize expected revenue. Under
interpretable conditions that parallel those assumed in the
theoretical auction literature, part (i) of our Proposition 2 establishes
existence and uniqueness of the solution to the FOC, while part (ii) shows
that the optimal reserve price is decreasing in the seller's Arrow--Pratt
measure of risk aversion. We use the latter monotonic relationship between the
optimal reserve price and risk aversion to establish identification of
parametric risk preferences that include CARA and CRRA utility functions as
special cases.

Our estimator of the risk-aversion parameter is a two-step parameter. In the first stage, nuisance functions, which are components of the expected revenue related to the distribution of bidder valuations, are estimated. The risk-aversion estimator is then defined by forcing the empirical FOC in the second stage. The nuisance functions consist of the derivative of the quantile function and the inverse of the quantile function. While only nonparametric restrictions on the bidder valuation distribution are required for identification, we further assume that the valuation quantile is linear in covariates for the purpose of practical estimation. This modeling choice represents a compromise between fully nonparametric and parametric alternatives, since the linear quantile specification allows for a flexible representation of conditional valuation distributions without suffering from the curse of dimensionality. In principle, other specifications for the distribution of bidder valuations can be adopted.

The use of quantiles in the econometrics of auctions is not new\footnote{An early insight can be found in \cite{haile2003nonparametric}. Nonparametric
estimation of quantiles is used by \cite{marmer2012quantile} and \cite{guerre2012uniform} in a first-price auction model, for example.}, but the
development of a quantile regression (QR) estimator in this context is
relatively recent. Methodologically, our approach builds on the estimators
used in \cite{gimenes2017econometrics} and Gimenes
and Guerre (2022, GG22 hereafter). Their strategy exploits auction-specific relations between quantiles of the bidder's valuation and
their bids, in which they use bid data to estimate the QR for bids and
transform it into a QR estimator for valuations. Specifically, \cite{gimenes2017econometrics} applies this strategy to estimate
ascending auctions with the classic Koenker-Bassett QR estimator, while GG22
develop the augmented quantile regression (AQR) estimator and apply it to
a first-price auction model.

In this paper, we use an AQR estimator to estimate the distribution of bidder valuations. Our quantile estimator is the AQR counterpart to the estimator used in \cite{gimenes2017econometrics}. An AQR estimator is defined as the maximizer of an integrated version of the classic \cite{koenker1978regression} objective function, combined with kernel smoothing and polynomial approximation of the quantile function. Similar to the local polynomial estimator of \cite{fan1992design}, advantages the AQR method has over standard QR are that it automatically yields estimators of both the quantile function and its derivative, and produces estimators with improved statistical properties near the boundaries. These features are crucial for our estimation problem, as the quantile derivative enters the FOC and all quantile levels of the AQR estimator can be used to estimate the risk-aversion parameter. By contrast, \cite{gimenes2017econometrics} does not consider the FOC and therefore does not require estimation of the nuisance functions that enter it.

Other than the fact that the auction-specific relationship between quantiles of bidder valuations and bids differs between ascending auctions and first-price auctions, our application of the AQR method differs from GG22’s analysis in non-trivial ways due to our goal of estimating the risk-aversion parameter. In particular, we view the following two aspects as secondary contributions of the paper, as they expand the scope of the AQR method beyond our application. 
First, we construct and derive
the asymptotic properties of a conditional CDF estimator defined as the inverse of an AQR estimator---see Lemma 3---which enters the FOC and does not appear in GG22. This estimator may be of independent interest as a flexible and low-dimensional approach to estimating conditional CDFs (see, e.g., \cite{koenker2013distributional} for a discussion on using
a quantile function to estimate CDF and vice versa). Second, we show AQR estimators can be used as preliminary estimators in a two-step semiparametric M-estimation problem. Specifically, our risk-aversion parameter is an estimand defined implicitly as a solution to the FOC, whereas the finite-dimensional estimands considered in GG22 are explicit functionals of quantile objects.\footnote{They motivate the study of such functionals with
economic quantities that have closed-form expressions in the quantile
functions (e.g., expected revenue (\cite{li2003semiparametric}) and bidder risk-aversion parameter in a
first-price auction (\cite{guerre2009nonparametric})).} Theorem 4 in GG22 shows that functionals of AQR estimators satisfy a central limit theorem (CLT). Our estimator does not take this form, but we show that it does so after suitable application of stochastic equicontinuity and linearization arguments, and that the CLT applies. This line of arguments is not new for developing limit theorems for semiparametric M-estimators. The novelty here is that we derive the linearized functional explicitly with AQR estimators as preliminary estimators, which has not been done previously. The econometrics and statistics literature is otherwise familiar with similar derivations when traditional nonparametric estimators, such as kernel and series estimators, are used---see, for example, \cite{andrews1994asymptotics}, \cite{newey1994asymptotic}, \cite{chen2003estimation}, and \cite{ichimura2010characterization}.

We perform a Monte Carlo study and find that our estimator behaves as
expected from the theory. We then apply our methodology to real estate
foreclosure auction data in S\~{a}o Paulo as an illustration. In this
application, the sellers are lenders (banks and private companies) and the Court of Justice of the State of S\~{a}o Paulo. Possible reasons why these sellers may be risk averse include downside risks associated with failure to sell. For example, there may be future costs of holding such properties (e.g., property taxes, insurance, and security), the value of foreclosure homes may depreciate over time, and sellers may be subject to directives to sell within a certain time frame.\footnote{Foreclosure assets are sometimes viewed as non-performing loans from an accounting perspective, and some central banks provide some guidelines on selling them accordingly. For example, see this ECB report: 
\url{https://www.bankingsupervision.europa.eu/ecb/pub/pdf/guidance_on_npl.en.pdf}.}

Our empirical results suggest that
sellers are risk-averse, and the model with a risk-averse seller fits the
data much better than one with a risk-neutral seller. Notably, we find the model with risk-neutral sellers
systematically overpredicts observed reserve prices relative to the model
with risk-averse sellers. This finding complements several empirical evidence
showing that observed reserve prices are well below the levels implied by
the risk-neutral benchmark in a variety of auction settings (see \cite{mcafee1992updating}, \cite{paarsch1997deriving}, \cite{mcafee2002set}, \cite{haile2003inference}, and \cite{tang2011bounds}). Indeed, seller risk
aversion has been proposed in the theoretical literature as one of the mechanisms that can  rationalize these discrepancies (\cite{maskin1984optimal}, \cite{matthews1987comparing}, \cite{hu2010risk}).

We end the introduction with a review of related literature on the econometrics of auctions. The rest of the paper is organized as follows. Section 2 introduces the model and establishes key identification results for bidder and seller primitives. Section 3 develops the estimation methodology and the asymptotic properties of the AQR estimators for observed bids and bidder valuations. Section 4 presents the estimation of the risk-aversion parameter and derives its asymptotic properties. Section 5 reports a Monte Carlo study of the estimators. Section 6 provides an empirical application to real estate auctions in S\~{a}o Paulo. Section 7
concludes. Proofs of results are collected in the Appendix.

\subsubsection*{Background Literature }
While there appear to be no prior econometric research on seller risk aversion,
there is a growing literature studying risk-averse bidders, particularly
in first-price auctions; see, for example \cite{lu2008estimating}, \cite{guerre2009nonparametric}, \cite{campo2011semiparametric}, \cite{li2015auctions}, \cite{zincenko2018nonparametric}, \cite{grundl2019identification}, and \cite{jun2022testing}. On the other hand, the empirical auction literature is built on the identification and inference of bidder valuation
distributions under bidder risk neutrality. This begins with the parametric
model of \cite{paarsch1992deciding} and grows with
Donald and Paarsch (1993,1996), \cite{laffont1995econometrics}, \cite{athey2001information} amongst others. 

Parametric models, however, can be computationally demanding to estimate and
are subject to model misspecification. Nonparametric approaches, beginning
with the seminal work of \cite{guerre2000optimal},
provide an alternative and have since been developed in a variety of
contexts; see, inter alia, \cite{athey2002identification}, \cite{lu2008estimating}, \cite{krasnokutskaya2011identification}, 
\cite{marmer2012quantile}, \cite{campo2011semiparametric}, \cite{marmer2013model}, \cite{enache2017quantile}, 
\cite{liu2017nonparametric}, \cite{luo2018integrated} and \cite{ma2019inference}. 

Purely nonparametric approaches are also not without limitations, as the convergence of nonparametric estimators deteriorates with the number of conditioning variables. Semiparametric approaches that aim to allow for more flexible modeling than parametric ones, without suffering from the curse of dimensionality, have been developed for mean regression (\cite{rezende2008econometrics}) and quantile regression (\cite{gimenes2017econometrics}, \cite{gimenes2022quantile}).  

\section{Model and identification}

Consider an ascending auction of an indivisible object with $I\geq 2$\
bidders. Bidders can raise prices continuously and without cost until only
one bidder remains. The object is sold to the
highest bidder for the price of his last bid, provided that it is at least as high as the
reservation price. We assume an independent private values (IPV) environment, where each bidder
knows only their own valuation and values are independently drawn across
bidders. 

Let the auctioned object have observable
characteristics $X\in \mathbb{R}^{D}$. Bidder $i$'s valuation of the object
is $V_{i}$, taking value in $\mathcal{V}=\left[ \underline{v},\overline{v}%
\right] $. For notational simplicity, we assume $\mathcal{V}$\ to be
independent of $X$. We denote the conditional CDF of $V_{i}$ by $F\left(
\cdot |X\right) $. To facilitate readers, as it may be instructive to
compare our assumptions and results with GG22, we use the same 
notations and terminologies as them when possible.

\subsection{Bidders' behavior}

The winning bid, denoted by $B=V^{I-1:I}\in 
\mathcal{V}$, is the $\left( I-1\right) $-th order statistic among the $I$
i.i.d. private values $\left\{ V_{i}\right\} _{i=1}^{I}$. This is the same equilibrium play for ascending auctions as used in \cite{aradillas2013identification} and \cite{gimenes2017econometrics}. We denote the
conditional CDF of $B$ by $G\left( \cdot |X\right) $. We impose the relations between these variables in the following assumption.

\bigskip

\textbf{Assumption M1.}

\textit{(i) }$F\left( \cdot |X\right) $\textit{\ is continuous and
strictly increasing almost surely on }$\mathcal{V}$\textit{;}

\textit{(ii) }$G\left( t|X\right) =\phi \left( F\left( t|X\right) \right) $%
\textit{\ a.s. for }$t\in \mathcal{V}$\textit{, where }$\phi \left( a\right)
=Ia^{I-1}-\left( I-1\right) a^{I}$\textit{\ for }$a\in \left[ 0,1\right] $%
\textit{.}

\bigskip

M1(i) imposes a minimal regularity condition on $F\left( \cdot
|X\right) $. M1(ii) is a structural assumption on the bidder's bidding behavior, following \cite{athey2002identification}, described using the relation between the CDF of the $%
\left( I-1\right) $-th order statistic and the underlying CDF that the
sample is drawn from.

We will take a quantile approach to model the bidder's private value
distribution. Let the $\alpha -$quantile of the valuation distribution be
denoted by $V\left( \alpha |X\right) =F^{-1}\left( \alpha |X\right) $ for $%
\alpha \in \left[ 0,1\right] $. As done in \cite{gimenes2017econometrics} and GG22, we take $V_{i}=V\left( A_{i}|X\right)$ where $A_{i}\sim Uni\left[ 0,1\right] $ can be viewed as the \textit{private rank} of the bidder, which is independent of $X$ and other
bidders' ranks.

We denote the $\alpha -$quantile of $B$\ by $B\left( \alpha |X\right)
=G^{-1}\left( \alpha |X\right) $ for $\alpha \in \left[ 0,1\right] $. $%
B\left( \cdot |X\right) $ exists, because $G\left( \cdot |X\right) $\ is
differentiable and strictly increasing, since $\phi :\left[ 0,1\right]
\rightarrow \left[ 0,1\right] $ is continuously differentiable and is
strictly increasing on $\left[ 0,1\right] $.

Proposition 1 shows that the quantile functions of the winning bid and the private valuation are
linked through a one-to-one relationship given in equation (\ref{B to V quantile}). This relation is the central identification argument on the bidder's side (\cite{gimenes2017econometrics}).

\bigskip

\textbf{Proposition 1.}\textit{\ }Suppose Assumption M1 holds, then%
\begin{equation}
V(\alpha |X)=B\left( \phi (\alpha )|X\right) \text{\ a.s.\ for all }\alpha
\in \left[ 0,1\right] .  \label{B to V quantile}
\end{equation}

\subsection{Seller's behavior}

The seller can influence her expected revenue in an auction by setting
reserve price. If all bids are below the reserve price, the seller keeps the object. If all but one bids are below the reserve price, the object is sold with the winner paying the reserve. If two or more bids are above the reserve price, the object is sold with the winner paying the second highest private value among all bidders.

We denote the seller's value of the sale object by $W$. The seller can set the reserve price optimally to maximize her expected utility. Specifically, if the seller has utility function $U\left( \cdot \right) $, the expected utility from setting reserve price to be $r\in \mathcal{V}$ is:
\begin{equation}
\widetilde{\Pi }(r,X,W)=U\left( W\right) F\left( r|X\right) ^{I}+U\left(
r\right) IF\left( r|X\right) ^{I-1}\left( 1-F\left( r|X\right) \right)
+\int_{r}^{\overline{v}}U\left( t\right) dG\left( t|X\right) .
\label{ER in r}
\end{equation}%
We denote the optimal reserve price by $R$. I.e., $R=\arg \max_{_{r\in \mathcal{V}}}\widetilde{\Pi }\left(
r,X,W\right) $, whose existence and uniqueness are guaranteed under the
conditions of Assumption M2 given below.

Analogously to writing the bidders' bids in terms of private ranks, the expected revenue above can be equivalently expressed as a function of the \textit{screening level} instead of the reserve price. For a screening level $\alpha$
that takes value in $\left[ 0,1\right] $, let us define 
\begin{equation}
\Pi (\alpha,X,W)=U\left( W\right) \alpha^{I}+U\left( V\left( \alpha|X\right) \right)
I\alpha^{I-1}\left( 1-\alpha\right) +I\left( I-1\right) \int_{\alpha}^{1}U\left( V\left(
t|X\right) \right) t^{I-2}\left( 1-t\right) dt,\text{ }  \label{ER in a}
\end{equation}
so that $\Pi (\alpha,X,W)=\widetilde{\Pi }\left( r,X,W\right) $ when $r=V\left(
\alpha|X\right) $. The optimal screening level is defined as $\alpha _{R}=\arg
\max_{_{\alpha\in \left[ 0,1\right] }}\Pi \left( \alpha,X,W\right) $.

\bigskip

\textbf{Assumption M2.}

\textit{(i) }$V_{i}$\textit{\ has a conditional PDF, denoted by }$f\left(
\cdot |X\right) $\textit{, that is bounded away from zero and infinity a.s.
on }$\mathcal{V}$\textit{; }

\textit{(ii) }$J(v|X)=v-\frac{1-F(v|X)}{f\left( v|X\right) }$\textit{,
defined for }$v\in \mathcal{V}$\textit{, is strictly increasing a.s. on }$%
\mathcal{V}$\textit{; }

\textit{(iii) }$W$ takes value in $\mathcal{W}\subseteq \mathcal{V}$\textit{%
; }

\textit{(iv) }$U\left( \cdot \right) $\textit{\ is a twice continuously
differentiable function with }$U^{\left( 1\right) }\left( \cdot \right) >0$%
\textit{\ and }$U^{\left( 2\right) }\left( \cdot \right) \leq 0$\textit{.}

\bigskip

Assumption M2 consists of standard conditions in the auction literature when
studying the seller's behavior. M2(i) is a regularity condition where the
bounding from below ensures $J\left( \cdot \right) $ in M2(ii)\ is well
defined. In his seminal paper, \cite{myerson1981optimal} calls $J\left( \cdot \right) $
the \textit{virtual valuation function}, as it represents the marginal
revenue contribution of a bidder with valuation $v$. He imposes monotonicity
of $J\left( \cdot \right) $, which holds when $V_{i}$ has an increasing
hazard rate, to prove incentive compatibility in the design of optimal
auctions. In M2(iii), the lower bound on $W$ rules out point mass of the
seller setting the reserve at $\underline{v}$, and the seller would be
better off not selling the object if $W>\overline{v}$. M2(iv) assumes the
utility function is smooth and allows the seller to be risk-averse as well
as risk-neutral. The monotonicity and concavity of $U\left( \cdot
\right) $\ in M2(iv) are standard assumptions in the risk aversion
literature where the differentiability conditions are imposed to facilitate
analytical tractability -- for example, it ensures Arrow-Pratt measure of
risk aversion to be defined.

For two utility functions that are strictly increasing and weakly concave, 
$U_{1}(\cdot )$\ and $U_{2}(\cdot )$, we say that $U_{2}(\cdot )$ represents
a strictly more risk-averse preference in the Arrow-Pratt sense if there exists a real-valued function $\zeta (\cdot )$ that is twice continuously
differentiable with $\zeta ^{\left( 1\right) }\left( \cdot \right)
>0$ and $\zeta ^{\left( 2\right) }\left( \cdot \right) <0$ such that $U_{2}(\cdot )=\zeta (U_{1}(\cdot ))$.\footnote{%
It is not necessary to define Arrow-Pratt risk aversion with differentiable $\zeta (\cdot )$, and strict monotonicity and concavity will suffice.
However, similarly to how we consider a smooth utility function in M2(iv),
differentiability is used to facilitate analytical tractability.} For differentiable utility functions, this
formulation is equivalent to saying that their Arrow-Pratt risk aversions satisfy $-\frac{U_{2}^{\left( 2\right) }\left( v\right) }{U_{2}^{\left( 1\right) }\left( v\right) } > -\frac{U_{1}^{\left( 2\right) }\left( v\right) }{U_{1}^{\left( 1\right) }\left( v\right) } $ for all $v$. The degree of risk aversion can take on a more compact form when utility functions belong to some parametric families. For example, when constant ARA (CARA) or RRA (CRRA) functions are used, ranking of Arrow-Pratt risk aversion between preferences is simply determined by the risk aversion parameters. For more background materials on Arrow-Pratt risk aversion, we refer the reader to Chapter 6.D in \cite{mas1995microeconomic}.

Under Assumption M2, Proposition 2(i) below says that the optimal reserve
price is characterized by the first-order condition obtained from
differentiating (\ref{ER in r}), and Proposition 2(ii) says that the optimal
reserve price decreases with the seller's risk aversion in the Arrow-Pratt
sense.

\bigskip

\textbf{Proposition 2.} Suppose Assumption M2 holds, then:

(i) For any $\left( X,W\right) $, the optimal reserve price exists and is
uniquely determined by $r^{\ast }\in \mathcal{V}$ that satisfies%
\begin{equation}
0=U\left( W\right) +U^{\left( 1\right) }\left( r^{\ast }\right) \frac{%
1-F(r^{\ast }|X)}{f\left( r^{\ast }|X\right) }-U\left( r^{\ast }\right) ;
\label{Seller's FOC}
\end{equation}

(ii) The optimal reserve price decreases with the seller's risk aversion in
the Arrow-Pratt sense.

\bigskip

The right hand side of equation (\ref{Seller's FOC}) is the partial
derivative of (\ref{ER in r}) with respect to $r$\ evaluated at $r^{\ast }$.
By putting $R$ in place of $r^{\ast }$\ in (\ref{Seller's FOC})\ and use the
identity that $R=V\left( \alpha _{R}|X\right) $, Proposition 2 confirms that 
$R$ is the unique maximizer of $\widetilde{\Pi }\left( \cdot ,X,W\right) $\
, and it satisfies 
\begin{equation}
0=U\left( W\right) +U^{\left( 1\right) }\left( R\right) V^{\left( 1\right)
}\left( \alpha _{R}|X\right) \left( 1-\alpha _{R}\right) -U\left( R\right) .
\label{FOC theory}
\end{equation}

Later on, we will assume the shape of $U\left( \cdot \right) $\ is known up
to the risk aversion parameter. For example, in our empirical application,
we use the CRRA utility function: 
\begin{equation}
U_{\theta }\left( v\right) =\left\{ 
\begin{array}{c}
\frac{v^{1-\theta }-1}{1-\theta } \\ 
\ln \left( v\right) 
\end{array}%
\begin{array}{c}
\theta \neq 1 \\ 
\theta =1%
\end{array}%
\right. ,  \label{CRRA}
\end{equation}%
defined for $v>0$\ and $\theta \in \mathbb{R}$, where higher $\theta $\
represents a higher degree of risk aversion. Under the CRRA specification, M2(iv) holds for $\theta \geq 0$. We show in the proof
of Lemma 4 how Proposition 2 can be used to identify an Arrow-Pratt risk
aversion parameter.

\section{Augmented quantile regression}

Given data on winning bids and auction characteristics, this section proposes estimators for the quantile function of private values and related functions under the linear quantile specification. Using Proposition 1, these estimators are obtained through appropriate transformations of the bid quantile function, which we estimate using the AQR method of GG22. 

Abstracting from the auction interpretation, our AQR estimator for bids can also be viewed as a general-purpose quantile regression estimator, whose inverse provides an estimator of the conditional CDF. We establish pointwise asymptotic properties for the AQR estimators of these functions, as well as their uniform convergence rates, in Lemmas 1 and 3, respectively.

The statistical properties of the estimators for the distribution of private values are stated as propositions. We note that only convergence rates for these estimators are required to derive the large-sample properties of the risk-aversion estimator in Section 4. Moreover, we use only bid data in this section, as seller-specific variables---namely the reserve price and seller's value of the auctioned object---are not used in the quantile estimation.

\subsection{Assumptions for AQR estimation}

We begin with some assumptions.

\textbf{Assumption Q. }

\textit{(i) The auction variables }$\left\{ \left( B_{l},X_{l}\right)
\right\} _{l=1}^{L}$\textit{\ are a random sample. For some }$F\left( \cdot
|X\right) $\textit{\ and }$G\left( \cdot |X\right) $\textit{\ that satisfy
Assumptions M1 and M2(i), }$B_{l}$\textit{\ takes value in }$\mathcal{V}$\textit{\ and
have conditional CDF }$G\left( \cdot |X\right) $\textit{. }$X_{l}$\textit{\
takes value in }$\mathcal{X}\subseteq R^{D}$\textit{\ such that }$\mathcal{X}$\textit{\ is
compact,}$\ $\textit{and the eigenvalues of }$E\left[ X_{l}X_{l}^{\top }%
\right] $\textit{\ are bounded away from zero and infinity. }

\textit{(ii) Let }$V\left( \alpha |X\right) =F^{-1}\left( \alpha |X\right) $%
\textit{\ and }$V\left( \alpha |X\right) =X_{1}^{\top }\gamma \left( \alpha
\right) =\gamma _{0}\left( \alpha \right) +X^{\top }\gamma _{1}\left( \alpha
\right) $\textit{\ where }$\gamma \left( \cdot \right) =\left[ \gamma
_{0}\left( \cdot \right) ,\gamma _{1}^{\top }\left( \cdot \right) \right]
^{\top }$\textit{\ is }$\left( s+1\right) -$\textit{times continuously
differentiable over }$\left[ 0,1\right] $\textit{\ for some }$s\geq 1$%
\textit{\ and }$X_{1}=\left[ 1,X^{\top }\right] ^{\top }$\textit{. }

\textit{(iii) The kernel function }$K\left( \cdot \right) $\textit{\ is
symmetric, continuously differentiable, and non-negative function on its
support, }$\left( -1,1\right) $\textit{. The bandwidth }$h$\textit{\ is
positive and satisfy }$h=o\left( 1\right) $\textit{\ and }$\log
^{2}L=o\left( Lh\right) $\textit{\ as }$L\rightarrow \infty $\textit{.}

\bigskip

Q(i) imposes standard regularity conditions and correct model specification.
In Q(ii), we assume that the quantile function of private values is linear
in covariates and satisfies standard smoothness conditions. Q(iii) specifies
the class of kernel functions and the conditions imposed on the bandwidth.

It is instructive to compare our assumptions with those found in Section 5.1
of GG22. First, we simplify their setting slightly by considering repeated
auctions with a fixed number of bidders, while they allow the number of
bidders to vary exogenously. It is straightforward for us to include this
feature with more notation. Our Q(i) and Q(ii) are analogous to their
Assumption A and Assumption S respectively. Our Q(iii) is the same with
their Assumption H other than they require $\log ^{2}L=o\left( Lh^{2}\right) 
$\textit{\ as }$L\rightarrow \infty $\textit{. }GG22$\ $imposes a more
stringent requirement on the bandwidth than us, because we are studying
different auction models. Specifically, we are estimating the quantile
function of private value using winning bids from ascending auctions,
whereas GG22 uses individual bids from first-price auctions. This matters,
as the quantile function of the bidder's private value depends on both the
quantile function of the optimal bid and its derivative in a first price
auction\footnote{%
If $B\left( \cdot \right) $\ and $B^{\left( 1\right) }\left( \cdot \right) $%
\ respectively were to denote the quantile function of the optimal first
price bid. Then it can be shown that: 
\begin{eqnarray*}
B\left( \alpha |X\right) &=&\frac{I-1}{\alpha ^{I-1}}\int_{0}^{\alpha
}a^{I-2}V\left( a|X\right) da\text{ with }\lim_{\alpha \rightarrow 0}B\left(
\alpha |X\right) =V\left( 0|X\right) , \\
V\left( \alpha |X\right) &=&B\left( \alpha |X\right) +\frac{\alpha B^{\left(
1\right) }\left( \alpha |X\right) }{I-1},
\end{eqnarray*}%
see equations (2.4)\ and (2.5) in GG22.}, which contrasts with equation (\ref%
{B to V quantile}) where the quantile functions of the bidder's private
value and winning bids have the same degree of
smoothness. Thus, GG22 cannot have the bandwidth decay too rapidly, since
the variance of the derivative of quantile estimator is inversely
proportional to the bandwidth while the bandwidth only appears in the higher
order terms for the variance of the level quantile estimator. We will impose
the same bandwidth condition as their Assumption H when we provide the
convergence rates of the derivative of the quantile function.

\subsection{Estimator of quantile function}

The winning bid's quantile function shares the linear quantile specification
as the private value's quantile function under Assumption Q. This follows
from combining Q(ii) with the identity in (\ref{B to V quantile}), which
gives:%
\begin{equation}
B\left( \alpha |X\right) =X_{1}^{\top }\beta \left( \alpha \right) \text{ \
with \ }\gamma \left( \alpha \right) =\beta \left( \phi (\alpha )\right) 
\text{ a.s. for }\alpha \in \left[ 0,1\right] ,  \label{Linear quantile spec}
\end{equation}%
recalling that $X_{1}=\left[ 1,X^{\top }\right] ^{\top }$ so $\beta \left(
\cdot \right) =\left[ \beta _{0}\left( \cdot \right) ,\beta _{1}^{\top
}\left( \cdot \right) \right] ^{\top }$.\ Since $\gamma \left( \cdot \right) 
$\ is a composite function of $\beta \left( \cdot \right) $ and $\phi \left(
\cdot \right) $, estimating $V\left( \cdot \right) $\ amounts to estimating $%
\beta \left( \cdot \right) $.

To motivate the AQR estimator for estimating $\beta \left( \cdot \right) $,
first recall that 
\begin{equation*}
B\left( \alpha |X\right) =\arg \min_{q}E\left[ \rho _{\alpha }\left(
B_{l}-q\right) |X\right] \text{ for }\alpha \in \left[ 0,1\right] ,
\end{equation*}%
where $\rho _{\alpha }\left( t\right) =t\left( \alpha -\mathbf{1}\left[ t<0%
\right] \right) $ is the check function. The minimizer of the sample
counterpart to the expectation above is the classic quantile regression
estimator of \cite{koenker1978regression}. This estimator is known to not
perform well when $\alpha $ is close to $0$ or $1$. Moreover,
it is piecewise linear (in $\alpha $) and different estimators for quantile derivatives are
required that complicates analysis of statistics that involve both quantile level
and its derivatives estimates. 

Instead, let us consider $B\left( \cdot |X\right) $
over $\left[ \alpha -h,\alpha +h\right] \cap \left[ 0,1\right] $, $\left\{
B\left( \tau |X\right) ,\tau \in \left[ \alpha -h,\alpha +h\right] \cap %
\left[ 0,1\right] \right\} $, which minimizes 
\begin{equation*}
\int_{0}^{1}E\left[ \rho _{a}\left( B_{l}-q\left( a,X\right) \right) |X%
\right] \frac{1}{h}K\left( \frac{a-\alpha }{h}\right) da,
\end{equation*}%
over any functions $q\left( \cdot ,X\right) $ since $K\left( \cdot \right) $
is non-negative. In the same spirit as a local polynomial estimator (e.g., \cite{fan1996local}), the AQR approach estimates the quantile coefficients and their derivatives simultaneously.
Specifically, with $B\left( \alpha +th|x\right)
=\sum\limits_{j=0}^{s}x_{1}^{\top }\beta ^{\left( j\right) }\left( \alpha
\right) \frac{\left( th\right) ^{j}}{j!}+O\left( h^{s+1}\right) $ in mind,
consider the following objective function,%
\begin{eqnarray*}
\widehat{\mathcal{R}}\left( b;\alpha \right) &=&\frac{1}{L}%
\sum\limits_{l=1}^{L}\int_{0}^{1}\rho _{a}\left( B_{l}-P\left(
X_{l},a-\alpha \right) ^{\top }b\right) \frac{1}{h}K\left( \frac{a-\alpha }{h%
}\right) da \\
&=&\frac{1}{L}\sum\limits_{l=1}^{L}\int_{-\frac{\alpha }{h}}^{\frac{1-\alpha 
}{h}}\rho _{\alpha +th}\left( B_{l}-P\left( X_{l},th\right) ^{\top }b\right)
K\left( t\right) dt,
\end{eqnarray*}%
where $b \in \mathbb{R}^{\left( s+1\right) \left( D+1\right) }$ and $P\left( x,t\right) =\pi \left( t\right) \otimes x_{1}$ with $\pi
\left( t\right) =\left[ 1,t,\ldots ,\frac{t^{s}}{s!}\right] ^{\top }$ and $%
x_{1}=\left[ 1,x^{\top }\right] ^{\top }$.\footnote{%
There is a subtle difference between our objective function
relative to GG22 here. Despite of us both assuming $\left( s+1\right)-$%
times continuous differentiability of $V\left( \cdot \right) $, we make a
polynomial approximation up to the $s-$th power term while GG22 goes up to
the $\left( s+1\right)-$th power term. This is because the quantile
function of the optimal first price auction bid has one more derivative than
the quantile function of the private value---see equation (2.4) in GG22,
which is given in the previous footnote.}

Let $b\left( \alpha \right) =\left[ \beta \left( \alpha \right) ^{\top
},\ldots ,\beta ^{\left( s\right) }\left( \alpha \right) ^{\top }\right]
^{\top }\in \mathbb{R}^{\left( s+1\right) \left( D+1\right) }$, so that $%
P\left( x,th\right) ^{\top }b\left( \alpha \right)
=\sum\limits_{j=0}^{s}x_{1}^{\top }\beta ^{\left( j\right) }\left( \alpha
\right) \frac{\left( th\right) ^{j}}{j!}$. Our estimator of $b\left( \alpha
\right) $\ is $\widehat{b}\left( \alpha \right) =\arg \min_{b}\widehat{%
\mathcal{R}}\left( b;\alpha \right) $. We estimate $\beta \left( \alpha
\right) $ by $\mathsf{S}_{0}\widehat{b}\left( \alpha \right) $, where $%
\mathsf{S}_{0}=S_{0}\otimes \mathrm{I}_{D+1}$ with $S_{0}=\left[ 1,\ldots ,0%
\right] \in \mathbb{R}^{s+1}$, so that 
\begin{equation*}
\widehat{B}\left( \alpha |x\right) =x_{1}^{\top }\mathsf{S}_{0}\widehat{b}%
\left( \alpha \right) ,
\end{equation*}%
is the estimator of $B\left( \alpha |x\right) $. We then estimate $V\left(
\alpha |x\right) $ by using equation (\ref{B to V quantile}) in Proposition
1:%
\begin{equation}
\widehat{V}\left( \alpha |x\right) =\widehat{B}\left( \phi \left( \alpha
\right) |x\right) \text{.}  \label{AQR V estimator}
\end{equation}

Since $\widehat{V}\left( \alpha
|x\right) $ inherits the properties of $\widehat{B}\left( \alpha
|x\right) $, we first provide the statistical properties of the latter as a lemma. Moreover, Lemma 1 contains pointwise properties for an AQR estimator, which is not given in GG22, as they only provide the uniform convergence rate---see Theorem D.1 in their paper.

\bigskip

\textbf{Lemma 1.} Suppose Assumption Q holds, there exists $\left(
J_{B}\left( \alpha ,x\right) ,J_{S}\left( \alpha ,x\right) ,J_{R}\left(
\alpha ,x\right) \right) $ such that for all $\alpha \in \left( 0,1\right) $
and $x\in \mathcal{X}$: 
\begin{eqnarray*}
\widehat{B}\left( \alpha |x\right) -B\left( \alpha |x\right) &=&J_{B}\left(
\alpha ,x\right) +J_{S}\left( \alpha ,x\right) +J_{R}\left( \alpha ,x\right)
,\text{ where} \\
J_{B}\left( \alpha ,x\right) &=&h^{s+1}\mathsf{Bias}_{h}\left( \alpha
,x\right) \mathsf{,} \\
E\left[ J_{S}\left( \alpha ,x\right) \right] &=&0\text{ and }\sqrt{L}%
J_{S}\left( \alpha ,x\right) \overset{d}{\rightarrow }N\left( 0,x_{1}^{\top
}\Sigma \left( \alpha \right) x_{1}\right) , \\
\sqrt{L}J_{R}\left( \alpha ,x\right) &=&o_{p}\left( 1\right) ,
\end{eqnarray*}%
for $\underline{t}_{\alpha ,h}=-\min \left( 1,\frac{\alpha }{h}\right) $, $%
\overline{t}_{\alpha ,h}=\max \left( 1,\frac{1-\alpha }{h}\right) $, 
\begin{eqnarray*}
\mathsf{Bias}_{h}\left( \alpha ,x\right) &=&B^{\left( s+1\right) }\left(
\alpha |x\right) S_{0}\Omega _{h}\left( \alpha \right) ^{-1}\int_{\underline{%
t}_{\alpha ,h}}^{\overline{t}_{\alpha ,h}}\frac{t^{s+1}\pi \left( t\right) }{%
\left( s+1\right) !}K\left( t\right) dt\text{ with} \\
\Omega _{h}\left( \alpha \right) &=&\int_{\underline{t}_{\alpha ,h}}^{%
\overline{t}_{\alpha ,h}}\pi \left( t\right) \pi \left( t\right) ^{\top
}K\left( t\right) dt,\text{ and } \\
\Sigma \left( \alpha \right) &=&\alpha \left( 1-\alpha \right) \mathbf{P}%
_{0}\left( \alpha \right) ^{-1}\mathbf{PP}_{0}\left( \alpha \right) ^{-1}%
\text{ with} \\
\mathbf{P}_{0}\left( \alpha \right) &=&E\left[ \frac{X_{l}X_{l}^{\top }}{%
B^{\left( 1\right) }\left( \alpha |X_{l}\right) }\right] \text{ and }\mathbf{%
P}=E\left[ X_{l}X_{l}^{\top }\right] \text{.}
\end{eqnarray*}%
Moreover, 
\begin{equation*}
\sup_{\left( \alpha ,x\right) \in \left[ 0,1\right] \times \mathcal{X}%
}\left\vert \widehat{B}\left( \alpha |x\right) -B\left( \alpha |x\right)
\right\vert =O_{p}\left( \sqrt{\frac{\log L}{L}}+h^{s+1}\right) .
\end{equation*}

The Appendix gives expressions for $\left(
J_{B}\left( \alpha ,x\right) ,J_{S}\left( \alpha ,x\right) ,J_{R}\left(
\alpha ,x\right) \right) $ in equations (\ref{quantile bias}) to (\ref%
{quantile remainder}). These terms respectively represent the bias, leading
stochastic term, and remainder term of $\widehat{B}\left( \alpha |x\right)
-B\left( \alpha |x\right) $. Note that the limiting distribution of $\sqrt{L}%
J_{S}\left( \alpha ,x\right) $ is the same as the standard quantile
regression estimator's without smoothing (for example, see Chapter 4 of
\cite{koenker2005quantile}), so that the AQR estimator has the same first order
asymptotic property as the Koenker and Bassett's estimator when $Lh^{2\left(s+1\right) }=o\left( 1\right) $.

Since $\widehat{V}\left( \alpha |x\right) $ is just a composite function of $%
\widehat{B}\left( \cdot \right) $ and a deterministic function $\phi \left(
\cdot \right) $, its statistical properties follow directly from Lemma 1.

\bigskip

\textbf{Proposition 3.} Suppose Assumption Q holds, for all $\alpha \in
\left( 0,1\right) $ and $x\in \mathcal{X}$: 
\begin{equation*}
\widehat{V}\left( \alpha |x\right) -V\left( \alpha |x\right) =J_{B}\left(
\phi \left( \alpha ,x\right) \right) +J_{S}\left( \phi \left( \alpha
,x\right) \right) +J_{R}\left( \phi \left( \alpha ,x\right) \right) ,
\end{equation*}%
for the same functions $\left( J_{B}\left( \alpha ,x\right) ,J_{S}\left(
\alpha ,x\right) ,J_{R}\left( \alpha ,x\right) \right) $ as in Lemma 1.
Moreover, 
\begin{equation*}
\sup_{\left( \alpha ,x\right) \in \left[ 0,1\right] \times \mathcal{X}%
}\left\vert \widehat{V}\left( \alpha |x\right) -V\left( \alpha |x\right)
\right\vert =O_{p}\left( \sqrt{\frac{\log L}{L}}+h^{s+1}\right) .
\end{equation*}

\subsection{Estimators of related functions}

To prepare for the estimation of the risk parameter in the next section, we
need to establish convergence rates for other functions related to the
quantile. Let us re-write the first-order condition in (\ref{FOC theory})
and replace $\alpha _{R}$\ with $R=V^{-1}\left( \alpha _{R}|X\right) $,
which gives:%
\begin{equation}
0=U\left( W\right) +U^{\left( 1\right) }\left( R\right) V^{\left( 1\right)
}\left( V^{-1}\left( R|X\right) |X\right) \left( 1-V^{-1}\left( R|X\right)
\right) -U\left( R\right) .  \label{FOC for SE}
\end{equation}%
Using the above equation for estimation requires estimators for $\left(
V^{\left( 1\right) }\left( \cdot \right) ,V^{-1}\left( \cdot \right) \right) 
$. Since the convergence rates for an estimator of $V^{\left( j\right)
}\left( \cdot \right) $\ is useful for semiparametric
estimation, we begin by providing convergence rates for them. In what
follows, we provide the relations between $\left( V^{\left( j\right) }\left(
\cdot \right) ,V^{-1}\left( \cdot \right) \right) $ and $\left( B^{\left(
j\right) }\left( \cdot \right) ,B^{-1}\left( \cdot \right) \right) $. Then,
we define our estimators as transformations of the AQR estimators of $\left(
B^{\left( j\right) }\left( \cdot \right) ,B^{-1}\left( \cdot \right) \right) 
$ and give their rates of convergence. 

\subsubsection{Derivatives of the quantile function}

We can differentiate (\ref{B to V quantile}) repeatedly to obtain the relationship between
the derivatives of the winning bid's and private value's quantile function.
While the relations are visually compact for lower order derivatives, such
as 
\begin{eqnarray*}
V^{\left( 1\right) }(\alpha |X) &=&\phi ^{\left( 1\right) }(\alpha
)B^{\left( 1\right) }\left( \phi (\alpha )|X\right) , \\
V^{\left( 2\right) }(\alpha |X) &=&\left( \phi ^{\left( 1\right) }(\alpha
)\right) ^{2}B^{\left( 2\right) }\left( \phi (\alpha )|X\right) +\phi
^{\left( 2\right) }(\alpha )B^{\left( 1\right) }\left( \phi (\alpha
)|X\right) ,
\end{eqnarray*}%
which hold a.s. for all $\alpha \in \left[ 0,1\right] $, they get cumbersome
quickly for higher derivatives. Since we are only interested in the uniform
convergence rates rather than pointwise properties here, it therefore suffices to know that:%
\begin{equation}
V^{\left( j\right) }(\alpha |X)=\sum_{k=1}^{j}B^{\left( k\right) }\left(
\phi (\alpha )|X\right) \mathcal{J}_{kj}\left( \phi ^{\left( 1\right)
}(\alpha ),\phi ^{\left( 2\right) }(\alpha ),\ldots ,\phi ^{\left(
j-k+1\right) }(\alpha )\right) \text{\ a.s. for all }\alpha \in \left[ 0,1%
\right] ,  \label{V - B deriv}
\end{equation}%
where $\mathcal{J}_{kj}$\ is a known continuous function that is uniformly
bounded over $\left[ 0,1\right] $\ for all $k$ and $j$.\footnote{%
This can be obtained by applying the Fa\`{a} di Bruno's formula\ for
computing chain rule to higher derivatives, for the $j$-th derivative, and $%
\mathcal{J}_{kj}\left( \cdot \right) $ is the exponential Bell
polynomial.}

The AQR approach readily estimates $B^{\left( j\right)
}(\alpha |x)$ for $j=1,\ldots ,s$, which gives%
\begin{eqnarray*}
\widehat{V}^{\left( j\right) }(\alpha |x) &=&\sum_{k=1}^{j}\widehat{B}%
^{\left( k\right) }\left( \phi (\alpha )|x\right) \mathcal{J}_{kj}\left(
\phi ^{\left( 1\right) }(\alpha ),\phi ^{\left( 2\right) }(\alpha ),\ldots
,\phi ^{\left( j-k+1\right) }(\alpha )\right) ,\text{\ where} \\
\widehat{B}^{\left( j\right) }\left( \alpha |x\right) &=&x_{1}^{\top }%
\mathsf{S}_{j}\widehat{b}\left( \alpha \right) ,\text{\ }
\end{eqnarray*}%
with $\mathsf{S}_{j}=S_{j}\otimes \mathrm{I}_{D+1}$ and $S_{j}$ is a row
vector of size $\left( s+1\right) $\ consists of $0$'s in every component
other than $1$ in its $\left( j+1\right) $-th entry. Lemma 2 gives the convergence rate for the AQR estimator of $\widehat{B}%
^{\left( j\right) }\left( \cdot \right) $.

\bigskip

\textbf{Lemma 2.} Suppose Assumption Q holds and $\lim_{L\rightarrow \infty }%
\frac{\log ^{2}L}{Lh^{2}}=0$, then for $j=1,2,\ldots ,s$:%
\begin{equation*}
\sup_{\left( \alpha ,x\right) \in \left[ 0,1\right] \times \mathcal{X}%
}\left\vert \widehat{B}^{\left( j\right) }\left( \alpha |x\right) -B^{\left(
j\right) }\left( \alpha |x\right) \right\vert =O_{p}\left( \sqrt{\frac{\log L%
}{Lh^{2j-1}}}+h^{s+1-j}\right) .
\end{equation*}

\bigskip

Notice that Lemma 2 imposes the same bandwidth condition as GG22, which
we alluded earlier. Indeed, GG22\ has given the same convergence rate as the
above when $j=1$, see equation (5.7) in their Theorem 2.\footnote{%
The order of their bias is written as $h^{s+1}$, which is a result of their
bid's quantile function having one more derivative than ours.} The component
of the convergence rate that corresponds to the stochastic term is $\sqrt{%
\frac{\log L}{Lh^{2j-1}}}$, which coincides with the usual rates of the $%
\left( j-1\right) $-th derivative of a kernel density estimator; this
finding is reassuring given the identity between the density and derivative
of the quantile. The worsening of the bias rate with higher derivatives also
mirrors standard local polynomial estimators. Since $\widehat{V}^{\left(
j\right) }\left( \cdot \right) $ is a smooth mapping from $\left\{ \widehat{B}^{\left(
k\right) }\left( \cdot \right) \right\} _{k=1}^{j}$, its rate of convergence
then follows that of $\widehat{B}^{\left( j\right) }\left( \cdot \right) $.

\bigskip

\textbf{Proposition 4.} Suppose Assumption Q holds and $\lim_{L\rightarrow
\infty }\frac{\log ^{2}L}{Lh^{2}}=0$, then for $j=1,2,\ldots ,s$:%
\begin{equation*}
\sup_{\left( \alpha ,x\right) \in \left[ 0,1\right] \times \mathcal{X}%
}\left\vert \widehat{V}^{\left( j\right) }\left( \alpha |x\right) -V^{\left(
j\right) }\left( \alpha |x\right) \right\vert =O_{p}\left( \sqrt{\frac{\log L%
}{Lh^{2j-1}}}+h^{s+1-j}\right) .
\end{equation*}

\subsubsection{Inverse of the quantile function}

The relation between the
inverse of private value and winning bid quantiles is given by the inverse
of composite functions formula applied to (\ref{B to V quantile}): 
\begin{equation}
V^{-1}(t|X)=\phi ^{-1}(B^{-1}\left( t|X\right) )\text{\ \ a.s. for all }t\in 
\mathcal{V}\text{.}  \label{V - B inverse}
\end{equation}%
We define our estimator for $V^{-1}\left( \cdot \right) $\ as follows,%
\begin{eqnarray*}
\widehat{V}^{-1}(t|x) &=&\phi ^{-1}(\widehat{B}^{-1}\left( t|x\right) ),%
\text{ for all }\left( t,x\right) \in \mathcal{V}\times \mathcal{X}\text{
where,} \\
\widehat{B}^{-1}\left( t|x\right) &=&\inf \left\{ a\in \left[ 0,1\right] :%
\widehat{B}\left( a|x\right) \geq t\right\} .
\end{eqnarray*}%
The statistical properties of $\widehat{V}^{-1}(t|x)$\ can be analyzed
through two applications of the Continuous Mapping Theorem.

First, $\widehat{B}^{-1}\left( t|x\right) $\ can be studied as the inverse
of $\widehat{B}\left( t|x\right) $ through the map $\Psi :\ell ^{\infty
}\left( \left[ 0,1\right] \right) \mapsto \ell ^{\infty }\left( \mathcal{V}%
\right) $, such that $\Psi \left( \pi \right) \left( t\right) =\inf \left\{
a\in \left[ 0,1\right] :\pi \left( a\right) \geq t\right\} $ for $t\in 
\mathcal{V}$ and $\pi \left( \cdot \right) \in \ell ^{\infty }\left( \left[
0,1\right] \right) $. We use $\ell ^{\infty }\left( \mathcal{A}\right) $%
\ to denote the space of bounded functions on $\mathcal{A}\subseteq \mathbb{R%
}$. It can be shown that the Hadamard derivative of $\Psi $\ exists and linearization methods apply (\cite{van1996weak}, Lemma 3.10.21). Particularly, the leading term in $\widehat{B}%
^{-1}\left( t|x\right) -B^{-1}\left( t|x\right) $ is, 
\begin{equation}
-\frac{\widehat{B}\left( B^{-1}\left( t|x\right) |x\right) -B\left(
B^{-1}\left( t|x\right) |x\right) }{B^{\left( 1\right) }\left( B^{-1}\left(
t|x\right) |x\right) },  \label{derivative of inverse}
\end{equation}%
when $B^{\left( 1\right) }\left( B^{-1}\left( t|x\right) |x\right) >0$.

Lemma 3 gives the statistical properties of $\widehat{B}^{-1}\left( t|x\right) $ for $t$ in the interior of $\mathcal{V}$, denoted by $int\left( 
\mathcal{V}\right) $. This result may be of independent interest, as $\widehat{B}^{-1}\left( t|x\right) $\ is a flexible yet low-dimensional general estimator for the conditional CDF.

\bigskip

\textbf{Lemma 3.} Suppose Assumption Q holds, for all $t\in int\left( 
\mathcal{V}\right) $ and $x\in \mathcal{X}$: 
\begin{eqnarray*}
\widehat{B}^{-1}\left( t|x\right) -B^{-1}\left( t|x\right) &=&H_{B}\left(
t,x\right) +H_{S}\left( t,x\right) +H_{R}\left( t,x\right) ,\text{ where} \\
H_{B}\left( t,x\right) &=&-\frac{h^{s+1}}{B^{\left( 1\right) }\left(
B^{-1}\left( t|x\right) |x\right) }\mathsf{Bias}_{h}\left( B^{-1}\left(
t|x\right) ,x\right) , \\
E\left[ H_{S}\left( t,x\right) \right] &=&0\text{ and }\sqrt{L}H_{S}\left(
t,x\right) \overset{d}{\rightarrow }N\left( 0,\frac{x_{1}^{\top }\Sigma
\left( B^{-1}\left( t|x\right) \right) x_{1}}{\left( B^{\left( 1\right)
}\left( B^{-1}\left( t|x\right) |x\right) \right) ^{2}}\right) , \\
\sqrt{L}H_{R}\left( t,x\right) &=&o_{p}\left( 1\right) ,
\end{eqnarray*}%
for the same $\mathsf{Bias}_{h}\left( \cdot \right) $ and $\Sigma \left(
\cdot \right) $\ as in Lemma 1. Moreover, 
\begin{equation*}
\sup_{\left( t,x\right) \in \mathcal{V}\times \mathcal{X}}\left\vert 
\widehat{B}^{-1}\left( t|x\right) -B^{-1}\left( t|x\right) \right\vert
=O_{p}\left( \sqrt{\frac{\log L}{L}}+h^{s+1}\right) .
\end{equation*}%

Second, we apply $\phi ^{-1}\left( \cdot \right) $ to $\widehat{B}%
^{-1}\left( \cdot \right) $. There is a potential complication when deriving
uniform convergence in this step, as $\phi ^{-1}\left( \cdot \right) $ is
continuously differentiable on $\left( 0,1\right) $ but not at the
boundaries. This can be seen from inspecting the derivative of $\phi
^{-1}\left( \cdot \right) $, which is $1/\phi ^{\left( 1\right) }\left( \phi
^{-1}\left( \cdot \right) \right) $, as we have $\phi ^{\left( 1\right)
}\left( 0\right) =0$ when $I>2$ and $\phi ^{\left( 1\right) }\left( 1\right) 
$ is $0$ for all $I$. It should be noted too that the bias and variance of $%
\widehat{B}^{-1}\left( \cdot \right) $\ go to zero as $t\rightarrow 
\underline{v}$, and also $t\rightarrow \overline{v}$\ for the variance.
These faster convergence rates may mitigate potential irregularity issues
for the de-meaned component of $\widehat{V}^{-1}\left( \cdot \right) $ as
well as the lower boundary bias. Nevertheless, we do not need these aspects
to derive the large sample properties of our risk aversion parameter in the
next section, and a comprehensive study on uniform properties of such
transformed AQR estimator is beyond the scope of this paper.

The next proposition gives the uniform convergence rate for $\widehat{V}%
^{-1}\left( \cdot \right) $\ on an inner interval of $\mathcal{V}$, denoted
by $\mathcal{V}_{\delta }$ that is defined as $\left[ \underline{v}+\delta ,%
\overline{v}-\delta \right] $ for $\delta \in \left( 0,\left( \overline{v}-%
\underline{v}\right) /2\right) $.

\textbf{Proposition 5.} Suppose Assumption Q holds, then%
\begin{equation*}
\sup_{\left( t,x\right) \in \mathcal{V}_{\delta }\times \mathcal{X}%
}\left\vert \widehat{V}^{-1}\left( t|x\right) -V^{-1}\left( t|x\right)
\right\vert =O_{p}\left( \sqrt{\frac{\log L}{L}}+h^{s+1}\right) .
\end{equation*}

\section{Risk-aversion estimator}

We now assume the utility function takes a parametric form: $\theta \mapsto
U_{\theta }\left( \cdot \right) $, for some $\theta \in \Theta \subset 
\mathbb{R}$ that represents an Arrow-Pratt measure of risk aversion. We can
then construct an objective function for estimating the risk parameter from
the first-order condition in (\ref{FOC for SE}). To do this, let us use $%
\mathcal{D}_{1}\left( \mathcal{A_{\delta }}\times \mathcal{X}\right) $\ and $%
\mathcal{D}_{2}\left( \mathcal{V_{\delta }}\times \mathcal{X}\right) $\ to
denote classes of functions whose images are $\mathcal{V}_{\delta }$\ and $%
\mathcal{A}_{\delta }$\ respectively. We denote candidates for the
derivative and inverse of the conditional quantile function of $V_{l}$ given 
$X_{l}$ by $\psi _{1}\left( \cdot \right) \in \mathcal{D}_{1}\left( \mathcal{%
A_{\delta }}\times \mathcal{X}\right) $ and $\psi _{2}\left( \cdot \right)
\in \mathcal{D}_{2}\left( \mathcal{V_{\delta }}\times \mathcal{X}\right) $
respectively. Here, we use $\mathcal{V}_{\delta }=\left[ \underline{v}%
+\delta ,\overline{v}-\delta \right] $ and $\mathcal{A}_{\delta }=\left[
\delta ,1-\delta \right] $ for small $\delta $. We restrict the support of
the quantile and valuation for the reasons discussed at the end of Section 3.
Note that $\delta $\ in\ $\mathcal{V}_{\delta }$ and $\mathcal{A}_{\delta }$
can generally be different. Moreover, the supports of (bidder's and
seller's) valuation (and the reserve price) can depend on $X$. Incorperating
these is conceptually straightforward. We forego the more general notations
for simplicity of presentation.

Consider the following real value function $q\left( z,\theta ,\psi \right) $
defined as follows: 
\begin{equation}
q\left( z,\theta ,\psi \right) =U_{\theta }\left( w\right) +U_{\theta
}^{\left( 1\right) }\left( r\right) \psi _{1}\left( \psi _{2}\left(
r,x\right) ,x\right) \left( 1-\psi _{2}\left( r,x\right) \right) -U_{\theta
}\left( r\right) ,  \label{summand}
\end{equation}%
where $z=\left( w,r,x\right) \in \mathcal{Z}=\mathcal{W}\times \mathcal{V}%
_{\delta }\times \mathcal{X}$, $\theta \in \Theta $, and $\psi \left( \cdot
\right) =\left( \psi _{1}\left( \cdot \right) ,\psi _{2}\left( \cdot \right)
\right) \in \mathcal{D}_{1}\left( \mathcal{A_{\delta }}\times \mathcal{X}%
\right) \times \mathcal{D}_{2}\left( \mathcal{V_{\delta }}\times \mathcal{X}%
\right) $. Henceforth, we compress the arguments of functions that are
parameters, i.e., $\psi \left( \cdot \right) $ to $\psi $, for notational
brevity.

Our structural assumption requires that the first order
condition in (\ref{FOC for SE}) coincides with $q\left( z,\theta _{0},\psi
_{0}\right) $ for some $\left( \theta _{0},\psi _{0}\right) $ for all $z$\
that is consistent with the auction model describe in Section 2. This is
suggestive for a minimum distance type objective function
for estimating $\theta _{0}$.

Let $P_{Z}$ denote a probability distribution of $Z=\left( W,R,X\right) $
and suppose $\left\{ Z_{l}\right\} _{l=1}^{L}$\ is a random sample drawn
from it. We define,%
\begin{equation*}
Q_{L}\left( \theta ,\psi \right) =\frac{1}{L}\sum\limits_{l=1}^{L}q\left(
Z_{l},\theta ,\psi \right) ^{2}\text{ \ and \ }Q\left( \theta ,\psi \right)
=\int q\left( Z,\theta ,\psi \right) ^{2}dP_{Z}\text{.}
\end{equation*}%
Given our usage of empirical process methods to proving the asymptotic
results, we write $Q$\ as an integral, making clear that only $Z$ is being
integrated out, so that $Q\left( \theta ,\psi \right) $ is a random variable
if either $\theta \ $or $\psi $\ is random. We add that, more precisely, $%
P_{Z}$\ can be understood as a conditional distribution for $R\in \mathcal{V}%
_{\delta }$ for the purpose of the proofs, although there is no data
truncation in practice. Related to the latter point, we emphasize that, we
are not at risk of identification loss with our minimum distance approach by
working on $\left( \mathcal{A}_{\delta },\mathcal{V}_{\delta }\right) $\
instead of $\left( \left[ 0,1\right] ,\mathcal{V}\right) $, which should be
contrasted with choosing moments in a conditional moment model (e.g., see \cite{dominguez2004consistent}), as a single value of $z\in \mathcal{Z}$\
identifies $\theta _{0}$ via (\ref{summand}).

Our estimation problem here is a semiparametric one, as we are interested in
the finite dimensional parameter $\theta _{0}$\ in the presence of infinite
dimensional nuisance functions. We denote the estimator for the latter by $%
\widehat{\psi }$, which consists of $\widehat{\psi }_{1}=\widehat{V}^{\left(
1\right) }$ and $\widehat{\psi }_{2}=\widehat{V}^{-1}$, respectively defined
as in 3.3.1 and 3.3.2 using a kernel function that satisfies Assumption
Q(iii). We then define our estimator for $\theta _{0}$, denoted by $\widehat{%
\theta }$, to be the $\arg\min_{\theta \in \Theta}
\left\lvert Q_L\bigl(\theta,\widehat{\psi}\bigr) \right\rvert$.

We impose the following conditions.

\bigskip

\textbf{Assumption S1. }

\textit{(i) }$q\left( Z,\theta _{0},\psi _{0}\right) =0$\textit{\ }$P_{Z}$%
-a.s. \textit{for some }$\theta _{0}\in \Theta $\textit{, }$\psi _{10}\in 
\mathcal{D}_{1}\left( \mathcal{A_{\delta }}\times \mathcal{X}\right) $%
\textit{,} \textit{and }$\psi _{20}\in \mathcal{D}_{2}\left( \mathcal{%
V_{\delta }}\times \mathcal{X}\right) $\textit{; }

\textit{(ii) The auction variables }$\left\{ \left( B_{l},Z_{l}\right)
\right\} _{l=1}^{L}$\textit{\ are a random sample such that }$Z_{l}\sim
P_{Z} $\textit{, and the distribution of }$\left( B_{l},Z_{l}\right) $%
\textit{\ satisfies conditions in Assumptions M1, M2(i), M2(ii), M2(iii),
and Q(i);}

\textit{(iii) }$\Theta $\textit{\ is compact, }$U_{\theta }$\textit{\ is
twice continuously differentiable with }$U_{\theta }^{\left( 1\right) }>0$%
\textit{\ and }$U_{\theta }^{\left( 2\right) }\leq 0$\textit{\ on }$\mathcal{%
V}_{\delta }$\textit{\ for all }$\theta \in \Theta $\textit{\ such that: (a)
for any }$\theta ^{\prime }>\theta $\textit{, there exists }$\zeta $ \textit{%
such that }$\zeta ^{\left( 1\right) }>0$\textit{,} $\zeta ^{\left( 2\right)
}<0$\textit{\ and }$U_{\theta ^{\prime }}=\zeta (U_{\theta })$\textit{; and
(b) }$\sup_{\theta \in \Theta }E\left[ \left\vert U_{\theta }^{\left(
j\right) }\left( R\right) \right\vert \right] <\infty $ for $j=0,1,2$\textit{%
;}

\textit{(iv) }$\mathcal{D}_{1}\left( \mathcal{A_{\delta }}\times \mathcal{X}%
\right) =\{$ $\frac{\partial }{\partial \alpha }\nu \left( \phi \left(
\alpha \right) ,x\right) $\textit{\ for }$\nu \in \mathcal{D}_{0}\left( 
\mathcal{A_{\delta }}\times \mathcal{X}\right) $\textit{\ }$\}$\textit{\ and 
}$\mathcal{D}_{2}\left( \mathcal{V_{\delta }}\times \mathcal{X}\right) =\{$%
\textit{\ }$\left( t,x\right) \mapsto \nu _{x}^{-1}\left( t\right) $\textit{%
\ for }$\left( t,x\right) \in \mathcal{V}_{\delta }\times \mathcal{X}$%
\textit{\ where }$\nu _{x}\left( \alpha \right) =\nu \left( \phi \left(
\alpha \right) ,x\right) $\textit{\ for }$\nu \in \mathcal{D}_{0}\left( 
\mathcal{A_{\delta }}\times \mathcal{X}\right) $\textit{\ }$\}$\textit{\
where }$\mathcal{D}_{0}\left( \mathcal{A_{\delta }}\times \mathcal{X}\right)
=\{$\textit{\ }$x_{1}^{\top }\mu \left( \alpha \right) $\textit{\ for }$x\in 
\mathcal{X}$\textit{\ and }$\mu :\mathcal{A}_{\delta }\rightarrow \mathbb{R}$%
\textit{\ is }$\left( s+1\right) $-\textit{times continuously differentiable
for some }$s\geq 1$ \textit{such that: (a) }$x_{1}^{\top }\mu \left( \alpha
\right) $\textit{\ takes value in }$\mathcal{V}_{\delta }$\textit{\ and is
strictly increasing in }$\alpha $\textit{; (b) }$x_{1}^{\top }\mu ^{\left(
1\right) }\left( \alpha \right) $\ \textit{is bounded away from zero and
from infinity uniformly; and (c) }$x_{1}^{\top }\mu ^{\left( 2\right)
}\left( \alpha \right) $\textit{\ is bounded away from infinity uniformly\ }$%
\}$\textit{, and the }$\phi \left( \alpha \right) $\textit{-th} \textit{%
quantile function of }$B_{l}$\textit{\ conditional on }$X_{l}$ for $\phi
\left( \alpha \right) \in \mathcal{A}_{\delta }$\textit{\ lies in }$\mathcal{%
D}_{0}\left( \mathcal{A}_{\delta }\times \mathcal{X}\right) $.

\bigskip

Assumption S1 consists of structural assumptions and regularity conditions
on the variables in the model. The condition $q\left( Z,\theta _{0},\psi _{0}\right) =0$ in S1(i) assumes
correct model specification and, together with S1(ii), they imply the
distribution of the data can be rationalized by the ascending auction model
described in Section 2. Particularly, it implies that\textit{\ }$\psi
_{10}\left( \alpha ,x\right) =V^{\left( 1\right) }\left( \alpha |x\right)
=\phi ^{\left( 1\right) }\left( \alpha \right) B^{\left( 1\right) }\left(
\phi \left( \alpha \right) |x\right) $ for $\left( \alpha ,x\right) \in 
\mathcal{A}_{\delta }\times \mathcal{X}$\ and $\psi _{20}\left( t,x\right)
=V^{-1}\left( t|x\right) =\phi ^{-1}\left( t\right) B^{-1}\left( \phi \left(
t\right) |x\right) $ for $\left( t,x\right) \in \mathcal{V}_{\delta }\times 
\mathcal{X}$. 

S1(iii) imposes a parametric assumption on the utility function that
satisfies M2(iv). Part (a) gives an interpretation for $\theta $\ to be an
Arrow--Pratt coefficient where higher $\theta $\ means higher degree of risk
aversion in the Arrow--Pratt sense as described in Section 2.2. For example,
the risk parameters in CARA and CRRA utility functions satisfy this
condition\ (\cite{ross1981some}). Part (b) is a regularity condition requiring
uniform squared integrability. In our application we use the CRRA utility
function, which is defined in (\ref{CRRA}). It should be noted that such utility
function is defined for all $\theta \in \mathbb{R}$, where $\theta >0$, $\theta=0,$ and $\theta <0$ represent risk-averse, risk-neutral, and risk-loving
preferences respectively. Under the CRRA preference, S1(iii) requires $\theta \geq 0$. The sole
purpose of this is to ensure we can apply Proposition 2, which we use to prove Lemma 4 that shows $\theta _{0}$ is identified as the minimizer of $Q$ under primitive conditions. Our estimation procedure does not restrict the parameter space to be non-negative. Importantly, suppose the implication of Lemma 4 holds as a high-level identification condition, the asymptotic theory for our estimator
applies for $\theta _{0}<0$ without any modification.\footnote{%
Our simulation study does not suggest any identification issue when $%
\theta _{0}<0$, and the estimator in the risk-loving case behaves in the
same was as the risk-neutral and risk-averse cases qualitatively. These
additional simulation results are available upon request.} We can therefore abstract away from the inference issues that arise when parameter is on
the boundary such as those discussed in \cite{andrews2001testing}. Note also that the parametric model can be enriched by embedding observables in the risk measure, for example, by replacing $\theta$ with a linear index of the seller's or auction characteristics.

S1(iv) ensures $\mathcal{D}_{1}\left( \mathcal{A}_{\delta }\times \mathcal{X}%
\right) $\ and $\mathcal{D}_{2}\left( \mathcal{V_{\delta }}\times \mathcal{X}%
\right) $\ are the correct classes of functions containing candidates of
derivative and inverse of quantile functions respectively. These classes of
functions are derived from $\mathcal{D}_{0}\left( \mathcal{A_{\delta }}%
\times \mathcal{X}\right) $ can represent quantile functions that are linear
in the regressors with additional regularity conditions: Part (a)\ ensures
quantile inverses (i.e., CDFs) exist with image in $\mathcal{A}_{\delta }$;
Part (b)\ requires the first derivatives of quantile functions are bounded
away from zero and infinity, ensuring the corresponding PDFs satisfy
Assumption M2(i);\ Part (c) imposes additional condition to ensure $\mathcal{%
Q}$\ is a Glivenko-Cantelli class of functions under $P_{Z}$.

We note that the uniform boundedness conditions imposed on $\mathcal{D}%
_{0}\left( \mathcal{A}_{\delta }\times \mathcal{X}\right) $, and
subsequently inherited by $\mathcal{D}_{1}\left( \mathcal{A}_{\delta }\times 
\mathcal{X}\right) $\ and $\mathcal{D}_{2}\left( \mathcal{V_{\delta }}\times 
\mathcal{X}\right) $, are very mild in practice. This is because the true
quantile function satisfies these conditions, and we have consistent
estimators for the derivative and inverse of the quantile function. We
therefore only need to consider $\mathcal{D}_{1}\left( \mathcal{A}_{\delta
}\times \mathcal{X}\right) $ and $\mathcal{D}_{2}\left( \mathcal{V_{\delta }}%
\times \mathcal{X}\right) $\ that contain functions in a neighborhood of $%
\left( \psi _{10},\psi _{20}\right) $. In a similar vein, the requirement
for the image of functions in $\mathcal{D}_{0}\left( \mathcal{A}_{\delta
}\times \mathcal{X}\right) $\ and $\mathcal{D}_{2}\left( \mathcal{V}_{\delta
}\times \mathcal{X}\right) $\ to respectively be $\mathcal{V}_{\delta }$\
and $\mathcal{A}_{\delta }$\ is not restrictive for a fixed $\delta $, as
our estimators for $\left( \psi _{10},\psi _{20}\right) $ converge to the
true functions uniformly over any inner subset of their respective supports.
Given the boundedness of functions involved, we shall use $\left\Vert \cdot
\right\Vert _{\infty }$ to generically denote the sup-norm of functions over
their supports.

Building on the result of Proposition 2, Lemma 4 says our population
objective function has a well separated minimum at $\theta _{0}$\ when $\psi
=\psi _{0}$ under S1.

\bigskip

\textbf{Lemma 4.} Suppose Assumption S1 holds, then for all $\epsilon >0$,
there exists $\delta >0$ such that $\inf_{\left\vert \theta -\theta
_{0}\right\vert >\epsilon }Q\left( \theta ,\psi _{0}\right) \geq Q\left(
\theta _{0},\psi _{0}\right) +\delta $.

\bigskip

Given the well-separated minimum condition on the population objective
function, it is well-known consistency of $\widehat{\theta }$\ will follow
if $Q_{L}\left( \theta ,\widehat{\psi }\right) $ converges to $Q\left(
\theta ,\psi _{0}\right) $ uniformly over $\Theta $ in probability (e.g., see Theorem 2.1 in \cite{newey1994large}). Lemma 5 states we have the
desired uniform convergence under the bandwidth conditions that ensure $%
\left\Vert \widehat{\psi }_{i}-\psi _{i0}\right\Vert _{\infty }=o_{p}\left(
1\right) $ for $i=1,2$.

\bigskip

\textbf{Lemma 5.} Suppose Assumption S1 holds and the bandwidth satisfies $%
h=o\left( 1\right) $ and $\log ^{2}L=o\left( Lh^{2}\right) $, then $%
\sup_{\theta \in \Theta }\left\vert Q_{L}\left( \theta ,\widehat{\psi }%
\right) -Q\left( \theta ,\psi _{0}\right) \right\vert =o_{p}\left( 1\right) $%
.

\bigskip

\textbf{Theorem 1.} Suppose Assumption S1 holds and the bandwidth satisfies $%
h=o\left( 1\right) $ and $\log ^{2}L=o\left( Lh^{2}\right) $, then $\widehat{%
\theta }=\theta _{0}+o_{p}\left( 1\right) $.

\bigskip

Our risk-aversion estimator satisfies $\frac{\partial }{\partial \theta }%
Q_{L}\left( \widehat{\theta },\widehat{\psi }\right) =0$. Its limiting distribution can be studied from the linearization of $\frac{\partial }{\partial \theta }Q_{L}\left( \widehat{\theta },\widehat{\psi }\right)$ around $\left( \theta _{0},\psi _{0}\right) $. We make the following additional assumptions to establish the limiting distribution of $\widehat{\theta}$.

\bigskip

\textbf{Assumption S2. }

\textit{(i) }$U_{\theta }$\textit{\ is three times continuously
differentiable with }$\sup_{\theta \in \Theta }E\left[ \left\vert U_{\theta
}^{\left( j\right) }\left( R\right) \right\vert ^{2}\right] <\infty $ for $%
j=0,1,2,3$\textit{;}

\textit{(ii) }$\mathcal{D}_{0}\left( \mathcal{A}_{\delta }\times \mathcal{X}%
\right) $\textit{\ is as described in S1(iv) other than }$\mu $ \textit{is }$%
\left( s+2\right) $-\textit{times continuously differentiable for some }$%
s\geq 1$ \textit{and there exists an enveloping function for }$\left( \alpha
,x\right) \mapsto x_{1}^{\top }\mu ^{\left( 3\right) }\left( \alpha \right) $%
\textit{\ that is }$L_{2}\left( P_{Z_{0}}\right) $\textit{-integrable;\ }

\textit{(iii) }$E\left[ \frac{\partial }{\partial \theta }q\left(
Z_{l},\theta _{0},\psi _{0}\right) ^{2}\right] $\textit{\ is invertible.}

\bigskip

The strengthened smoothness and moment conditions in S2(i) and S2(ii) ensure
that $\mathcal{Q}$\ and the related class of functions are $P_{Z}-$Donsker
and allow us to bound various moments in the proof. S2(iii) is the
invertible Hessian condition.

A key step in deriving the distribution theory of our estimator involves taking the pathwise derivative
of $\frac{\partial }{\partial \theta }Q_{L}\left(
\theta _{0},\widehat{\psi }\right) $ at $\psi _{0}$\ in direction $\left[ 
\widehat{\psi }-\psi _{0}\right] $. Two main ingredients for
obtaining a $\sqrt{L}-$consistent semiparametric estimator are that (i) the
higher-order terms in the linearization are negligible at the $L^{-1/2}$
rate, and (ii) the leading term of the linearization is asymptotically
normal. With a nuisance function estimated by kernel smoothing, the former
can be achieved by appropriate bandwidth choice to control the bias. For
asymptotic normality involving nuisance functions, one can obtain a
parametric convergence rate by averaging nonparametric estimators, thereby
increasing the convergence speed; this intuition can be made transparent via the
Riesz representer of the pathwise derivative as an integral (e.g., see \cite{newey1994asymptotic} and \cite{chen2003estimation}). The next lemma
states sufficient conditions on the bandwidth and on the integral
representation of the pathwise derivative under which the estimator is $%
\sqrt{L}-$ asymptotically normal.

\bigskip

\textbf{Lemma 6.} Suppose Assumption S1 and S2 hold, and the bandwidth
satisfies $Lh^{4s}=o\left( 1\right) $ and $\log ^{2}L=o\left( Lh^{3}\right) $%
, then the linearization of $\frac{\partial }{\partial \theta }Q\left(
\theta _{0},\widehat{\psi }\right) $ at $\psi _{0}$\ in direction $\left[ 
\widehat{\psi }-\psi _{0}\right] $ is 
\begin{eqnarray*}
&&\int c_{Y_{1}}\left( \widehat{\psi }_{1}-\psi _{10}\right) \circ \psi
_{20}+c_{Y_{2}}\left( \widehat{\psi }_{2}-\psi _{20}\right)
dP_{Z}+o_{p}\left( \frac{1}{\sqrt{L}}\right) \text{ where,} \\
&&\sqrt{L}\int c_{Y_{1}}\left( \widehat{\psi }_{1}-\psi _{10}\right) \circ
\psi _{20}+c_{Y_{2}}\left( \widehat{\psi }_{2}-\psi _{20}\right) dP_{Z}%
\overset{d}{\rightarrow }N\left( 0,\sigma _{0}^{2}\right) \text{ for some }%
\sigma _{0}^{2}\left. >\right. 0\text{,}
\end{eqnarray*}%
where $c_{Y_{1}}$\ and $c_{Y_{2}}$\ are functions of $Z$, and $\circ $\
denotes the composition of functions. See equations (\ref{cY1}) and (\ref{cY2}) in
the Appendix respectively for the explicit forms of $c_{Y_{1}}$\ and $c_{Y_{2}}$.

\bigskip

The integral in the display of Lemma 6 is precisely the pathwise derivative of $\frac{%
\partial }{\partial \theta }Q\left( \theta _{0},\widehat{\psi }\right) $ at $%
\psi _{0}$\ in direction $\left[ \widehat{\psi }-\psi _{0}\right] $. As
mentioned earlier, $P_{Z}$\ only integrates out $Z$ and the integral
functional is a random variable due to $\widehat{\psi }$. The bandwidth
restriction in the lemma ensures the higher order term from linearizing $%
\frac{\partial }{\partial \theta }Q\left( \theta _{0},\widehat{\psi }\right) 
$ is $o\left( L^{-1/2}\right) $. We show in the Appendix that the leading
higher order term is due to $\left\Vert \widehat{\psi }_{1}^{\left( 1\right)
}-\psi _{10}^{\left( 1\right) }\right\Vert _{\infty }\times \left\Vert 
\widehat{\psi }_{2}-\psi _{20}\right\Vert _{\infty }=O_{p}\left( \frac{\log L%
}{Lh^{3/2}}+h^{2s}\right) $.

It is worth noting that our bandwidth requirement of $\log ^{2}L=o\left(
Lh^{3}\right) $\ coincides with the condition in Theorem 4 of GG22, which
establishes asymptotic normality for a similar integral functional that contains the
bid's quantile function and its derivative. This is reassuring, as they estimate their functional using AQR estimators that have the
same convergence rates as $\widehat{\psi }_{20}$ and $\widehat{\psi }_{10}$. For example, when the bandwidth decays at the rate $\log
L^{1/2}L^{-\varsigma }$\ for some $\varsigma >0$, Lemma 6 requires $\frac{1}{4s}%
<\varsigma <\frac{1}{3}$ for it to hold.

\bigskip

\textbf{Theorem 2.} Suppose Assumptions S1 and S2 hold, the bandwidth
satisfies $Lh^{4s}=o\left( 1\right) $ and $\log ^{2}L=o\left( Lh^{3}\right) $%
, and for the same $\sigma _{0}^{2}$ in Lemma 6,%
\begin{equation*}
\sqrt{L}\left( \widehat{\theta }-\theta _{0}\right) \overset{d}{\rightarrow }%
N\left( 0,\frac{\sigma _{0}^{2}}{\left( E\left[ \frac{\partial }{\partial
\theta }q\left( Z_{l},\theta _{0},\psi _{0}\right) ^{2}\right] \right) ^{2}}%
\right) .
\end{equation*}

In practice, inference on $\theta _{0}$ be performed
using a nonparametric bootstrap. \cite{chen2003estimation} provide
high-level conditions under which the asymptotic distribution of a two-step
semiparametric estimator can be consistently bootstrapped. We conjecture
that our setup is amenable to such a result. Indeed, we apply this in our Monte
Carlo study and find that bootstrap standard errors for $\widehat{\theta }$\
perform very well. A formal proof of bootstrap validity when AQR estimators
are used to estimate nuisance functions in a general two-step semiparametric
procedure is, however, beyond the scope of this paper.

\section{Simulation}

In this section, we consider some finite sample properties of our AQR and
semiparametric estimators proposed in the paper. 

\subsection{Simulation design}

Taking inspirations from \cite{gimenes2017econometrics} and GG22, the private-value quantile function is specified as follows: 
\begin{eqnarray*}
V(\alpha |X) &=& \gamma _{0}(\alpha )+\gamma _{1}(\alpha )X_{1}+\gamma
_{2}(\alpha )X_{2} \text{, where} \\
\gamma_0(\alpha) &=& -\log (1-(1-1/e) \alpha), \\
\gamma_1(\alpha) &=& 1, \\
\gamma_2(\alpha) &=& 1-\exp (-5\alpha).
\end{eqnarray*}

The quantile functions $\gamma _{0}$, $\gamma _{1}$, $%
\gamma _{2}$ are all strictly increasing, while $\gamma _{0}$ is convex, $\gamma _{1}$ is linear, and $\gamma _{2}$
is concave. The covariates $X_{1}$ and $X_{2}$ are independent and
uniformly distributed on $[0,1]$. For the outside option, we let $W(X)=V(\beta |X)$, where $\beta$ is an independent draw from a uniform distribution on $[0.05,0.5]$.

We use the CRRA utility function and consider $\theta _{0}=0,0.5,1$. Under this specification, $\theta_{0} >0$ means the seller is risk-averse and $\theta_{0}=0$ means the seller is risk-neutral (and $\theta_{0} <0$ means the seller is risk-loving).

We consider $L$ auctions with $3$ bidders, where $L=250,500,1000$. In the estimation, AQR quantile functions are computed over the estimation grid of $\alpha$ taking values $0.02,0.04,...,0.98$. Following GG22, we set the AQR polynomial order to be 2 and estimate it using the Epanechnikov kernel: $K(t)=0.75(1-t^{2})\mathbf{1}(t\in \lbrack -1,1])$. Three different bandwidths are used in the simulation: $h=sL^{-1/5},sL^{-1/6},sL^{-1/7}$, where $s$ is the sample standard deviation of the winning bids. The number of replications is 1000 in all experiments. 

\subsection{Simulation results}

Figure 1 collects the simulation results for the estimation of the private-value quantile function and its derivative for different $L$. The black solid line is the true function, the red dashed line is the mean of our estimator, and the red dotted lines represent its $2.5$th and $97.5$th percentiles. In all these figures, we set $X_1=X_2=0.5$. Here we only report the estimation results of $V(\alpha|X)$ and $V^{(1)}(\alpha|X)$ using $h=sL^{-1/6}$. The results using $h=sL^{-1/5}$ and $h=sL^{-1/7}$ are similar.

We can see our estimator of $V$, on average, lies very close to the true, within the $95\%$ central quantile interval, and with decreasing variance as $L$ increases. The estimator of $V^{(1)}$ shares these properties but has higher estimation error, which is what we expect from the theory. To illustrate the differences more quantitatively, we calculate the integrated mean squared error (IMSE) for $\widehat{V}$ and $\widehat{V}^{(1)}$ when $X_1=X_2=0.5$. These can be found in the Table 1.

\begin{figure}[H]
\centering
\includegraphics[width=7.5cm,height=6cm]{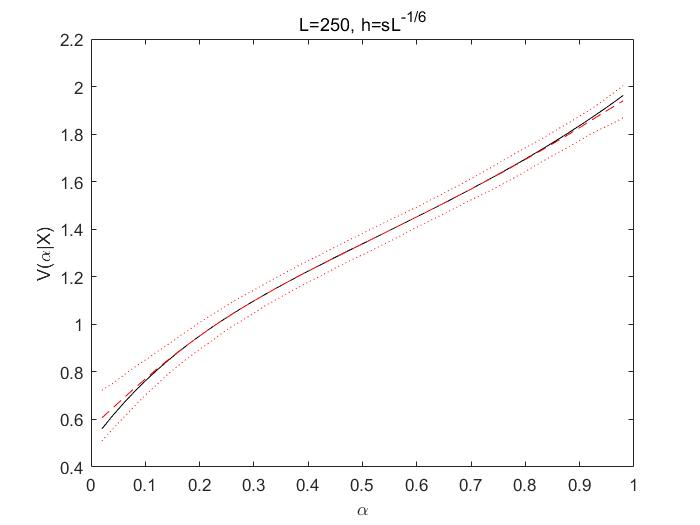}
\includegraphics[width=7.5cm,height=6cm]{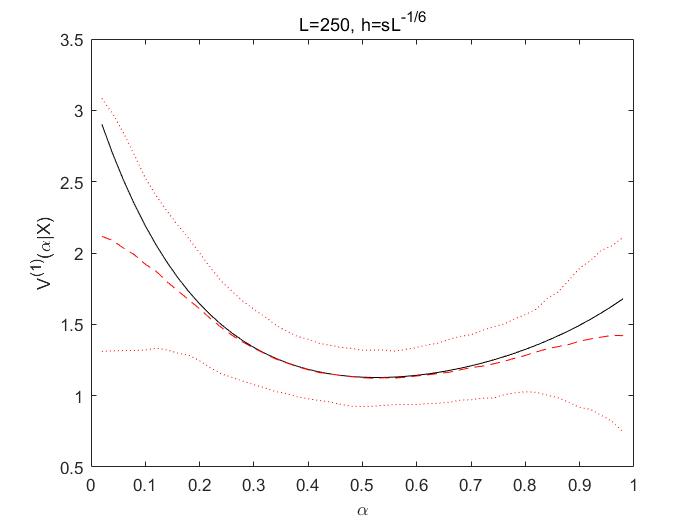}
\includegraphics[width=7.5cm,height=6cm]{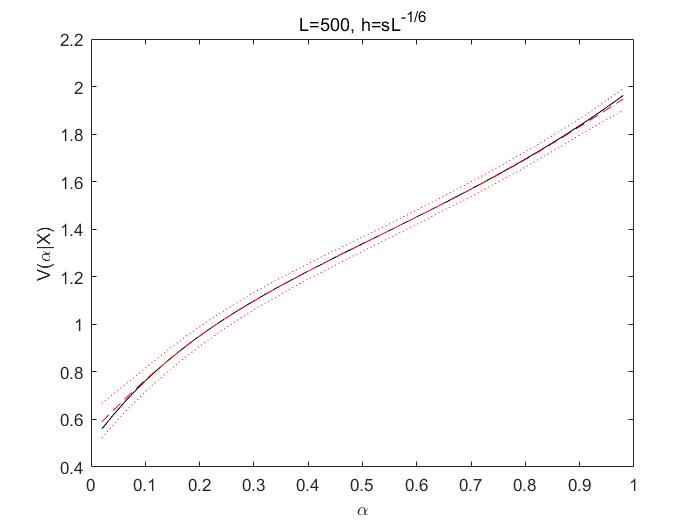}
\includegraphics[width=7.5cm,height=6cm]{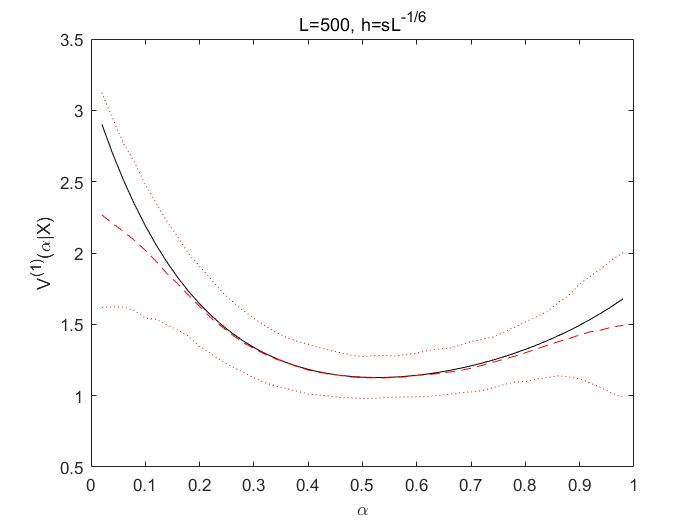}
\includegraphics[width=7.5cm,height=6cm]{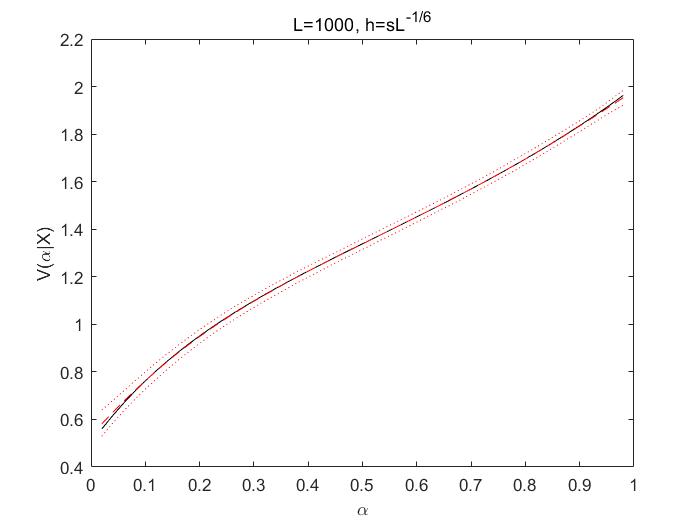}
\includegraphics[width=7.5cm,height=6cm]{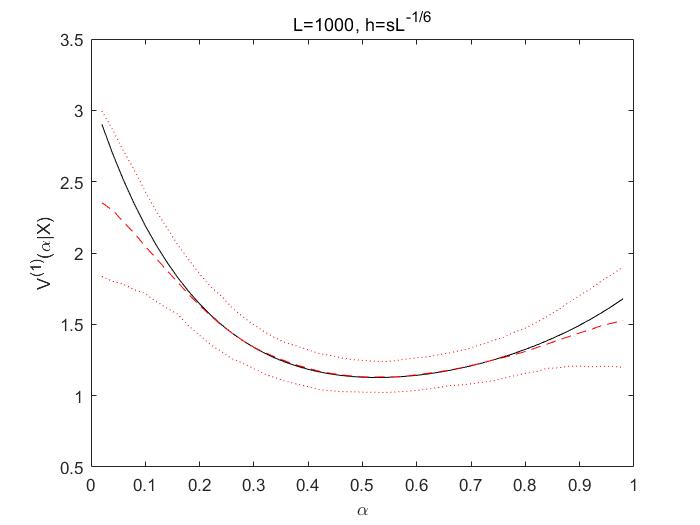}
\caption{Simulation results for $\widehat{V}$ and $\widehat{V}^{(1)}$.}
\end{figure}

\begin{table}[H]
\begin{center}
\caption{IMSE for $\widehat{V}$ and $\widehat{V}^{(1)}$.}
\begin{tabular}{cccc}
$h$ & $L$ & IMSE for $\widehat{V}$ & IMSE for $\widehat{V}^{(1)}$ \\
\hline \hline
\multirow{3}{*}{$sL^{-1/5}$}
& 250 & $8.88 \times 10^{-4}$ & 0.0808 \\
& 500 & $4.19 \times 10^{-4}$ & 0.0475 \\
& 1000 & $2.13 \times 10^{-4}$ & 0.0315 \\
\hline
\multirow{3}{*}{$sL^{-1/6}$}
& 250 & $8.85 \times 10^{-4}$ & 0.0733 \\
& 500 & $4.19 \times 10^{-4}$ & 0.0422 \\
& 1000 & $2.11 \times 10^{-4}$ & 0.0267 \\
\hline
\multirow{3}{*}{$sL^{-1/7}$}
& 250 & $8.84 \times 10^{-4}$ & 0.0694 \\
& 500 & $4.21 \times 10^{-4}$ & 0.0391 \\
& 1000 & $2.13 \times 10^{-4}$ & 0.0246 \\
\hline
\end{tabular}
\end{center}
\end{table}

For the statistical performance of $\widehat{\theta}$, this is tabulated in Table 2, which contains the bias (\texttt{bias}), median bias (\texttt{mbias}), standard deviation (\texttt{std}), bootstrap standard error (\texttt{b-se}), mean squared error (\texttt{mse}), and the scaled interquartile range (\texttt{iqr}) of the estimator. In particular, we use the nonparametric bootstrap to estimate the bootstrap standard error. We do this by resampling each simulated dataset with replacement and re-do the estimation procedure $99$ times. And we calculate \texttt{iqr} by dividing the interquartile range of our studentized estimates by 1.349.

The results give evidence that our $\widehat{\theta}$ is a consistent estimator for $\theta_0$, as the bias and standard deviation, and subsequently mean squared error, are decreasing with $L$. The reported \texttt{iqr} being close to $1$ indicates that our estimator has a normal-like tail behavior. The bootstrap standard error also approximates the standard deviation well. These comments apply to all bandwidths considered and the performances across bandwidths are comparable. Our simulation study thus supports our theoretical results and the recommendation that inference on the risk parameter can be done by using the nonparametric bootstrap.

\begin{table}[H]
\caption{Estimation results for $\widehat{\theta}$.}
\begin{center}
\begin{tabular}{ccccccccc}
$\theta_{true}$ & $h$ & $L$ & bias & mbias & std & b-se & mse & iqr  \\
\hline \hline
\multirow{9}{*}{$0$} 
& \multirow{3}{*}{$sL^{-1/5}$} 
& 250 & 0.1675 & 0.1881 & 0.2969 & 0.2789 & 0.1161 & 0.9857 \\
& & 500 & 0.1063 & 0.1121 & 0.2193 & 0.2163 & 0.0593 & 0.9668 \\
& & 1000 & 0.0686 & 0.0751 & 0.1672 & 0.1632 & 0.0326 & 0.9976 \\
\cline{2-9}
& \multirow{3}{*}{$sL^{-1/6}$} 
& 250 & 0.1547 & 0.1811 & 0.2910 & 0.2731 & 0.1086 & 0.9764 \\
& & 500 & 0.1016 & 0.1041 & 0.2123 & 0.2100 & 0.0553 & 0.9459 \\
& & 1000 & 0.0706 & 0.0776 & 0.1615 & 0.1575 & 0.0311 & 0.9934 \\
\cline{2-9}
& \multirow{3}{*}{$sL^{-1/7}$} 
& 250 & 0.1501 & 0.1738 & 0.2859 & 0.2683 & 0.1042 & 0.9722 \\
& & 500 & 0.0990 & 0.1037 & 0.2044 & 0.2046 & 0.0515 & 0.9520 \\
& & 1000 & 0.0728 & 0.0776 & 0.1558 & 0.1522 & 0.0296 & 0.9944 \\
\hline
\multirow{9}{*}{$0.5$} 
& \multirow{3}{*}{$sL^{-1/5}$} 
& 250 & 0.0994 & 0.1182 & 0.2781 & 0.2728 & 0.0871 & 0.9894 \\
& & 500 & 0.0588 & 0.0535 & 0.1994 & 0.2043 & 0.0432 & 0.9766 \\
& & 1000 & 0.0376 & 0.0360 & 0.1504 & 0.1508 & 0.0240 & 0.9888 \\
\cline{2-9}
& \multirow{3}{*}{$sL^{-1/6}$} 
& 250 & 0.0979 & 0.1194 & 0.2701 & 0.2646 & 0.0825 & 0.9812 \\
& & 500 & 0.0610 & 0.0577 & 0.1929 & 0.1970 & 0.0409 & 0.9881 \\
& & 1000 & 0.0444 & 0.0459 & 0.1441 & 0.1450 & 0.0227 & 0.9904 \\
\cline{2-9}
&\multirow{3}{*}{$sL^{-1/7}$} 
& 250 & 0.1007 & 0.1177 & 0.2638 & 0.2581 & 0.0797 & 0.9731 \\
& & 500 & 0.0640 & 0.0614 & 0.1852 & 0.1907 & 0.0384 & 0.9638 \\
& & 1000 & 0.0496 & 0.0495 & 0.1384 & 0.1394 & 0.0216 & 1.0292 \\
\hline
\multirow{9}{*}{$1$} 
& \multirow{3}{*}{$sL^{-1/5}$} 
& 250 & 0.0487 & 0.0547 & 0.2782 & 0.2918 & 0.0797 & 0.9706 \\
& & 500 & 0.0186 & 0.0107 & 0.1934 & 0.2093 & 0.0377 & 0.9834 \\
& & 1000 & 0.0153 & 0.0154 & 0.1440 & 0.1516 & 0.0210 & 0.9727 \\
\cline{2-9}
& \multirow{3}{*}{$sL^{-1/6}$} 
& 250 & 0.0559 & 0.0674 & 0.2680 & 0.2792 & 0.0749 & 0.9753 \\
& & 500 & 0.0269 & 0.0205 & 0.1879 & 0.1997 & 0.0360 & 0.9700 \\
& & 1000 & 0.0249 & 0.0195 & 0.1378 & 0.1441 & 0.0196 & 0.9963 \\
\cline{2-9}
& \multirow{3}{*}{$sL^{-1/7}$} 
& 250 & 0.0636 & 0.0728 & 0.2614 & 0.2693 & 0.0723 & 0.9723 \\
& & 500 & 0.0348 & 0.0287 & 0.1818 & 0.1917 & 0.0342 & 0.9397 \\
& & 1000 & 0.0315 & 0.0266 & 0.1327 & 0.1378 & 0.0186 & 1.0023 \\
\hline 
\end{tabular}
\end{center}
\end{table}

\section{Empirical illustration}

This section applies our estimator to real estate auction
data from S\~{a}o Paulo. Real estate auctions constitute a large and active market in Brazil. Our sample consists of foreclosure apartments that we webscrapped from the website of a single large auctioneer, Zukerman (\texttt{https://www.zukerman.com.br}), which is recognized as the largest real estate auction platform in Brazil. The sellers are typically private and public banks, private companies that provide funds for borrowers, and the Court of Justice of the State of S\~{a}o Paulo (Tribunal de Justi\c{c}a do Estado de S\~{a}o Paulo, TJ-SP).

Properties can be auctioned off for several legal reasons: (i) default on mortgage payments for more than six months; (ii) default on condominium maintenance fees; (iii) labor-related lawsuits; and (iv) other unpaid debt obligations. Auctions in categories (i) and (ii) are the most prevalent. Type (i) cases are classified as extrajudicial because the auction does not require judicial, or court, approval and are most often initiated by financial institutions when a property serving as collateral under fiduciary alienation is repossessed after borrower default. All other types generally require authorization by a court and typically consider the cases of unpaid loans, bankruptcy, or overdue condominium fees.

Our application involves auctions that were completed over two rounds. The second round takes place if the property is not sold initially, in which case the reserve price is reduced as judicial auctions typically applied a 50\% discount on the reserve price and extrajudicial auctions often set the second-round reserve price to the outstanding debt. To apply our methodology, we take the reserve price in the first round to be the seller's optimal reserve price. This is motivated by the fact that the first reserve is set based on actual appraisals of the property value, while the discounted reserve price in the second round is guided by either a judicial mandate or the property debt.

All sales follow the English auction format, in which bids are placed electronically in ascending order. The auction details including property information and auction schedule are publically available. The auction platform publishes all submitted bids in real time. We note that the second rounds, on average, take place 20 days after the first round. Our sample consists only of auctions with at least two bidders. We assume that all potential bidders submitted bids, which means we observe $N$ that corresponds to the number of different bidders in each auction.\footnote{The assumptions on observing $N$ is a common one in practice, although models with endogenous bidders' entry or unknown number of bidders have been studied in the literature.}

The data we use contains 754 observations covering the period from 2017 to 2023, with the majority corresponding to auctions of type (i) and (ii). We did not collect data for cases of type (iii), as labor courts manage these auctions separately under distinct rules. The sample also excluded auctions deemed as outliers in terms of the reserve price, which we define as having the ratio of the reserve price to the size of the apartment in each auction that is larger than the 99th percentile or smaller than the 1st percentile of the sample. The price of the apartment is in Brazillian reais, and we measure it in R\$$100,000$s.\footnote{All prices are expressed in constant January 2017 R\$.} We use the size of the apartment (total area measured in sqm) as the covariate, $X$. To compute the value of the seller's outside option, $W$, we subtract from the evaluation value of the property the debt registered in its sale report. We only have these information for 341 properties of the total sample. Since we do not use these information to estimate the quantile function of bidder's private value, the quantiles are estimated using the whole sample. We then use the subsample with evaluation and debt values to estimate the seller's risk parameter. Throughout this exercise, we assume the seller has a CRRA utility function.

Our estimation procedure then takes place over two stages. First, we estimate the bidder's private value quantile function for the property. In a second step, we evaluate the seller's CRRA risk parameter. We  estimate $V(\alpha|X)$ using the AQR method for  $\alpha=0.01,0.02,...,0.99$.\footnote{In a very small proportion of auctions, the observed reserve price is larger
than $\widehat{V}(0.99|X)$ or smaller than $\widehat{V}(0.01|X)$. These auctions are removed when estimating $\theta$ and during the model fit analysis.} We consider three different bandwidths in our estimation: $h=0.1$, $h=0.15$, and $h=0.2$. The results show that our estimation methodology is robust for different bandwidth choices. To derive the standard error and some quantile levels of $\widehat{\theta}$, we use the nonparametric bootstrap. The bootstrap size is set at 99.

\subsection{Descriptive statistics}

We provide some descriptive statistics of our data. The variables involved are the reserve price ($R$), the winning bid ($B$), the size of the
property ($X$), the outside option value ($W$), and the number of
bidders in each auction ($N$).
The summary statistics of the data containing the mean, standard deviation, and the quartiles are shown in Table 3.
\begin{table}[H]
\caption{Descriptive statistics of the real estate data.}
\begin{center}
\begin{tabular}{ccccccc}
variable & mean & median & 25-th pct& 75-th pct& std & observations \\ \hline \hline
$R$ & 5.2990 & 3.5952 & 2.2305 & 6.1361 & 5.7411 & 754 \\ 
$B$ & 3.9331 & 2.7178 & 1.6736 & 4.7758 & 4.0066 & 754 \\ 
$W$ & 3.8883 & 2.5513 & 1.4385 & 4.4816 & 4.5412 & 341 \\ 
$X$ & 146.48 & 113.09 & 83.40 & 165.56 & 109.41 & 754 \\ 
$N$ & 5.8753 & 5 & 3 & 8 & 4.1901 & 754 \\ \hline
\end{tabular}
\end{center}
\end{table}

Note that the reserve prices in the data tend to be higher than the winning bids, which is due to sales occurring in the second auction with lowered reserves. Observing bids below the reserve facilitates identification of the bidder's valuations in the same manner to auctions with a secret or hidden reserve price (\cite{elyakime1994first}, \cite{andreyanov2022secret}).

We provide the scatter plots of $X$ against $B$ and $R$ in Figure 2. The plots indicate a general trend that both winning bid and reserve price increase with the size of the property.
\begin{figure}[H]
\centering
\includegraphics[width=7.5cm,height=6cm]{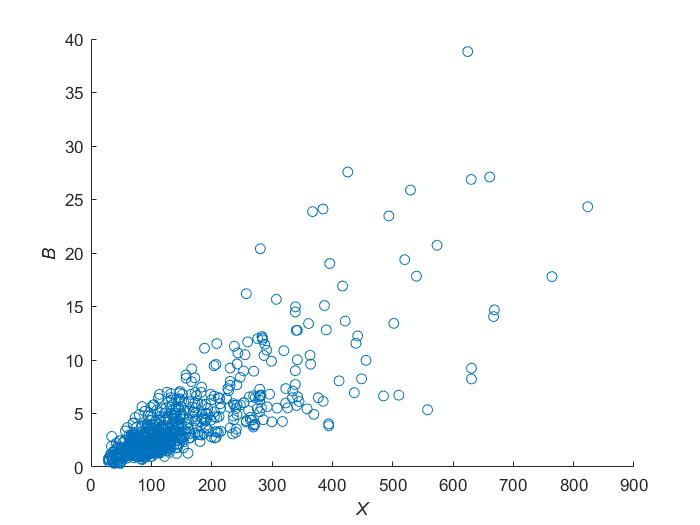}
\includegraphics[width=7.5cm,height=6cm]{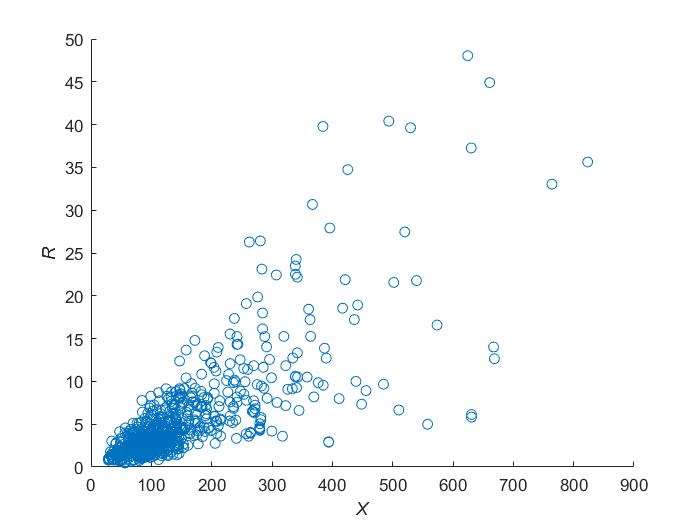}
\caption{Scatter figures of $X$, $B$, and $R$.}
\end{figure}

\subsection{Estimation results}

We start by presenting figures of the estimates of the private value quantile function conditioning for property size at the three quartiles for different bandwidths. 
\begin{figure}[H]
\centering
\includegraphics[width=5.5cm,height=4.4cm]{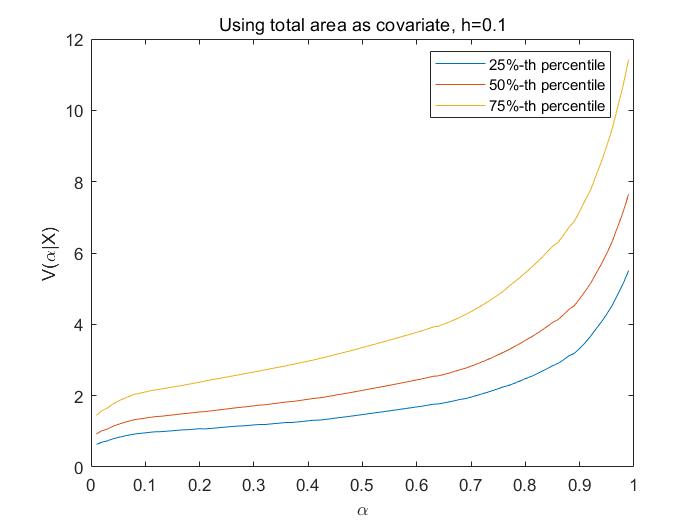}
\includegraphics[width=5.5cm,height=4.4cm]{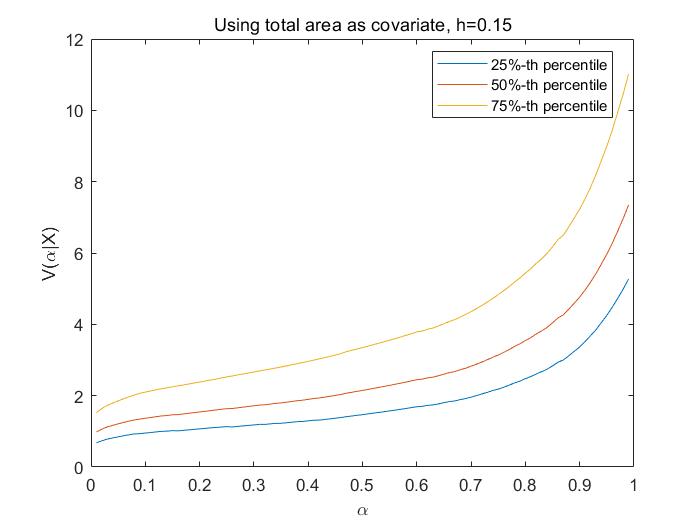}
\includegraphics[width=5.5cm,height=4.4cm]{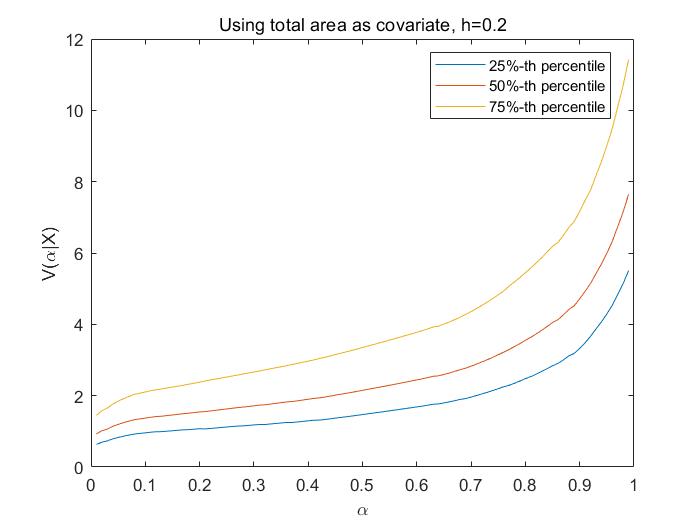}
\caption{Plots of $\widehat{V}(\cdot|X)$.}
\end{figure}
\noindent From the figures, we can see that $\widehat{V}(\alpha |X)$ is
increasing in $\alpha$, slightly concave for small $\alpha $, and convex for large $\alpha 
$ for different $X$'s. The results using different $h$ are similar.

The estimation results for the risk-aversion parameter are given in Table 4, containing the bootstrap standard errors and the 2.5th and 97.5th percentiles.

\begin{table}[H]
\caption{Results on $\widehat{\theta}$.}
\begin{center}
\begin{tabular}{ccccc}
bandwidth & $\widehat{\theta}$ & b-se & 2.5-th pct & 97.5-th pct\\ 
\hline \hline
$h=0.1$ & 1.6025 & 0.2284 & 1.2373 & 2.0907 \\ 
$h=0.15$ & 1.5399 & 0.2538 & 1.1994 & 2.1625 \\ 
$h=0.2$ & 1.5242 & 0.2549 & 1.1387 & 2.0073 \\ 
\hline
\end{tabular}
\end{center}
\end{table}
The results are qualitatively the same for all bandwidths. The $95\%$ bootstrapped coverage does not contain zero and the studentized statistic rejects the risk neutrality assumption in favor of risk aversion at any reasonable significant level.

\subsection{Model fit}

It is also instructive to consider the model fit under risk aversion. To do this, we simulate the winning bid and estimate the
optimal reserve price using our estimated parameters and compute their CDFs
to compare with the observed data. For the winning bid, we use the estimated $\widehat{V}(\alpha |X)$ to simulate the winning bid for each observed pair of ($N$, $X$) 1000 times, then combine the simulated winning bids across all
different pairs of ($N,X$) in the sample to get a simulated CDF. To
obtain the sample CDF, we use the empirical distribution of the observed
winning bid data in the sample. The following figures show the comparison
between the sample winning bid distribution and the simulated winning bid
distribution using different bandwidths:
\begin{figure}[H]
\centering
\includegraphics[width=5.5cm,height=4.4cm]{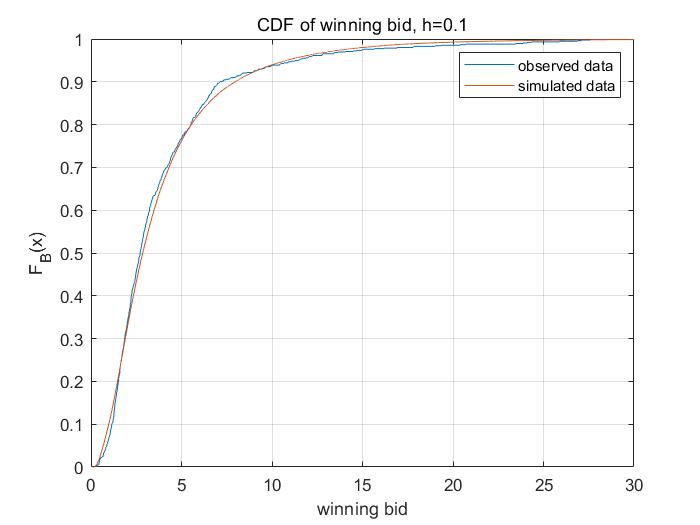}
\includegraphics[width=5.5cm,height=4.4cm]{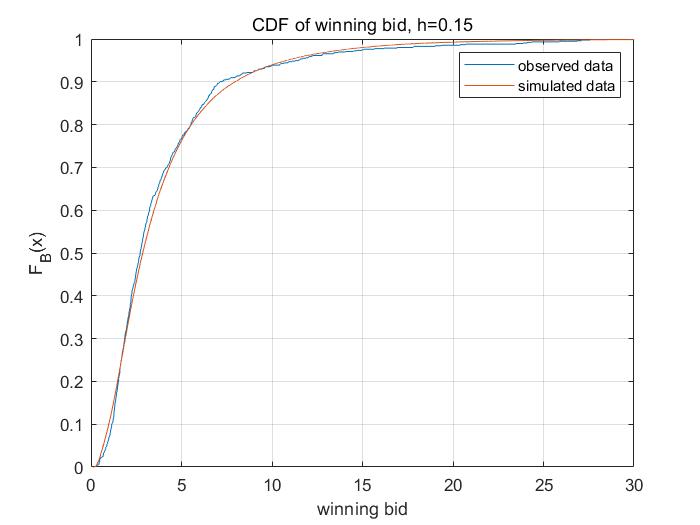}
\includegraphics[width=5.5cm,height=4.4cm]{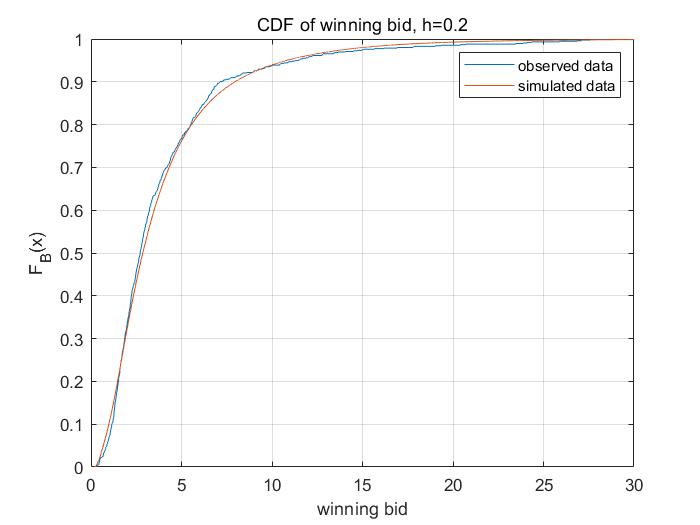}
\caption{Model fit of winning bid distribution.}
\end{figure}
Visually, we can see that the simulated winning bid distributions fit the
sample winning bid distributions very well for all bandwidths. This is complemented by the statistics in Table 5, which contains the bias
(mean of simulated winning bids minus mean of sample winning bids), the
percentage bias (bias divided by mean of sample winning bids), and the IMSE which is defined by 
\begin{equation*}
IMSE_{B}=\int_{supp(B)} \left[F_{B}(x)-\widehat{F}_{B}(x)\right]^{2}dx,
\end{equation*}
where $F_{B}(\cdot)$ is the CDF of the sample winning bids and $\widehat{F}_{B}(\cdot)$ is the CDF of the simulated winning bids. Table 5 shows that the percentage bias and the IMSE are all small, which suggest the fit of winning bid distribution is very good.
\begin{table}[H]
\caption{Model fit of $\widehat{F}_{B}$.}
\begin{center}
\begin{tabular}{cccc}
$h$ & bias & percentage bias & IMSE$_B$ \\ 
\hline \hline
0.1 & -0.0192 & -0.49\% & 0.0045 \\ 
0.15 & -0.0192 & -0.49\% & 0.0044 \\ 
0.2 & -0.0137 & -0.35\% & 0.0045 \\ 
\hline
\end{tabular}
\end{center}
\end{table}

We can also construct the model implied distribution of the reserve price. We use $\widehat{V}(\alpha |X)$, $\widehat{\theta}$, and the observed ($N,X,W$) to calculate the seller's
expected utility for $\alpha = 0.01,0.02,...,0.99$, then find the
optimal screening level $\alpha_{R }$ as the one that maximizes the
seller's expected utility. Finally, we use $\widehat{V}(\alpha_{R }|X)$ as
the estimated optimal reserve price. Similar to the description above, we compare the CDF
of observed reserve price and the CDF of estimated reserve price. The results
are shown in the Figure 5.
\begin{figure}[H]
\centering
\includegraphics[width=5.5cm,height=4.4cm]{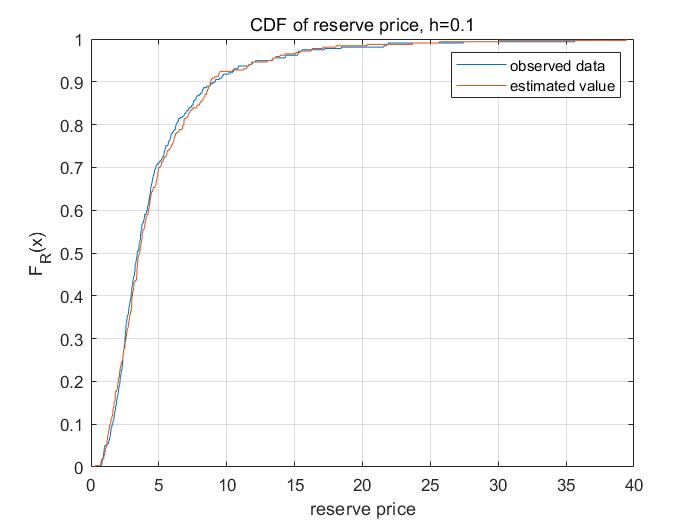}
\includegraphics[width=5.5cm,height=4.4cm]{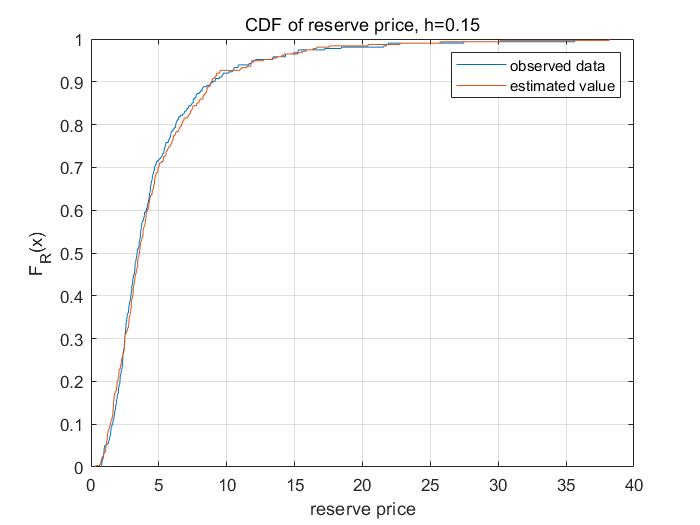}
\includegraphics[width=5.5cm,height=4.4cm]{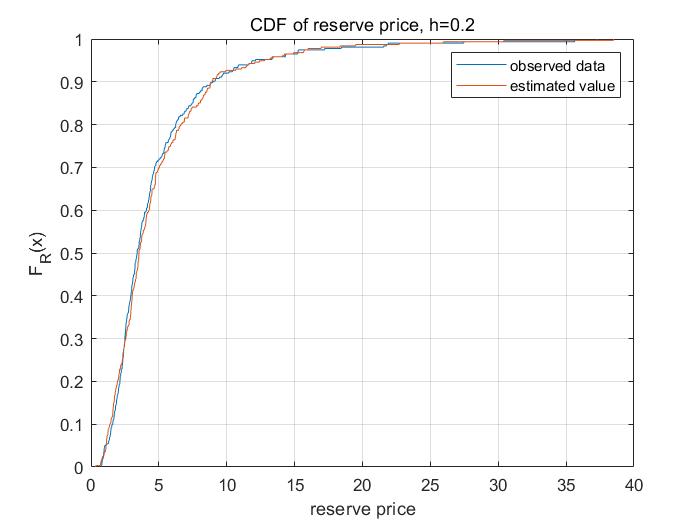}
\caption{Model fit of reserve price distribution.}
\end{figure}
Table 6 gives the bias, percentage bias, and IMSE of the estimated reserve price distribution, where the definition of bias, percentage bias, and IMSE is similar to that for the winning bid.
\begin{table}[H]
\caption{Model fit of $\widehat{F}_R$.}
\begin{center}
\begin{tabular}{cccc}
$h$ & bias & percentage bias & IMSE$_R$ \\ 
\hline \hline
0.1 & 0.0589 & 1.23\% & 0.0038 \\ 
0.15 & 0.0556 & 1.18\% & 0.0039 \\ 
0.2 & 0.0760 & 1.61\% & 0.0057 \\ 
\hline
\end{tabular}
\end{center}
\end{table}
The figures and table presented above suggest our model generates the reserve price that fits the data well. Importantly, the results are qualitatively the same and are stable for all bandwidths considered.

\subsection{Counterfactual analysis}

As a simple counterfactual exercise, we construct CDFs of the reserve price distribution for a risk-neutral seller, which were constructed analogously as those in Figure 5. We provide these in Figure 6.

\begin{figure}[H]
\centering
\includegraphics[width=5.5cm,height=4.4cm]{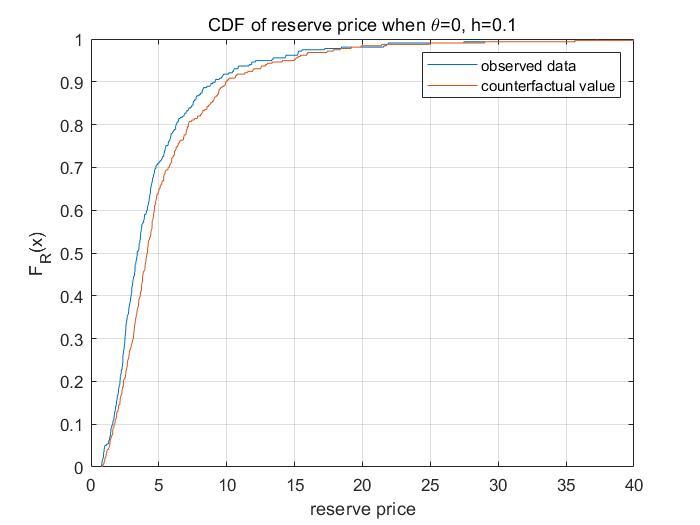}
\includegraphics[width=5.5cm,height=4.4cm]{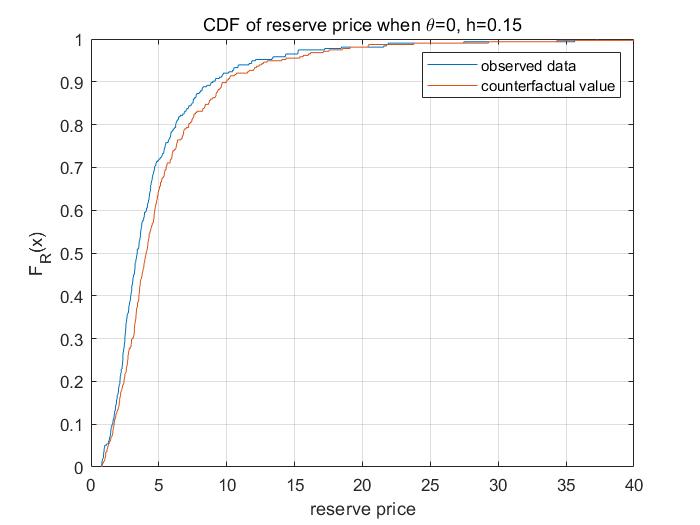}
\includegraphics[width=5.5cm,height=4.4cm]{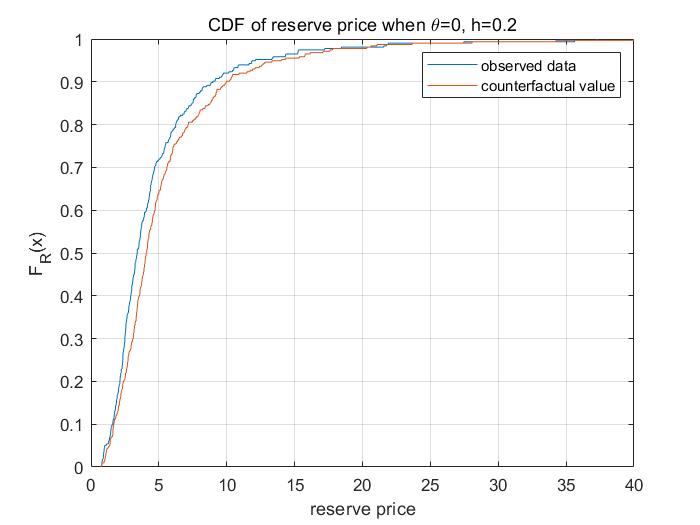}
\caption{Counterfactual reserve price distribution.}
\end{figure}
We see that the distribution of the observed reserve price is almost
first order stochastic dominated by the distribution of the counterfactual
reserve price when the seller is risk neutral, which implies the risk-neutral seller's reserve price tend to be higher than the risk-averse counterpart. To provide some quantitative comparisons, we
calculate the average of amount increase and the percentage increase in the
counterfactual reserve price for the whole sample and for three subsamples around each quartile of the apartment size:
small group (with area between $20$th and $30$th percentiles), medium group
(with area between $45$th and $55$th percentiles), and large group (with area between $70$th and $80$th percentiles). The results are summarized in Table 7.
\begin{table}[H]
\caption{Counterfactual reserve price.}
\begin{center}
\begin{tabular}{cccc}
$h$ & sample & increase in $R$ & percentage increase \\ \hline \hline
\multirow{4}{*}{$0.1$} & overall & 0.6852 & 14.36\% \\ 
& small & 0.6609 & 26.19\% \\ 
& medium & 0.7251 & 20.61\% \\ 
& large & 0.7498 & 13.23\% \\ \hline
\multirow{4}{*}{$0.15$} & overall & 0.6581 & 13.93\% \\ 
& small & 0.6704 & 27.73\% \\ 
& medium & 0.7001 & 20.00\% \\ 
& large & 0.6432 & 11.40\% \\ \hline
\multirow{4}{*}{$0.2$} & overall & 0.6690 & 14.16\% \\ 
& small & 0.6854 & 28.35\% \\ 
& medium & 0.6625 & 18.92\% \\ 
& large & 0.6533 & 11.58\% \\ \hline
\end{tabular}
\end{center}
\end{table}
Overall, the seller's reserve price would increase by 13\% to 15\% under the risk neutrality assumption, confirming that the seller's optimal reserve price is decreasing in the degree of risk aversion.

\section{Conclusion}

In this paper, we propose a framework to identify and estimate the seller’s risk-aversion parameter in ascending auctions. The model is semiparametric, with risk preferences assumed to come from a parametric utility family, while bidder valuations satisfy a linear quantile specification. We show that the risk-aversion parameter is identified under mild conditions that are commonly assumed in the theoretical literature. We then propose a two-step semiparametric estimator for it. This procedure uses the AQR approach of GG22 to estimate bidder valuation quantiles from winning bids in the first stage. We establish that the estimator is consistent and asymptotically normal under standard regularity conditions. Our estimator performs well in a simulation study. We then apply our methodology to real estate auction data in Brazil and find statistical evidence that sellers are risk-averse.

Our study makes a useful contribution to empirical auction research, as optimal auction design can depend crucially on whether the seller is risk averse or risk neutral. Moreover, any supply-side welfare analysis requires correct specification of the seller’s risk preference. Currently, empirical studies assume seller risk neutrality. Learning about seller-side primitives is, however, necessarily more demanding in terms of data requirements than learning about bidder-side primitives, although for the latter---when bid data alone are sometimes sufficient---identification and estimation still depend on the auction format, assumptions about bidders’ information structure, and data availability (e.g., all bids, bids below the reserve, or winning bids). Our application, which uses data from two-stage auction settings, is intended to serve as an illustration of the proposed methodology. More natural settings for its application include auctions with secret or hidden reserve prices, which are common in practice and for which the relevant data are observed by auction platforms and policymakers. While such data are less widely available to researchers, they have been used, for instance, in the first-price auction context (see \cite{elyakime1994first} and \cite{andreyanov2022secret}).

It should be noted that seller risk aversion is not the only mechanism that can rationalize low observed reserve prices. Alternative explanations include endogenous entry (\cite{levin1994equilibrium}), affiliated types (\cite{levin1996optimal}), and risk-averse bidders with interdependent values (\cite{hu2019low}). We do not attempt to distinguish between these possibilities. To our knowledge, no existing empirical work has done so, making this an interesting direction for future studies.

Since bidders' behavior in a second-price auction is strategically equivalent to that of an ascending auction under IPV, the estimation strategy in this paper applies to second-price auctions. Our general approach to model and estimate the seller’s risk aversion through the revenue maximization condition can be applied to other auction settings. Beyond considering an alternative auction format, the IPV assumption may be relaxed. Various extensions from this framework have been proposed in the empirical auction literature, such as unobserved heterogeneity (\cite{krasnokutskaya2011identification}, \cite{hu2013identification}, and \cite{luo2023identification}), endogenous entry (\cite{marmer2013model}, \cite{gentry2014identification}, \cite{chen2025identification}), and interdependent values (\cite{gimenes2020nonparametric}). Even within the narrower domain of AQR applications, which have thus far been applied to ascending and first-price auctions, some practical aspects will benefit from further studies. For instance, while convergence rates for AQR estimators and their corresponding optimal bandwidths have been derived, neither our paper nor GG22 provide practical guidance on bandwidth selection, and both rely on resampling methods for inference without formal justification. Despite encouraging finite-sample performance in Monte Carlo experiments and stable estimates with real data across bandwidth choices, further studies on these aspects will be useful for applied research.

\section*{Appendix}

The appendix is organized into three subappendices. Appendix A gives the
proof of Proposition 2. Appendix B gives the proofs of results in Section
3. Appendix C gives the proofs of results in Section 4.

\subsection*{A. Characterization of the optimal reserve price}

In this proof, we omit $X$ for notational simplicity as the argument holds
conditionally on $X$.

\textbf{Proof of Proposition 2.}

For part (a), omitting the auction covariates, equation (\ref{ER in r})
simplifies to%
\begin{equation*}
\widetilde{\Pi }(r,w)=U\left( w\right) F\left( r\right) ^{I}+IU\left(
r\right) F\left( r\right) ^{I-1}\left( 1-F\left( r\right) \right)
+I(I-1)\int_{r}^{\overline{v}}U\left( t\right) F(t)^{I-2}\left(
1-F(t)\right) f(t)dt,
\end{equation*}%
for any $r$\ and $w$. Taking partial derivative with respect to $r$\ gives,%
\begin{equation*}
\frac{\partial }{\partial r}\widetilde{\Pi }(r,w)=IF(r)^{I-1}f(r)\left\{
U\left( w\right) +\frac{1}{f(r)}U^{\left( 1\right) }(r)\left( 1-F(r)\right)
-U(r)\right\} .
\end{equation*}%
Let us define, 
\begin{equation}
h(r,w)=U(w)+\frac{1}{f(r)}U^{\left( 1\right) }(r)\left( 1-F(r)\right) -U(r).
\label{h(r,w)}
\end{equation}%
By M2(i), $f(r)>0$, thus the sign of $\frac{\partial }{\partial r}\widetilde{%
\Pi }(r,w)$ is determined by the sign of $h(r,w)$. Taking partial derivative
with respect to $r$\ gives%
\begin{equation*}
\frac{\partial }{\partial r}h(r,w)=-2U^{\left( 1\right) }(r)-U^{\left(
1\right) }(r)\frac{f^{\left( 1\right) }(r)\left( 1-F(r)\right) }{f(r)^{2}}%
+U^{\left( 2\right) }(r)\frac{1-F(r)}{f(r)}.
\end{equation*}%
By M2(iv), $U^{\left( 2\right) }(\cdot )\leq 0$, thus we have 
\begin{equation*}
\frac{\partial }{\partial r}h(r,w)\leq -2U^{\left( 1\right) }(r)-U^{\left(
1\right) }(r)\frac{f^{\left( 1\right) }(r)\left( 1-F(r)\right) }{f(r)^{2}}%
=-U^{\left( 1\right) }(r)\left( 2+\frac{f^{\left( 1\right) }(r)\left(
1-F(r)\right) }{f(r)^{2}}\right) .
\end{equation*}%
By M2(ii), for all $v\in \lbrack \underline{v},\bar{v}]$, 
\begin{equation*}
J^{\left( 1\right) }(v)=1-\frac{-f(v)^{2}-\left( 1-F(v)\right) f(v)}{f(v)^{2}%
}=2+\frac{\left( 1-F(v)\right) f(v)}{f(v)^{2}}>0.
\end{equation*}%
Since $U^{\left( 1\right) }(\cdot )>0$, there must be $\frac{\partial }{%
\partial r}h(r,w)\leq -U^{\left( 1\right) }(r)J^{\left( 1\right) }(r)<0$,
which implies $h(\cdot ,w)$ is strictly decreasing in $[\underline{v},\bar{v}%
]$. Next, we check the boundary. Since $w\in \lbrack \underline{v},\bar{v})$%
, 
\begin{equation*}
\begin{aligned} h(\underline{v},w) &= U(w)+\frac{1}{f(\underline{v})}
U'(\underline{v})-U(\underline{v}) > 0, \\ h(\bar{v},w) &= U(w)-U(\bar{v}) <
0. \\ \end{aligned}
\end{equation*}%
Therefore, there exists a unique $r\in (\underline{v},\bar{v})$ that
maximizes $\Pi (r,w)$.

Now we consider part (b). As already shown, the optimal reserve price $%
r^{\ast }$ is the unique solution to $h(r^{\ast },w)=0$ and $\frac{\partial 
}{\partial r}h(r,w)<0$ for all $r\in (\underline{v},\bar{v})$ and $w$. By
the Implicit Function Theorem, the optimal reserve price $r^{\ast }$ can be
written as a map $w\mapsto r^{\ast }\left( w\right) $. It follows that $%
r^{\ast }\left( w\right) $ strictly increases with $w$, since 
\begin{equation*}
\frac{dr^{\ast }\left( w\right) }{dw}=-\frac{\partial h(r,w)/\partial w}{%
\partial h(r,w)/\partial r},
\end{equation*}%
and $\frac{\partial }{\partial w}h(r,w)=U^{\left( 1\right) }(w)>0$. To show $%
r^{\ast }$ strictly decreases with the seller's risk aversion, consider $%
U_{1}(\cdot )$ and $U_{2}(\cdot )$ where there exists a strictly increasing
and strictly concave function $\zeta (\cdot )$ such that $U_{2}(\cdot
)=\zeta (U_{1}(\cdot ))$. Denote the optimal reserve price for $U_{1}(\cdot )
$ and $U_{2}(\cdot )$ by $r_{1}$ and $r_{2}$ respectively. By the result of
part (a), $r_{1}$ and $r_{2}$ satisfy 
\begin{equation*}
\begin{aligned} U_1(w)+\frac{1}{f(r_1)} U_1^{(1)}(r_1)(1-F(r_1))-U(r_1) &=
0, \\ U_2(w)+\frac{1}{f(r_2)} U_2^{(1)}(r_2)(1-F(r_2))-U(r_2) &= 0, \\
\end{aligned}
\end{equation*}%
which in turn implies 
\begin{equation*}
U_{2}(r_{2})-\frac{1-F(r_{2})}{f(r_{2})}U_{2}^{\left( 1\right)
}(r_{2})=\zeta \left( U_{1}(r_{1})-\frac{1-F(r_{1})}{f(r_{1})}U_{1}^{\left(
1\right) }(r_{1})\right) .
\end{equation*}%
Since $\zeta (\cdot)$ is strictly concave, 
\begin{equation*}
\begin{aligned} \zeta\left(U_1(r_1)-\frac{1-F(r_1)}{f(r_1)}
U_1^{(1)}(r_1)\right) &< \zeta(U_1(r_1)) -
\zeta^{(1)}(U_1(r_1))\frac{1-F(r_1)}{f(r_1)} U_1^{(1)}(r_1) \\ &=
U_2(r_1)-\frac{1-F(r_1)}{f(r_1)}U_2^{(1)}(r_1). \end{aligned}
\end{equation*}%
Therefore, there must be 
\begin{equation*}
U_{2}(r_{2})-\frac{1-F(r_{2})}{f(r_{2})}U_{2}^{\left( 1\right)
}(r_{2})<U_{2}(r_{1})-\frac{1-F(r_{1})}{f(r_{1})}U_{2}^{\left( 1\right)
}(r_{1}).
\end{equation*}%
As shown in part (a), we know $U_{2}(r)-\frac{1-F(r)}{f(r)}U_{2}^{\left(
1\right) }(r)$ is strictly increasing, thus $r_{2}<r_{1}$.$\blacksquare $

\subsection*{B. Asymptotic Theory for Quantile Function Estimators}

Since we define $\widehat{V}\left( \alpha |x\right) =\widehat{B}\left( \phi
\left( \alpha \right) |x\right) $\ for all $\left( \alpha ,x\right) $, the
statistical properties of $\widehat{V}\left( \cdot \right) $ follow from $%
\widehat{B}\left( \cdot \right) $. Before getting into the technical details, it is instructive to clarify how
our objects of interest differ from those in GG22. In a first-price auction,
the quantile function of a bidder's valuation depends on~$B\left( \cdot
\right) $ and $B^{\left( 1\right) }\left( \cdot \right) $ -- the quantile of
an individual's bid and its derivative. GG22 show that the estimation error
of $B\left( \cdot \right) $ converges to zero at a faster rate than the
counterparts of the estimator of $B^{\left( 1\right) }\left( \cdot \right) $%
, so they derive leading bias and variance expressions for the quantile
derivative estimator of the observed bids, but they only provide a convergence
rate for the quantile estimator itself. In contrast, in an ascending
auction, only the quantile of the winning bids is needed to recover the
quantile of the bidder's valuation. Our results provide bias and variance
expressions for the quantile estimator in this setting.~

Our quantile estimation setup below is a simplification of GG22's general
framework. We use what they call augmented quantile regression (AQR), which
is presented in their main text. In contrast, their appendix provides proofs
for augmented sieve quantile regression (ASQR), where the linear quantile
specification includes an increasing number of additive terms, leading to a
nonparametric specification. As a result, our appendix is lighter in
notation and the proofs are simplified.

Our presentation in this sub-appendix aims to be as concise as possible. We outline the key arguments and omit steps that follow directly from GG22. We refer readers to the original source for those steps and focus only on the elements that differ from their work. This approach avoids unnecessary repetition of GG22’s appendix, which is a carefully crafted piece spanning several sub-appendices and would remain lengthy even without the AQR components.

In what follows, we refer to GG22's supplementary material as GG-SM. This
document can found online at \texttt{%
https://doi.org/10.1016/j.jeconom.2021.02.009}. The relevant sections are
most of Appendices B, C, D, and E. To facilitate readers, we use the
same notation as in GG22 when possible.

\subsubsection*{B.1 Outline}

\noindent \noindent We start by defining the sample and population objective
functions with a reparameterized parameter. As standard in the asymptotic
analysis local polynomial regression, for example see Fan and Gijbels
(1996), we rescale the parameters to avoid degeneracy of the
\textquotedblleft design\textquotedblright\ matrix, $E\left[ P\left(
X_{l},th\right) P\left( X_{l},th\right) ^{\top }\right] $, as $h\rightarrow
0 $.\ We do this by changing the parameter to $\mathsf{b}=Hb$, where $%
H=diag\left( 1,\ldots ,h^{s}\right) \otimes \mathrm{I}_{D+1}$, so that $%
P\left( X_{l},th\right) ^{\top }b=P\left( X_{l},t\right) ^{\top }\mathsf{b}$.

Then, we define $\widehat{\mathsf{b}}\left( \alpha \right) =\arg \min_{b}%
\widehat{\mathsf{R}}\left( \mathsf{b};\alpha \right) $, where%
\begin{equation}
\widehat{\mathsf{R}}\left( \mathsf{b};\alpha \right) =\frac{1}{L}%
\sum\limits_{l=1}^{L}\int_{-\frac{\alpha }{h}}^{\frac{1-\alpha }{h}}\rho
_{\alpha +ht}\left( B_{l}-P\left( X_{l},t\right) ^{\top }\mathsf{b}\right)
K\left( t\right) dt,  \label{R-hat rep}
\end{equation}%
so that $\widehat{b}\left( \alpha \right) =H^{-1}\widehat{\mathsf{b}}\left(
\alpha \right) $. We use $b\left( \alpha \right) $\ to denote the vector of
true coefficients of the quantile slopes and their derivatives, $b\left(
\alpha \right) =\left[ \beta \left( \alpha \right) ^{\top },\beta ^{\left(
1\right) }\left( \alpha \right) ^{\top },\ldots ,\beta ^{\left( s+1\right)
}\left( \alpha \right) ^{\top }\right] ^{\top }$ and define $\mathsf{b}%
\left( \alpha \right) =Hb\left( \alpha \right) $. Since we are interested in 
$B\left( \alpha |X\right) =X_{1}^{\top }\beta \left( \alpha \right) $, and
later on its derivatives, it will be useful to define row vectors of size $%
\left( s+1\right) $, denoted by $S_{j}$, so that of $S_{j}\ $has $0$ in each
of its entry other than $1$ in the $\left( j+1\right) $-th entry for $%
j=0,1,\ldots ,s$. For example, $S_{0}=\left[ 1,\ldots ,0\right] $, $S_{1}=%
\left[ 0,1,\ldots ,0\right] $, and so on. Then let $\mathsf{S}%
_{j}=S_{j}\otimes \mathrm{I}_{D+1}$, so we can estimate $B\left( \alpha
|x\right) $ by $\widehat{B}\left( \alpha |x\right) =x_{1}^{\top }\mathsf{S}%
_{0}\widehat{\mathsf{b}}\left( \alpha \right) $ and $\widehat{B}^{\left(
j\right) }\left( \alpha |x\right) =x_{1}^{\top }\mathsf{S}_{j}\widehat{%
\mathsf{b}}\left( \alpha \right) /h^{j}$ for $j=1,\ldots ,s$.

The properties of $\widehat{\mathsf{b}}\left( \alpha \right) $\ can be
conveniently analyzed using its Bahadur representation:%
\begin{equation}
\widehat{\mathsf{b}}\left( \alpha \right) =\overline{\mathsf{b}}\left(
\alpha \right) +\widehat{\mathsf{e}}\left( \alpha \right) +\widehat{\mathsf{d%
}}\left( \alpha \right) ,  \label{Bahadur Rep}
\end{equation}%
where $\overline{\mathsf{b}}\left( \alpha \right) =\arg \min_{\mathsf{b}}%
\overline{\mathsf{R}}\left( \mathsf{b};\alpha \right) $,\ $\widehat{\mathsf{e%
}}\left( \alpha \right) =-\left[ \overline{\mathsf{R}}^{\left( 2\right)
}\left( \overline{\mathsf{b}}\left( \alpha \right) ;\alpha \right) \right]
^{-1}\widehat{\mathsf{R}}^{\left( 1\right) }\left( \overline{\mathsf{b}}%
\left( \alpha \right) ;\alpha \right) $ with $\overline{\mathsf{R}}^{\left(
2\right) }\left( \overline{\mathsf{b}}\left( \alpha \right) ;\alpha \right)
=E\left[ \widehat{\mathsf{R}}^{\left( 2\right) }\left( \overline{\mathsf{b}}%
\left( \alpha \right) ;\alpha \right) \right] $ that can be shown to exist
and have full rank for small enough $h$\footnote{%
Lemma B.3(i) shows in GG-SM shows $\widehat{\mathsf{R}}\left( \mathsf{b}%
;\alpha \right) $ and $E\left[ \widehat{\mathsf{R}}\left( \mathsf{b};\alpha
\right) \right] $\ are twice continuously differentiable for $\mathsf{b}\in 
\mathcal{B}_{\infty }\left( \mathsf{b}\left( \alpha \right) ,C_{0}h\right) $
for some $C_{0}$ and small enough $h$. Moreover, $\overline{\mathsf{R}}%
^{\left( 2\right) }\left( \mathsf{b}\left( \alpha \right) ;\alpha \right) $\
has full rank for all $\alpha \in \left[ 0,1\right] $\ by Lemma B.3(ii).
This is relevant since $\left\Vert \overline{\mathsf{b}}\left( \alpha
\right) -\mathsf{b}\left( \alpha \right) \right\Vert =o\left( h^{s+1}\right) 
$, see equation (C.2) which is shown in the proof of Theorem C.3, and Lemma
B.3(i) shows $\sup_{\alpha \in \left[ 0,1\right] \mathsf{b}^{1},\mathsf{b}%
^{0}\in \mathcal{B}_{\infty }\left( \mathsf{b}\left( \alpha \right)
,C_{0}h\right) }\frac{\left\Vert \overline{\mathsf{R}}^{\left( 2\right)
}\left( \mathsf{b}^{1};\alpha \right) -\overline{\mathsf{R}}^{\left(
2\right) }\left( \mathsf{b}^{0};\alpha \right) \right\Vert }{\left\Vert 
\mathsf{b}^{1}-\mathsf{b}^{0}\right\Vert /\left( \alpha \left( 1-\alpha
\right) +h\right) }=O\left( 1\right) $ for some $C_{0}$ and small enough $h$.%
}, and $\widehat{\mathsf{d}}\left( \alpha \right) =\widehat{\mathsf{b}}%
\left( \alpha \right) -\overline{\mathsf{b}}\left( \alpha \right) -\widehat{%
\mathsf{e}}\left( \alpha \right) $.

Let $\widehat{\mathsf{b}}^{\ast }\left( \alpha \right) $ denote $\overline{%
\mathsf{b}}\left( \alpha \right) +\widehat{\mathsf{e}}\left( \alpha \right) $%
. $\widehat{\mathsf{b}}^{\ast }\left( \alpha \right) $ is the unique
minimizer of the following quadratic approximation of $\widehat{\mathsf{R}}%
\left( \mathsf{b};\alpha \right) $:%
\begin{equation*}
\widehat{\mathsf{R}}\left( \overline{\mathsf{b}}\left( \alpha \right)
;\alpha \right) +\left( \mathsf{b}-\overline{\mathsf{b}}\left( \alpha
\right) \right) ^{\top }\widehat{\mathsf{R}}^{\left( 1\right) }\left( 
\overline{\mathsf{b}}\left( \alpha \right) ;\alpha \right) +\frac{1}{2}%
\left( \mathsf{b}-\overline{\mathsf{b}}\left( \alpha \right) \right) ^{\top }%
\overline{\mathsf{R}}^{\left( 2\right) }\left( \overline{\mathsf{b}}\left(
\alpha \right) ;\alpha \right) \left( \mathsf{b}-\overline{\mathsf{b}}\left(
\alpha \right) \right) .
\end{equation*}%
The validity of the quadratic approximation will be confirmed by the
convergence rate of $\widehat{\mathsf{d}}$.

In the AQR framework, $\overline{\mathsf{b}}\left( \alpha \right) $ can be
interpreted as the pseudo-true parameter. Then $x_{1}^{\top }\mathsf{S}%
_{0}\left( \overline{\mathsf{b}}\left( \alpha \right) -\mathsf{b}\left(
\alpha \right) \right) $\ captures the polynomial approximation bias of $%
B\left( \alpha |x\right) $. It can be shown that $\widehat{\mathsf{e}}\left(
\alpha \right) $\ is a zero mean vector, and suitably scaled components of $%
\widehat{\mathsf{e}}\left( \alpha \right) $\ satisfy the CLT.

\subsubsection*{B.2 Proofs of results}

We only provide the proofs of Lemmas 1--3. Propositions 3--5 immediately
follow from these lemmas.

\bigskip

\textbf{Proof of Lemma 1.}

Consider the following decomposition:%
\begin{eqnarray}
\widehat{B}\left( \alpha |x\right) -B\left( \alpha |x\right) &=&J_{B}\left(
\alpha ,x\right) +J_{S}\left( \alpha ,x\right) +J_{R}\left( \alpha ,x\right)
,\text{ where}  \notag \\
J_{B}\left( \alpha ,x\right) &=&x_{1}^{\top }\mathsf{S}_{0}\left( \overline{%
\mathsf{b}}\left( \alpha \right) -\mathsf{b}\left( \alpha \right) \right) ,
\label{quantile bias} \\
J_{S}\left( \alpha ,x\right) &=&x_{1}^{\top }\mathsf{S}_{0}\widehat{\mathsf{e%
}}\left( \alpha \right) ,  \label{quantile stochastic} \\
J_{R}\left( \alpha ,x\right) &=&x_{1}^{\top }\mathsf{S}_{0}\widehat{\mathsf{d%
}}\left( \alpha \right) ,\   \label{quantile remainder}
\end{eqnarray}%
where $\left( J_{B}\left( \alpha ,x\right) ,J_{S}\left( \alpha ,x\right)
,J_{R}\left( \alpha ,x\right) \right) $ are respectively the bias, leading
stochastic term, and remainder term.

We start with the bias. Let us show that $x_{1}^{\top }\mathsf{S}_{0}\left( 
\overline{\mathsf{b}}\left( \alpha \right) -\mathsf{b}\left( \alpha \right)
\right) =h^{s+1}$\textsf{$B$}$\mathsf{ias}_{h}\left( \alpha \right) +o\left(
1\right) $. To study the composition of $\overline{\mathsf{b}}\left( \alpha
\right) -\mathsf{b}\left( \alpha \right) $, we differentiate (\ref{R-hat rep}%
) and take expectation, respectively leading to:%
\begin{eqnarray*}
\widehat{\mathsf{R}}^{\left( 1\right) }\left( \mathsf{b};\alpha \right) &=&%
\frac{1}{L}\sum\limits_{l=1}^{L}\int_{\underline{t}_{\alpha ,h}}^{\overline{t%
}_{\alpha ,h}}\left( \mathbf{1}\left[ B_{l}\leq P\left( X_{l},t\right)
^{\top }\mathsf{b}\right] -\left( \alpha +ht\right) \right) P\left(
X_{l},t\right) K\left( t\right) dt \\
E\left[ \widehat{\mathsf{R}}^{\left( 1\right) }\left( \mathsf{b};\alpha
\right) \right] &=&\int_{\underline{t}_{\alpha ,h}}^{\overline{t}_{\alpha
,h}}\left( E\left[ \mathbf{1}\left[ B_{l}\leq P\left( X_{l},t\right) ^{\top }%
\mathsf{b}\right] \right] -\left( \alpha +ht\right) \right) P\left(
X_{l},t\right) K\left( t\right) dt \\
&=&\int \left( \int_{\underline{t}_{\alpha ,h}}^{\overline{t}_{\alpha
,h}}\left( G\left( P\left( x,t\right) ^{\top }\mathsf{b}|x\right) -\left(
\alpha +ht\right) \right) P\left( x,t\right) K\left( t\right) dt\right)
f\left( x\right) dx,
\end{eqnarray*}%
where $f\left( \cdot \right) $\ is the PDF of $X_{l}$, $G\left( \cdot
\right) $\ is the conditional CDF of $B_{l}$ given\ $X_{l}$, $\underline{t}%
_{\alpha ,h}=-\min \left( 1,\frac{\alpha }{h}\right) $ and $\overline{t}%
_{\alpha ,h}=\max \left( 1,\frac{1-\alpha }{h}\right) $, noting that $%
\mathcal{T}_{\alpha ,h}=\left[ \underline{t}_{\alpha ,h},\overline{t}%
_{\alpha ,h}\right] $ serve as the effective range of integration since $%
K\left( \cdot \right) $ is truncated.

GG-SM prove, in Step 1 of the proof of their Theorem C.3, that there exists $%
\overline{\mathsf{b}}\left( \alpha \right) \in $\textbf{B}$_{\infty
}\left( \mathsf{b}\left( \alpha \right) ,C_{0}h\right) $ for some $C_{0}$
and small enough $h$, that $E\left[ \widehat{\mathsf{R}}^{\left( 1\right)
}\left( \overline{\mathsf{b}}\left( \alpha \right) ;\alpha \right) \right]
=0 $. Let $\Psi \left( t|x,\mathsf{b}\right) $ denote $P\left(
X_{l},t\right) ^{\top }\mathsf{b}$ for $t\in \mathcal{T}_{\alpha ,h}$, which
can be shown to be invertible when $\mathsf{b}$ is in a vicinity of $\mathsf{%
b}\left( \alpha \right) $, which applies here for $\mathsf{b}\in $\textbf{B%
}$_{\infty }\left( \mathsf{b}\left( \alpha \right) ,C_{0}h\right) $.
Expanding the first-order condition, $E\left[ \widehat{\mathsf{R}}^{\left(
1\right) }\left( \overline{\mathsf{b}}\left( \alpha \right) ;\alpha \right) %
\right] =0 $, and follow the arguments in Step 2 in the proof of Theorem C.3
in GG-SM yields:%
\begin{eqnarray*}
0 &=&\int \left( \int_{\underline{t}_{\alpha ,h}}^{\overline{t}_{\alpha ,h}}%
\overline{g}\left( \alpha |t,x\right) \left( \Psi \left( t|x,\overline{%
\mathsf{b}}\left( \alpha \right) \right) -\Psi \left( t|x,\mathsf{b}\left(
\alpha \right) \right) \right) P\left( x,t\right) K\left( t\right) dt\right)
f\left( x\right) dx \\
&&+\int \left( \int_{\underline{t}_{\alpha ,h}}^{\overline{t}_{\alpha ,h}}%
\overline{g}\left( \alpha |t,x\right) \left( \Psi \left( t|x,\mathsf{b}%
\left( \alpha \right) \right) -B\left( \alpha +ht\right) \right) P\left(
x,t\right) K\left( t\right) dt\right) f\left( x\right) dx,
\end{eqnarray*}%
where $\overline{g}\left( \alpha |t,x\right) =\int_{0}^{1}g\left( \Psi
\left( t|x,\overline{\mathsf{b}}\left( \alpha \right) \right) +u\left(
B\left( \alpha +ht|x\right) -\Psi \left( t|x,\mathsf{b}\left( \alpha \right)
\right) \right) \right) du$. Applying Taylor's expansion in the second term,
viz., parts (iii) and (iv)\ in Lemma B.2 of GG-SM, gives%
\begin{eqnarray}
\mathsf{\check{R}}^{\left( 2\right) }\left( \alpha \right) \left( \overline{%
\mathsf{b}}\left( \alpha \right) -\mathsf{b}\left( \alpha \right) \right) &=&%
\mathsf{\check{R}}^{\left( 2\right) }\left( \alpha \right) h^{s+1}\mathsf{%
\check{b}}_{s+1}\left( \alpha \right) +o\left( h^{s+1}\right) ,\text{ where}
\label{AQR leading bias term} \\
\mathsf{\check{b}}_{s+1}\left( \alpha \right) &=&\left[ \mathsf{\check{R}}%
^{\left( 2\right) }\left( \alpha \right) \right] ^{-1}\int \left( \int_{%
\underline{t}_{\alpha ,h}}^{\overline{t}_{\alpha ,h}}\overline{g}\left(
\alpha |t,x\right) \frac{t^{s+1}B^{\left( s+1\right) }\left( \alpha
|x\right) }{\left( s+1\right) !}P\left( x,t\right) K\left( t\right)
dt\right) f\left( x\right) dx,  \notag \\
\mathsf{\check{R}}^{\left( 2\right) }\left( \alpha \right) &=&\int \left(
\int_{\underline{t}_{\alpha ,h}}^{\overline{t}_{\alpha ,h}}\overline{g}%
\left( \alpha |t,x\right) P\left( x,t\right) P\left( x,t\right) ^{\top
}K\left( t\right) dt\right) f\left( x\right) dx.  \notag
\end{eqnarray}%
Note that we perform a Taylor's expansion upto the $\left( s+1\right) $-th
term while GG-SM do it to the $\left( s+2\right) $-th term. This is because
the quantile function of optimal first price auction bids has one more
derivative than the quantile function of the underlying bidder's valuation.

GG-SM show, as part of the proof of their Lemma C.2, that%
\begin{equation*}
\max_{\left( \alpha ,x\right) \in \left[ 0,1\right] \times \mathcal{X}%
}\max_{t\in \left[ \underline{t}_{\alpha ,h}+C_{1}\epsilon _{1},\overline{t}%
_{\alpha ,h}-C_{1}\epsilon _{1}\right] }\left\vert \overline{g}\left( \alpha
|t,x\right) -\frac{1}{B^{\left( 1\right) }\left( \alpha |x\right) }%
\right\vert =O\left( h\right) ,
\end{equation*}%
for some positive constant $C_{1}$ and $\epsilon _{1}=o\left( h\right) $,
and 
\begin{equation*}
\max_{\left( \alpha ,x\right) \in \left[ 0,1\right] \times \mathcal{X}%
}\left\Vert \mathsf{\check{R}}^{\left( 2\right) }\left( \alpha \right)
-\Omega _{h}\left( \alpha \right) \otimes \mathbf{P}_{0}\left( \alpha
\right) \right\Vert =O\left( h\right) ,\text{ where}
\end{equation*}%
\begin{equation*}
\Omega _{h}\left( \alpha \right) =\int_{\underline{t}_{\alpha ,h}}^{%
\overline{t}_{\alpha ,h}}\pi \left( t\right) \pi \left( t\right) ^{\top
}K\left( t\right) dt\text{ and }\mathbf{P}_{0}\left( \alpha \right) =E\left[ 
\frac{X_{l}X_{l}^{\top }}{B^{\left( 1\right) }\left( \alpha |X_{l}\right) }%
\right] .
\end{equation*}%
Therefore, since $h=o\left( 1\right) $ and dominated convergence applies, we
have:%
\begin{eqnarray}
&&\mathsf{\check{b}}_{s+1}\left( \alpha \right)  \label{b-cap} \\
&=&\left[ \Omega _{h}\left( \alpha \right) \otimes \mathbf{P}_{0}\left(
\alpha \right) +o\left( 1\right) \right] ^{-1}\int \left( \int_{\underline{t}%
_{\alpha ,h}}^{\overline{t}_{\alpha ,h}}\left( \frac{t^{s+1}B^{\left(
s+1\right) }\left( \alpha |x\right) }{\left( s+1\right) !B^{\left( 1\right)
}\left( \alpha |x\right) }+o\left( 1\right) \right) P\left( x,t\right)
K\left( t\right) dt\right) f\left( x\right) dx  \notag \\
&=&\left[ \Omega _{h}\left( \alpha \right) ^{-1}\otimes \mathbf{P}_{0}\left(
\alpha \right) ^{-1}\right] \left[ \int_{\underline{t}_{\alpha ,h}}^{%
\overline{t}_{\alpha ,h}}\frac{t^{s+1}\pi \left( t\right) }{\left(
s+1\right) !}K\left( t\right) dt\otimes \int \frac{x_{1}x_{1}^{\top }}{%
B^{\left( 1\right) }\left( \alpha |x\right) }f\left( x\right) dx\beta
^{\left( s+1\right) }\left( \alpha \right) \right] +o\left( 1\right)  \notag
\\
&=&\Omega _{h}\left( \alpha \right) ^{-1}\int_{\underline{t}_{\alpha ,h}}^{%
\overline{t}_{\alpha ,h}}\frac{t^{s+1}\pi \left( t\right) }{\left(
s+1\right) !}K\left( t\right) dt\otimes \beta ^{\left( s+1\right) }\left(
\alpha \right) +o\left( 1\right) .  \notag
\end{eqnarray}%
Since $\mathsf{S}_{0}\left[ \Omega _{h}\left( \alpha \right) ^{-1}\int_{%
\underline{t}_{\alpha ,h}}^{\overline{t}_{\alpha ,h}}\frac{t^{s+1}K\left(
t\right) }{\left( s+1\right) !}\pi \left( t\right) dt\otimes \beta ^{\left(
s+1\right) }\left( \alpha \right) \right] =S_{0}\Omega _{h}\left( \alpha
\right) ^{-1}\int_{\underline{t}_{\alpha ,h}}^{\overline{t}_{\alpha ,h}}%
\frac{t^{s+1}K\left( t\right) }{\left( s+1\right) !}\pi \left( t\right)
dt\beta ^{\left( s+1\right) }\left( \alpha \right) $, the proof is completed
as%
\begin{eqnarray*}
x_{1}^{\top }\mathsf{S}_{0}\left( \overline{\mathsf{b}}\left( \alpha \right)
-\mathsf{b}\left( \alpha \right) \right) &=&h^{s+1}x_{1}^{\top }\beta
^{\left( s+1\right) }\left( \alpha \right) S_{0}\Omega _{h}\left( \alpha
\right) ^{-1}\int_{\underline{t}_{\alpha ,h}}^{\overline{t}_{\alpha ,h}}%
\frac{t^{s+1}\pi \left( t\right) }{\left( s+1\right) !}K\left( t\right)
dt+o\left( h^{s+1}\right) \\
&=&h^{s+1}B^{\left( s+1\right) }\left( \alpha |x\right) S_{0}\Omega
_{h}\left( \alpha \right) ^{-1}\int_{\underline{t}_{\alpha ,h}}^{\overline{t}%
_{\alpha ,h}}\frac{t^{s+1}\pi \left( t\right) }{\left( s+1\right) !}K\left(
t\right) dt+o\left( h^{s+1}\right) .
\end{eqnarray*}

Next, we consider $J_{S}\left( \alpha ,x\right) $. We start with the
variance expression for $\widehat{\mathsf{e}}\left( \alpha \right) $ given
in GG-SM. In particular, when proving their Lemma B.6 in Appendix F.3.3,
they show that $Var\left( \widehat{\mathsf{e}}\left( \alpha \right) \right)
=\left( \mathsf{V}_{e}\left( \alpha \right) +o\left( h\right) \right) /L$,%
\footnote{%
There are a couple of typos in GG-SM related to $Var\left( \widehat{\mathsf{e%
}}\left( \alpha \right) \right) $. First, at the
bottom of their page 134, they were meant to say $Var\left( \widehat{\mathsf{%
e}}\left( \alpha \right) \right) =\left( \mathsf{V}_{e}+o\left( h\right)
\right) /L$\ instead of $Var\left( \widehat{\mathsf{e}}\left( \alpha \right)
\right) =\mathsf{V}_{e}/L+o\left( h\right) $. Second, the last component of $%
\mathsf{V}_{e}$ given in the second display of page 135 should be $%
S_{0}^{\top }S_{1}+S_{1}^{\top }S_{0}$ and not $S_{0}S_{1}^{\top
}+S_{1}S_{0}^{\top }$.} where%
\begin{eqnarray}
\mathsf{V}_{e}\left( \alpha \right)  &=&\alpha \left( 1-\alpha \right) \left[
S_{0}^{\top }S_{0}\right] \otimes \left[ \mathbf{P}_{0}\left( \alpha \right)
^{-1}\mathbf{PP}_{0}\left( \alpha \right) ^{-1}\right]   \label{Variance} \\
&&+h\alpha \left( 1-\alpha \right) \left[ S_{1}^{\top }S_{0}\right] \otimes %
\left[ \mathbf{P}_{0}\left( \alpha \right) ^{-1}\mathbf{P}_{1}\mathbf{P}%
_{0}\left( \alpha \right) ^{-1}\mathbf{PP}_{0}\left( \alpha \right) ^{-1}%
\right]   \notag \\
&&+h\alpha \left( 1-\alpha \right) \left[ S_{0}^{\top }S_{1}\right] \otimes %
\left[ \mathbf{P}_{0}\left( \alpha \right) ^{-1}\mathbf{PP}_{0}\left( \alpha
\right) ^{-1}\mathbf{P}_{1}\mathbf{P}_{0}\left( \alpha \right) ^{-1}\right] 
\notag \\
&&+h\left[ \Omega _{h}^{-1}\mathbf{\Pi }_{m}\Omega _{h}^{-1}-\left(
S_{0}^{\top }S_{1}+S_{1}^{\top }S_{0}\right) \right] \otimes \left[ \mathbf{P%
}_{0}\left( \alpha \right) ^{-1}\mathbf{PP}_{0}\left( \alpha \right) ^{-1}%
\right] ,  \notag \\
\mathbf{\Pi }_{m}\left( \alpha \right)  &=&\int_{\underline{t}_{\alpha ,h}}^{%
\overline{t}_{\alpha ,h}}\int_{\underline{t}_{\alpha ,h}}^{\overline{t}%
_{\alpha ,h}}\min \left( t_{1},t_{2}\right) \pi \left( t\right) \pi \left(
t\right) ^{\top }K\left( t_{1}\right) K\left( t_{2}\right) dt_{1}dt_{2}, 
\notag \\
\mathbf{P} &=&E\left[ X_{l}X_{l}^{\top }\right] \text{ and }\mathbf{P}%
_{0}\left( \alpha \right) =E\left[ \frac{X_{l}X_{l}^{\top }}{B^{\left(
1\right) }\left( \alpha |X_{l}\right) }\right] .  \notag
\end{eqnarray}%
It follows that $Var\left( \mathsf{S}_{0}\widehat{\mathsf{e}}\left( \alpha
\right) \right) =\alpha \left( 1-\alpha \right) \mathbf{P}_{0}\left( \alpha
\right) ^{-1}\mathbf{PP}_{0}\left( \alpha \right) ^{-1}/L+O\left( h/L\right) 
$, and $Var\left( \sqrt{L}\mathsf{S}_{0}\widehat{\mathsf{e}}\left(
\alpha \right) \right) =\Sigma _{h}\left( \alpha \right) +o\left( 1\right) $
as desired.

The CLT result follows from the same argument used
in the proof of Theorem 3 (and A.3)\ in GG-SM that can be found in their
Appendix E.2. Write $\left( \frac{L}{x_{1}^{\top }\Sigma _{h}\left( \alpha
\right) x_{1}}\right) ^{1/2}x_{1}^{\top }\mathsf{S}_{0}\widehat{\mathsf{e}}%
\left( \alpha \right) =\sum\limits_{l=1}^{L}r_{l}\left( \alpha |x\right) $,
so that%
\begin{eqnarray*}
r_{l}\left( \alpha |x\right) &=&\left( \frac{1}{Lx_{1}^{\top }\Sigma
_{h}\left( \alpha \right) x_{1}}\right) ^{1/2}x_{1}^{\top }\mathsf{S}_{0}%
\left[ \overline{\mathsf{R}}^{\left( 2\right) }\left( \overline{\mathsf{b}}%
\left( \alpha \right) ;\alpha \right) \right] ^{-1} \\
&&\times \int_{\underline{t}_{\alpha ,h}}^{\overline{t}_{\alpha ,h}}\left( 
\mathbf{1}\left[ B_{l}\leq P\left( X_{l},t\right) ^{\top }\overline{\mathsf{b%
}}\left( \alpha \right) \right] -\left( \alpha +ht\right) \right) P\left(
X_{l},t\right) K\left( t\right) dt.
\end{eqnarray*}%
Since $\left\vert Var\left( r_{l}\left( \alpha |x\right) \right)
-1\right\vert =o\left( 1\right) $, we can bound $\left\vert E\left[
r_{l}^{3}\left( \alpha |x\right) \right] \right\vert $ by%
\begin{equation*}
\left\vert r_{l}\left( \alpha |x\right) \right\vert Var\left( r_{l}\left(
\alpha |x\right) \right) \leq O\left( \frac{1}{L}\right) =o\left( 1\right) ,
\end{equation*}%
thus CLT applies.

For the remainder term, the proof that $\sqrt{L}J_{R}\left( \alpha ,x\right)
=o_{p}\left( 1\right) $ follows immediately from applying Theorem D.1 in
GG-SM, where they show that%
\begin{equation*}
\sup_{\alpha \in \left[ 0,1\right] }\left\Vert \frac{Lh^{1/2}}{\left(
h+\alpha \left( 1-\alpha \right) \right) ^{1/2}\log L}\widehat{\mathsf{d}}%
\left( \alpha \right) \right\Vert =O_{p}\left( 1\right) ,
\end{equation*}%
under the condition $\frac{\log ^{2}L}{Lh}=o\left( 1\right) $.

For the uniform rate, the result will follow from the triangle inequality
once we verify:%
\begin{eqnarray*}
\sup_{\left( \alpha ,x\right) \in \left[ 0,1\right] \times \mathcal{X}%
}\left\vert J_{B}\left( \alpha ,x\right) \right\vert &=&O\left(
h^{s+1}\right) , \\
\sup_{\left( \alpha ,x\right) \in \left[ 0,1\right] \times \mathcal{X}%
}\left\vert J_{S}\left( \alpha ,x\right) \right\vert &=&O_{p}\left( \sqrt{%
\frac{\log L}{L}}\right) , \\
\sup_{\left( \alpha ,x\right) \in \left[ 0,1\right] \times \mathcal{X}%
}\left\vert J_{R}\left( \alpha ,x\right) \right\vert &=&o_{p}\left( \sqrt{%
\frac{\log L}{L}}\right) .\ 
\end{eqnarray*}

Since $\mathcal{X}$ is compact, the pointwise bias rate holds uniformly as
other components in $\mathsf{Bias}_{h}\left( \cdot \right) $ are bounded.

GG-SM have shown the required rate for $J_{S}\left( \cdot \right) $ in their
Lemma B.6(ii).

Lastly, for $J_{R}\left( \cdot \right) $, this follows from 
\begin{equation*}
\sup_{\alpha \in \left[ 0,1\right] }\left\Vert \widehat{\mathsf{d}}\left(
\alpha \right) \right\Vert =O_{p}\left( \frac{\log L}{Lh^{1/2}}\right) ,
\end{equation*}%
which is shown in the proof of Theorem D.1 in GG-SM -- as implied by their
equation (D.4). By compactness of $\mathcal{X}$, $\sup_{\left( \alpha
,x\right) \in \left[ 0,1\right] \times \mathcal{X}}\left\vert x_{1}^{\top }%
\mathsf{S}_{0}\widehat{\mathsf{d}}\left( \alpha \right) \right\vert
=O_{p}\left( \frac{\log L}{Lh^{1/2}}\right) =O_{p}\left( \sqrt{\frac{\log L}{%
L}}\sqrt{\frac{\log L}{Lh}}\right) =o_{p}\left( \sqrt{\frac{\log L}{L}}%
\right) $ since $\log L=o\left( Lh\right) $ under our bandwidth condition.$%
\blacksquare $

\bigskip

\textbf{Proof of Lemma 2.} From the Bahadur representation, (\ref{Bahadur
Rep}), we have:%
\begin{equation}
\widehat{B}^{\left( j\right) }\left( \alpha |x\right) -B^{\left( j\right)
}\left( \alpha |x\right) =\frac{x_{1}^{\top }\mathsf{S}_{j}\left( \overline{%
\mathsf{b}}\left( \alpha \right) -\mathsf{b}\left( \alpha \right) \right) }{%
h^{j}}+\frac{x_{1}^{\top }\mathsf{S}_{j}\widehat{\mathsf{e}}\left( \alpha
\right) }{h^{j}}+\frac{x_{1}^{\top }\mathsf{S}_{j}\widehat{\mathsf{d}}\left(
\alpha \right) }{h^{j}}.\   \notag
\end{equation}%
The three terms above respectively represent the bias, leading stochastic
term, and remainder term. It suffices to show the following:%
\begin{eqnarray*}
\sup_{\left( \alpha ,x\right) \in \left[ 0,1\right] \times \mathcal{X}%
}\left\vert \frac{x_{1}^{\top }\mathsf{S}_{j}\left( \overline{\mathsf{b}}%
\left( \alpha \right) -\mathsf{b}\left( \alpha \right) \right) }{h^{j}}%
\right\vert &=&O\left( h^{s+1-j}\right) , \\
\sup_{\left( \alpha ,x\right) \in \left[ 0,1\right] \times \mathcal{X}%
}\left\vert \frac{x_{1}^{\top }\mathsf{S}_{j}\widehat{\mathsf{e}}\left(
\alpha \right) }{h^{j}}\right\vert &=&O_{p}\left( \sqrt{\frac{\log L}{%
Lh^{2j-1}}}\right) , \\
\sup_{\left( \alpha ,x\right) \in \left[ 0,1\right] \times \mathcal{X}%
}\left\vert \frac{x_{1}^{\top }\mathsf{S}_{j}\widehat{\mathsf{d}}\left(
\alpha \right) }{h^{j}}\right\vert &=&o_{p}\left( \sqrt{\frac{\log L}{%
Lh^{2j-1}}}\right) .
\end{eqnarray*}

The form of the bias can be obtained from the expansion of $\overline{%
\mathsf{b}}\left( \alpha \right) -\mathsf{b}\left( \alpha \right) =h^{s+1}%
\mathsf{\check{b}}_{s+1}\left( \alpha \right) +o\left( h^{s+1}\right) $, as
shown in (\ref{AQR leading bias term}), where $\mathsf{\check{b}}%
_{s+1}\left( \alpha \right) $ is given in (\ref{b-cap}). Then,%
\begin{equation*}
x_{1}^{\top }\mathsf{S}_{j}\left( \overline{\mathsf{b}}\left( \alpha \right)
-\mathsf{b}\left( \alpha \right) \right) =h^{s+1-j}x_{1}^{\top }\beta
^{\left( s+1\right) }\left( \alpha \right) S_{j}\Omega _{h}\left( \alpha
\right) ^{-1}\int_{\underline{t}_{\alpha ,h}}^{\overline{t}_{\alpha ,h}}%
\frac{t^{s+1}\pi \left( t\right) }{\left( s+1\right) !}K\left( t\right)
dt+o\left( h^{s+1-j}\right) .
\end{equation*}%
It follows from the inverse function theorem that $\beta ^{\left( s+1\right)
}\left( \cdot \right) $\ is continuous on $\left[ 0,1\right] $, as $\gamma
\left( \cdot \right) $, which has $s+1$ continuous derivatives, is a
composite function of $\beta \left( \cdot \right) $ and a differentiable and
strictly increasing function, $\phi \left( \cdot \right) $. Since $\mathcal{X%
}$ is compact, the bias is $O\left( h^{s+1-j}\right) $ uniformly.

We next consider the leading stochatic term. GG-SM prove this in their Lemma
B.6 when $j=1$, by showing that%
\begin{equation*}
\sup_{\left( \alpha ,x\right) \in \left[ 0,1\right] \times \mathcal{X}%
}\left\vert \frac{x_{1}^{\top }\mathsf{S}_{j}\widehat{\mathsf{e}}\left(
\alpha \right) }{h^{1/2}}\right\vert =O_{p}\left( \sqrt{\frac{\log L}{L}}%
\right) ,
\end{equation*}%
see their equation (F.3). The same proof in fact applies to cases when $j>1$
as long as $Var\left( \mathsf{S}_{j}\widehat{\mathsf{e}}\left( \alpha
\right) /h^{1/2}\right) $ is uniformly bounded for $\alpha \in \left[ 0,1%
\right] $. This is indeed the case, as it can be seen from (\ref{Variance})
that: 
\begin{equation*}
Var\left( \mathsf{S}_{j}\widehat{\mathsf{e}}\left( \alpha \right) \right) =h%
\left[ S_{j}\Omega _{h}^{-1}\mathbf{\Pi }_{m}\Omega _{h}^{-1}S_{j}^{\top }%
\right] \otimes \left[ \mathbf{P}_{0}\left( \alpha \right) ^{-1}\mathbf{PP}%
_{0}\left( \alpha \right) ^{-1}\right] \text{ for all }j>0;
\end{equation*}%
the expression follows from $S_{j}S_{j^{\prime }}^{\top }=0$ whenever $%
j^{\prime }\neq j$.

For the remainder term, we make use of of the fact that%
\begin{equation*}
\sup_{\alpha \in \left[ 0,1\right] }\left\Vert \widehat{\mathsf{d}}\left(
\alpha \right) \right\Vert _{\infty }=O_{p}\left( \frac{\log L}{Lh^{1/2}}%
\right) ,
\end{equation*}%
which was shown in the proof of Theorem D.1 in GG-SM -- as implied by their
equation (D.4). Then,%
\begin{eqnarray*}
\sup_{\left( \alpha ,x\right) \in \left[ 0,1\right] \times \mathcal{X}%
}\left\vert \frac{x_{1}^{\top }\mathsf{S}_{j}\widehat{\mathsf{d}}\left(
\alpha \right) }{h^{j}}\right\vert &=&O_{p}\left( \frac{\log L}{Lh^{j+1/2}}%
\right) \\
&=&O_{p}\left( \left( \frac{\log L}{Lh^{2j-1}}\right) ^{1/2}\left( \frac{%
\log L}{Lh^{2}}\right) ^{1/2}\right) \\
&=&o_{p}\left( \left( \frac{\log L}{Lh^{2j-1}}\right) ^{1/2}\right) ,
\end{eqnarray*}%
since $\log L=o\left( Lh^{2}\right) $ is implied by our bandwidth condition.$%
\blacksquare $

\bigskip

\textbf{Proof of Lemma 3.}

Given the linearization: 
\begin{equation*}
\widehat{B}^{-1}\left( t|x\right) -B^{-1}\left( t|x\right) =-\frac{\widehat{B%
}\left( B^{-1}\left( t|x\right) |x\right) -B\left( B^{-1}\left( t|x\right)
|x\right) }{B^{\left( 1\right) }\left( B^{-1}\left( t|x\right) |x\right) }%
+o_{p}\left( \sup_{\left( \alpha ,x\right) \in \left[ 0,1\right] \times 
\mathcal{X}}\left\vert \widehat{B}\left( \alpha |x\right) -B\left( \alpha
|x\right) \right\vert \right) ,
\end{equation*}%
the results trivially follow from Lemma 1.$\blacksquare $

\subsection*{C. Asymptotic Theory for Risk-Aversion Estimator}

\subsubsection*{C.1 Preliminaries}

The following class of functions feature in our objective function and is
central to the analysis of our semiparametric estimator: 
\begin{equation}
\mathcal{Q}=\left\{ 
\begin{array}{c}
q_{\theta ,\psi }\left( w,r,x\right) =U_{\theta }\left( w\right) +U_{\theta
}^{\left( 1\right) }\left( r\right) \psi _{1}\left( \psi _{2}\left(
r,x\right) ,x\right) \left( 1-\psi _{2}\left( r,x\right) \right) -U_{\theta
}\left( r\right) \text{ } \\ 
\text{for }\left( \theta ,\psi \right) \in \Theta \times \mathcal{D}%
\end{array}%
\right\} ,  \label{Q class of functions}
\end{equation}%
where $\theta \in \Theta $ is a compact subset of $\mathbb{R}\ $and $\psi
=\left( \psi _{1},\psi _{2}\right) \in \mathcal{D}=\mathcal{D}_{1}\times 
\mathcal{D}_{2}$ where $\mathcal{D}_{1}\mathcal{\ }$and $\mathcal{D}_{2}$
respectively abbreviate $\mathcal{D}_{1}\left( \mathcal{A_{\delta }}\times 
\mathcal{X}\right) \times \mathcal{D}_{2}\left( \mathcal{V_{\delta }}\times 
\mathcal{X}\right) $.\ 

Before proceeding further, it will be useful to state some conventions on
notations. First, to simplify, we will frequently suppress the pointwise
arguments of the functions\ unless its serves to clarify matters. Second,
since composite functions are a prominent feature in $\mathcal{Q}$, to
facilitate the readers, we use $\circ $ to indicate it in our proofs. Third,
given the class of functions we consider are uniformly bounded, we will
generically use $\left\Vert \cdot \right\Vert _{\infty }$\ to denote the
sup-norm over all arguments of a real value function over its support. For
example, $\sup_{\left( r,x\right) \in \mathcal{V}_{\delta }\times \mathcal{X}%
}\left\vert \psi _{1}\left( \psi _{2}\left( r,x\right) ,x\right) \right\vert 
$ will be abbrebiated to $\left\Vert \psi _{1}\circ \psi _{2}\right\Vert
_{\infty }$. And, to emphasize we are performing calculations with respect
to the parameters when applying empirical process theory, we highlight the
parameters by making them the arguments of the functions, e.g., $q_{\theta
,\psi }\left( \cdot \right) $ becomes $q\left( \theta ,\psi \right) $. We
will use $\lesssim $\ to denote the less than or equal to relation up to a
universal constant.

We will show that $\mathcal{Q}$\ and related class of functions are $P_{Z}-$%
Glivenko-Cantelli under S1 and $P_{Z}-$Donsker under S2. We focus our proofs
on the Donsker case, which is the more difficult of the two. Let $%
N_{[]}\left( \varepsilon ,\mathcal{Q},L_{2}\left( P_{Z}\right) \right) $
denote the covering numbers with bracketing with respect to the $L_{2}$-norm
(see Definition 2.1.6 in VW23). We define the covering numbers for other
class of functions using a similar notation. It will be convenient to use $%
Z_{0}$ to denote a generic subset from $\left( W,R,X\right) $ and $P_{Z_{0}}$
to denote the probability distribution of $Z_{0}$. We will abbreviate $%
L_{2}\left( P_{Z_{0}}\right) $\ to $L_{2}$ when there is no risk of
ambiguity. Correspondingly, we use $\left\Vert \cdot \right\Vert _{L_{2}}$\
to denote the $L_{2}$-norm.

We will prove the Donsker property for a class of functions by showing it
has an $L_{2}$-integrable envelope\ and satisfies the following integral
bracketing entropy condition:%
\begin{equation*}
\int_{0}^{\infty }\sqrt{\log N_{[]}\left( \varepsilon ,\mathcal{Q}%
,L_{2}\right) }d\varepsilon <\infty ,
\end{equation*}%
see Theorem 2.5.2 in VW23. GC property holds under a simpler to verify
requirement that $N_{[]}\left( \varepsilon ,\mathcal{Q},L_{1}\right) <\infty 
$ for all $\varepsilon >0$, see Theorem 2.4.1 in VW23.

The main bracketing number calculations can be traced to the functions in $%
\mathcal{D}_{j}$, which are assumed to have at least two continuous
derivatives under Assumption S2(ii). We denote the class of functions that
have $2$\ continuous derivatives on $\mathcal{A}_{\delta }$ with an $L_{2}$
integrable envelope by $\mathcal{C}^{2}$.

In what follows, we define $c_{\mathcal{Q}^{0}}$\ to be an envelope for
functions in $\mathcal{Q}$:%
\begin{eqnarray}
\left\vert q_{\theta ,\psi }\left( w,r,x\right) \right\vert &\leq &c_{%
\mathcal{Q}^{0}}\left( w,r,x\right) ,\text{ where}  \label{Q - envelope} \\
c_{\mathcal{Q}^{0}}\left( w,r,x\right) &=&\left\vert c_{\Theta ^{0}}\left(
w\right) \right\vert +\left\vert c_{\Theta ^{0}}\left( r\right) \right\vert
+C_{\mathcal{D}_{1}^{0}}\left\vert c_{\Theta ^{1}}\left( r\right)
\right\vert ,  \notag
\end{eqnarray}%
such that $c_{\Theta ^{j}}\left( w\right) =\sup_{\theta \in \Theta
}\left\vert U_{\theta }^{\left( j\right) }\left( w\right) \right\vert $ and $%
\sup_{\psi _{i}\in \mathcal{D}_{i}^{j}}\left\Vert \psi _{i}^{\left( j\right)
}\right\Vert _{\infty }<C_{\mathcal{D}_{i}^{j}}$ for some $C_{\mathcal{D}%
_{i}^{j}}\ $for$\ i=1,2$ and $j=0,1,2$. Under Assumptions S1(iii)-(iv) and
S2(i)-(ii), $E\left[ c_{\Theta ^{j}}\left( W,R,X\right) ^{i}\right] <\infty $
and $C_{\mathcal{D}_{i}^{j}}<\infty $ for $i=1,2$ and $j=0,1,2$.

\subsubsection*{C.2 Additional Lemmas}

\indent\textbf{Lemma C.1.} $\mathcal{D}_{1}$ and $\mathcal{D}_{2}$ defined
in Assumption S2(ii) are $P_{Z}-$Donsker.\ 

\bigskip

\textbf{Proof of Lemma C.1.}

We start by showing the class of functions that $\mathcal{D}_{1}$ and $%
\mathcal{D}_{2}$ are derived from are Donsker. Let $\mathcal{M}_{j}=\{$ $%
x_{j}\mu \left( \alpha \right) $ for $x_{j}\in \mathcal{X}_{j}$ and $\mu \in 
\mathcal{C}^{2}$ $\}$ where $\mathcal{X}_{j}\subseteq \mathbb{R}$\ is the
support of the $j$-th component in the random vector $X$. It is well known
that $\mathcal{C}^{2}$ is a Donsker class of functions. $\mathcal{M}_{j}$ is
a class of $\mathcal{C}^{2}$\ functions with a bounded fixed multiplier due
to compactness of $\mathcal{X}_{j}$. Let $\overline{\mu }$\ be an envelope
on $\mathcal{C}^{2}$ with $P_{Z_{0}}\overline{\mu }^{2}<\infty $, then $C_{%
\mathcal{X}_{j}}\overline{\mu }$ is an envelope for $\mathcal{M}_{j}$\ where 
$C_{\mathcal{X}_{j}}=\max_{x_{j}\in \mathcal{X}_{j}}\left\vert
x_{j}\right\vert $. It follows that $\mathcal{M}_{j}$ is Donsker since
brackets $\left\{ \left[ l_{\jmath },u_{\jmath }\right] \right\} _{\jmath
=1}^{N}$ that cover $\mathcal{C}^{2}$\ leads to $\left\{ \left[
x_{j}l_{\jmath },x_{j}u_{\jmath }\right] \right\} _{\jmath =1}^{N}$ which
covers $\mathcal{M}_{j}$. Then, $N_{[]}\left( \varepsilon ,x_{j}\mathcal{C}%
_{2},L_{2}\left( P_{Z_{0}}\right) \right) \leq N_{[]}\left( \varepsilon /C_{%
\mathcal{X}_{j}},\mathcal{C}_{2},L_{2}\left( P_{Z_{0}}\right) \right) $, so
the bracketing entropy condition is satisfied since $\log N_{[]}\left(
\varepsilon /C_{\mathcal{X}_{j}},\mathcal{C}_{2},L_{2}\left(
P_{Z_{0}}\right) \right) \lesssim \varepsilon ^{-1}$ (e.g., see Corollary
2.7.2 in VW23). Let $\mathcal{M}=\{$ $m_{0}\left( \alpha \right)
+\sum\limits_{j=1}^{D}m_{j}\left( \alpha ,x_{j}\right) $ for $x\in \mathcal{X%
}$ such that $m_{0}\in \mathcal{C}^{2}\left( \mathcal{A}_{\delta }\right) $\
and $m_{j}\in \mathcal{M}_{j}$\ $\}$. $\left( 1+\sum\limits_{j=1}^{D}C_{%
\mathcal{X}_{j}}\right) \overline{\mu }$\ is an $L_{2}$-integrable envelope
for $\mathcal{M}$. $\mathcal{M}$ is also Donsker due to the preservation of
Donsker properties of a finite sum of classes of Donsker functions (VW23,
Theorem 2.10.8).

Let $\mathcal{M}^{1}=\{$ $\phi ^{\left( 1\right) }\left( \alpha \right)
m\left( \phi \left( \alpha \right) ,x\right) $\ for\textit{\ }$m\in \mathcal{%
M}\ \}$. A function in $\mathcal{M}^{1}$ takes the form $\frac{\partial }{%
\partial \alpha }\nu \left( \phi \left( \alpha \right) ,x\right) $ for $\nu
\in \mathcal{D}_{0}$\textit{\ }defined in Assumption S1(iv). Thus, $\mathcal{%
D}_{1}$ is a subset of $\mathcal{M}^{1}$ when Assumption S2(ii) holds, and $%
\mathcal{D}_{1}$ will be Donsker if $\mathcal{M}^{1}$\ is Donsker (VW23,
Theorem 2.10.8). $\mathcal{M}^{1}$\ is indeed Donsker. This follows as $%
\mathcal{M}^{1}$ is a class of Donsker functions with a bounded fixed
multiplier, $\phi ^{\left( 1\right) }$ -- whose image is compact on $%
A_{\delta }$. Note that the composition of $\phi $\ in the first argument of 
$m\in \mathcal{M}\ $does not affect the bracketing entropy calculation. This
is because $\phi $\ is strictly increasing and the composition serves as a
change of probability\ measure that defines the $L_{2}$\ norm in the
bracketing number calculation, and the bracketing number can be chosen to be
independent of the probability\ measure.

Under Assumption S1(iv), $\mathcal{D}_{2}$\ is defined as $\left( t,x\right)
\mapsto \nu _{x}^{-1}\left( t\right) $\textit{\ }where $\nu _{x}\left(
\alpha \right) =\nu \left( \phi \left( \alpha \right) ,x\right) $\textit{\ }%
for $\nu \in \mathcal{D}_{0}$. Noting that $\mathcal{D}_{0}$\ is a subset of 
$\mathcal{M}$. By definition of $\mathcal{D}_{0}$, the partial inverse of $%
\nu _{x}$\ exists and we denote it by $\nu _{x}^{-1}:\mathcal{V}_{\delta
}\mapsto \mathcal{A}_{\delta }$. Consider the relation $\nu \left( \phi
\left( \nu _{x}^{-1}\left( t\right) \right) ,x\right) =t$. The Implicit
Function Theorem gives,%
\begin{equation*}
\frac{\partial }{\partial x}\nu _{x}^{-1}\left( t\right) =\frac{\mu \left(
\phi \left( \nu _{x}^{-1}\left( t\right) \right) \right) }{\phi ^{\left(
1\right) }\left( \nu _{x}^{-1}\left( t\right) \right) x_{1}^{\top }\frac{%
\partial }{\partial \alpha }\mu \left( \phi \left( \nu _{x}^{-1}\left(
t\right) \right) \right) }.
\end{equation*}%
Under S1(iv), $\sup_{\left( t,x\right) \in \mathcal{V}_{\delta }\times 
\mathcal{X}}\frac{\partial }{\partial x}\nu _{x}^{-1}\left( t\right) <\infty 
$ uniformly for $\nu \in \mathcal{D}_{0}$, which implies functions in $%
\mathcal{D}_{2}$ is uniformly Lipschitz with respect to the argument on $%
\mathcal{X}$. We denote the corresponding Lipschitz constant by $K_{\mathcal{%
X}}$. We will use this Lipschitz property to build suitable brackets
covering $\mathcal{D}_{2}$ and show the bracketing entropy is integrable.

For any $\varepsilon >0$, take a $\eta $-net, $\left\{ x_{\ell }\right\}
_{\ell =1}^{N_{\mathcal{X}}}$, of $\mathcal{X}$ such that $\eta =\varepsilon
/3K_{\mathcal{X}}$ and $N_{\mathcal{X}}\lesssim \varepsilon ^{-D}$. For each 
$x_{\ell }$, we can find $N_{\ell }\left( \varepsilon \right) $\ brackets to
cover all functions $t\mapsto \nu _{x_{\ell }}^{-1}\left( t\right) $\ that
are uniformly bounded and monotone on $\mathcal{V}_{\delta }$, $\left\{ %
\left[ l_{\ell \jmath },u_{\ell \jmath }\right] \right\} _{\jmath
=1}^{N_{\ell }\left( \varepsilon \right) }$, such that $\left\Vert u_{\jmath
\ell }-l_{\jmath \ell }\right\Vert _{L_{2}\left( P_{Z_{0}}\right) }\leq
\varepsilon /3$ and $\log N_{\ell }\left( \varepsilon \right) \lesssim
\varepsilon ^{-1}$ (VW23, Theorem 2.7.9). We can then extend the brackets at 
$\left\{ x_{\ell }\right\} _{\ell =1}^{N_{\mathcal{X}}}$ to brackets on $%
\mathcal{V}_{\delta }\times \mathcal{X}$. To do this, for any $x$ such that $%
\left\Vert x-x_{\ell }\right\Vert \leq \eta $, define $\underline{l}_{\ell
\jmath }\left( t,x\right) =l_{\ell \jmath }\left( t\right) -\varepsilon /3$
and $\overline{u}_{\ell \jmath }\left( t,x\right) =u_{\ell \jmath }\left(
t\right) +\varepsilon /3$. By combining the Lipschitz continuity (over $x$)\
together with the brackets for $\nu _{x_{\ell }}^{-1}$, we have $\underline{l%
}_{\ell \jmath }\left( t,x\right) \leq \nu _{x}^{-1}\left( t\right) \leq 
\overline{u}_{\ell \jmath }\left( t,x\right) $ for all $t$ and $\left\Vert
x-x_{\ell }\right\Vert \leq \eta $.\ Thus, $\left\{ \left[ \underline{l}%
_{\ell \jmath },\overline{u}_{\ell \jmath }\right] \right\} _{\jmath
=1}^{N_{\ell }\left( \varepsilon \right) }$ are brackets for $\nu _{x}^{-1}$%
\ on the $\eta $-ball around $x_{\ell }$. Since the union of the $\eta $%
-balls around $\left\{ x_{\ell }\right\} _{\ell =1}^{N_{\mathcal{X}}}$ cover 
$\mathcal{X}$, we have a cover for $\mathcal{D}_{2}$. Note that $\left\Vert 
\overline{u}_{\ell \jmath }-\underline{l}_{\ell \jmath }\right\Vert
_{L_{2}\left( P_{Z_{0}}\right) }\leq \left\Vert u_{\ell \jmath }-l_{\ell
\jmath }\right\Vert _{L_{2}\left( P_{Z_{0}}\right) }+2\varepsilon /3\leq
\varepsilon $.\ Therefore, $N_{[]}\left( \varepsilon ,\mathcal{D}%
_{2},L_{2}\left( P_{Z_{0}}\right) \right) \leq \sum\limits_{\ell =1}^{N_{%
\mathcal{X}}}N_{\ell }\left( \varepsilon \right) $ and $\log N_{[]}\left(
\varepsilon ,\mathcal{D}_{2},L_{2}\left( P_{Z_{0}}\right) \right) \lesssim
\varepsilon ^{-1}+\log \left( \varepsilon ^{-1}\right) $, so that $%
\int_{0}^{1}\sqrt{\log N_{[]}\left( \varepsilon ,\mathcal{D}_{2},L_{2}\left(
P_{Z_{0}}\right) \right) }d\varepsilon <\infty $.$\blacksquare $

\bigskip

\textbf{Lemma C.2.} Under Assumption S2(ii), $\mathcal{Q}$ is $P_{Z}-$%
Donsker.\ 

\bigskip

\textbf{Proof of Lemma C.2.}

An element in $\mathcal{Q}$\ is a sum of three terms. Two of them belong
to a low complexity class of functions, $\left\{ U_{\theta }:\text{ }\theta
\in \Theta \right\} $, that are Lipschitz over $\Theta $ with an $%
L_{2}\left( P_{Z_{0}}\right) $\ integrable envelope. Such a class of
functions satisfies the entropy bracketing condition\ (VW23, Theorem 2.7.17)
and is Donsker. The remaining (middle) term is a product of three functions: $\left(
r,x\right) \mapsto U_{\theta }^{\left( 1\right) }\left( r\right) \psi
_{1}\left( \psi _{2}\left( r,x\right) ,x\right) \left( 1-\psi _{2}\left(
r,x\right) \right) $ where $\left( \psi _{1},\psi _{2}\right) \in \mathcal{D}%
_{1}\times \mathcal{D}_{2}$. $\left\{ U_{\theta }^{\left( 1\right) }:\theta
\in \Theta \right\} $ is Donsker, as it is also a Lipschitz-in-parameter
class of functions with an $L_{2}\left( P_{Z_{0}}\right) $\ integrable
envelope. We have already shown in Lemma C.1 that $\mathcal{D}_{2}$\ is
Donsker, so that $\left( r,x\right) \mapsto 1-\psi _{2}\left( r,x\right) $
is Donsker. We will show below that $\mathcal{D}_{1\circ 2}=\{$$\left(
r,x\right) \mapsto \psi _{1}\left( \psi _{2}\left( r,x\right) ,x\right) $
for $\left( \psi _{1},\psi _{2}\right) \in \mathcal{D}_{1}\times \mathcal{D}%
_{2}$$\}$ is also Donsker. Together, this ensures $\mathcal{Q}$ is Donsker
as permanence properties of Donsker classes apply to finite sums and
products of Donsker classes of functions.

$\mathcal{D}_{1\circ 2}$\ is a class of composition of functions with the
outer map belonging to $\mathcal{D}_{1}$. The $L_{2}\left( P_{Z_{0}}\right) $
integrable envelope for $\mathcal{D}_{1}$ then applies for functions in $%
\mathcal{D}_{1\circ 2}$.\ Note that the standard permanence Donsker theorem
employed previously (VW23, Theorem 2.10.8) assumes the outer map of the
composition function is fixed, so it does not apply to $\mathcal{D}_{1\circ
2}$. We need to verify Donsker permanence holds uniformly over $\mathcal{D}%
_{1}$\ as well. To do this, take any $\psi _{1},\psi _{1}^{\prime }$ in $%
\mathcal{D}_{1}$ and $\psi _{2}\in \mathcal{D}_{2}$, we have%
\begin{eqnarray*}
\left\Vert \psi _{1}\circ \psi _{2}-\psi _{1}^{\prime }\circ \psi
_{2}\right\Vert _{L_{2}\left( P_{Z_{0}}\right) }^{2} &=&\int \left( \psi
_{1}\left( \psi _{2}\left( R,X\right) ,X\right) -\psi _{1}^{\prime }\left(
\psi _{2}\left( R,X\right) ,X\right) \right) ^{2}dP_{Z_{0}} \\
&=&\int \left( \psi _{1}\left( A,X\right) -\psi _{1}^{\prime }\left(
A,X\right) \right) ^{2}dP_{\widetilde{Z}_{0}\left( \psi _{2}\right) }.
\end{eqnarray*}%
The second equality above follows from a change of measure that is valid due
to $\psi _{2}$\ being strictly increasing. Let $\lambda _{R,X}$ be a
dominating measure of $P_{Z_{0}}$, defined as the product of a Lebesgue
measure on $\mathcal{V}_{\delta }$ and a general measure on $X$ (possibly
containing a counting measure if components of $X$ is discrete). We denote $%
\frac{dP_{Z_{0}}}{d\lambda _{R,X}}$, a Radon--Nikodym derivative, by $%
p_{R,X} $. Then, $\frac{dP_{\widetilde{Z}_{0}\left( \psi _{2}\right) }}{%
d\lambda _{A,X}}$\ is a map $\left( \alpha ,x\right) \mapsto \frac{%
p_{R,X}\left( \psi _{2}^{-1}\left( \alpha ,x\right) ,x\right) }{\psi
_{2}^{\left( 1\right) }\left( \psi _{2}^{-1}\left( \alpha ,x\right)
,x\right) }$, where $\lambda _{A,X}$\ is a dominating measure defined
similarly to $\lambda _{R,X}$\ with a measure on $\mathcal{A}_{\delta }$\
instead of $\mathcal{V}_{\delta }$. Under conditions of S1(iv), $\frac{dP_{%
\widetilde{Z}_{0}\left( \psi _{2}\right) }}{d\lambda _{A,X}}$ is (pointwise)
bounded above by some constant $C_{P}<\infty $ uniformly over $\mathcal{D}%
_{2}$. Hence, 
\begin{equation*}
\left\Vert \psi _{1}\circ \psi _{2}-\psi _{1}^{\prime }\circ \psi
_{2}\right\Vert _{L_{2}\left( P_{Z_{0}}\right) }=\left\Vert \psi _{1}-\psi
_{1}^{\prime }\right\Vert _{L_{2}\left( P_{\widetilde{Z}_{0}\left( \psi
_{2}\right) }\right) }\leq \sqrt{C_{P}}\left\Vert \psi _{1}-\psi
_{1}^{\prime }\right\Vert _{L_{2}\left( \lambda _{A,X}\right) }.
\end{equation*}%
Since $\mathcal{D}_{1\circ 2}$\ consists of composition functions with
functions from $\mathcal{D}_{1}$\ as the outer map, the brackets that cover $%
\mathcal{D}_{1}$\ can be used to cover $\mathcal{D}_{1\circ 2}$.
Subsequently, the inequality above implies: 
\begin{equation*}
N_{[]}\left( \varepsilon ,\mathcal{D}_{1\circ 2},L_{2}\left(
P_{Z_{0}}\right) \right) \leq N_{[]}\left( \varepsilon /\sqrt{C_{P}},%
\mathcal{D}_{1},L_{2}\left( \lambda _{A,X}\right) \right) .
\end{equation*}%
The bound on the bracketing numbers used to establish Donsker property of $%
\mathcal{D}_{1}$\ in Lemma C.1 applies to any $L_{2}$ norm defined with
finite measure. Thus, we have $\int_{0}^{1}\sqrt{\log N_{[]}\left(
\varepsilon ,\mathcal{D}_{1},L_{2}\left( \lambda _{A,X}\right) \right) }%
d\varepsilon <\infty $.$\blacksquare $

\bigskip

\textbf{Lemma C.3.} Under Assumption S1(iv), $\mathcal{Q}$ is $P_{Z}-$%
Glivenko-Cantelli.

\bigskip

\textbf{Proof of Lemma C.3.}

The arguments used in proving Lemmas C.1 and C.2 are applicable with the
change of $L_{2}-$integrability of envelopes to $L_{1}-$integrability and checking that
bracketing numbers are finite. There are also permanence results for
Glivenko--Cantelli classes, analogous to those established for Donsker
classes, for example, see Theorem 2.10.5 in VW23. We omit further details to
avoid repetition.$\blacksquare $

\subsubsection*{C.3 Proofs of results}

\indent\textbf{Proof of Lemma 4.}

Since $\Theta $ is compact and $\theta \mapsto Q\left( \theta ,\psi
_{0}\right) $ is continuous, it suffices to show that $Q\left( \theta
_{0},\psi _{0}\right) \leq Q\left( \theta ,\psi _{0}\right) $ and the
equality holds if and only if $\theta =\theta _{0}$. For this, as done in
the proof of Proposition 2, let us first omit $X$ and consider the
parameterized version of (\ref{h(r,w)}), 
\begin{equation*}
h_{\theta }(r,w)=U_{\theta }(w)+U_{\theta }^{\left( 1\right) }(r)\frac{1-F(r)%
}{f(r)}-U_{\theta }(r).
\end{equation*}%
By Proposition 2, for any $w$, suppose $r$ is the corresponding optimal
reserve price when the seller's risk aversion is $\theta _{0}$, there must
be 
\begin{equation*}
h_{\theta _{0}}(r,w)=0.
\end{equation*}%
We first show that $h_{\theta }(r,w)\neq 0$ whenever $\theta \neq \theta _{0}
$. Consider $\theta >\theta _{0}$, there exists a function $\zeta (\cdot )$
satisfying condition (a) of S1(iii) su that $U_{\theta }(\cdot )=\zeta
(U_{\theta _{0}}(\cdot ))$. Therefore, we have 
\begin{equation*}
\begin{aligned} U_{\theta}(r)-U_{\theta}^{(1)}(r)\frac{1-F(r)}{f(r)} &=
\zeta(U_{\theta_0}(r)) -
\zeta^{(1)}(U_{\theta_0}(r))U_{\theta_0}^{(1)}(r)\frac{1-F(r)}{f(r)} \\ &>
\zeta\left( U_{\theta_0}(r)-U_{\theta_0}^{(1)}(r)\frac{1-F(r)}{f(r)} \right)
\\ &= \zeta(U_{\theta_0}(w)) \\ &= U_{\theta}(w), \\ \end{aligned}
\end{equation*}%
which implies 
\begin{equation*}
h_{\theta }(r,w)=U_{\theta }(w)+U_{\theta }^{\left( 1\right) }(r)\frac{1-F(r)%
}{f(r)}-U_{\theta }(r)<0.
\end{equation*}%
A similar argument can be applied to the case $\theta <\theta _{0}$, so that 
\begin{equation*}
\begin{aligned} U_{\theta_0}(w) &=
U_{\theta_0}(r)-U_{\theta_0}^{(1)}(r)\frac{1-F(r)}{f(r)} \\ &=
\zeta(U_{\theta}(r)) -
\zeta^{(1)}(U_{\theta}(r))U_{\theta}^{(1)}(r)\frac{1-F(r)}{f(r)} \\ &>
\zeta\left( U_{\theta}(r)-U_{\theta}^{(1)}(r)\frac{1-F(r)}{f(r)} \right).
\end{aligned}
\end{equation*}%
We then have,%
\begin{equation*}
U_{\theta }(r)-U_{\theta }^{\left( 1\right) }(r)\frac{1-F(r)}{f(r)}<\zeta
^{-1}(U_{\theta _{0}}(w))=U_{\theta }(w),
\end{equation*}%
which implies 
\begin{equation*}
h_{\theta }(r,w)=U_{\theta }(w)+U_{\theta }^{\left( 1\right) }(r)\frac{1-F(r)%
}{f(r)}-U_{\theta }(r)>0.
\end{equation*}%
Inputting back the dependence on $X$, and noting that $\frac{1-F(r|x)}{f(r|x)%
}=\psi _{10}\left( \psi _{20}\left( r,x\right) ,x\right) \left( 1-\psi
_{20}\left( r,x\right) \right) $, the argument above implies that $q\left(
z,\theta ,\psi _{0}\right) \neq 0$ when $\theta \neq \theta _{0}$ for $z\in 
\mathcal{Z}$, which means $Q\left( \theta ,\psi _{0}\right) >0$ when $\theta
\neq \theta _{0}$.$\blacksquare $

\bigskip

\textbf{Proof of Lemma 5.}

From the triangle inequality, we have%
\begin{equation*}
\left\vert Q_{L}\left( \theta ,\widehat{\psi }\right) -Q\left( \theta ,\psi
_{0}\right) \right\vert \leq \left\vert Q_{L}\left( \theta ,\widehat{\psi }%
\right) -Q\left( \theta ,\widehat{\psi }\right) \right\vert +\left\vert
Q\left( \theta ,\widehat{\psi }\right) -Q\left( \theta ,\psi _{0}\right)
\right\vert .
\end{equation*}%
Under the assumed bandwidth condition, $\widehat{\psi }_{i}\in \mathcal{D}%
_{i}$ w.p.a. $1$ for $i=1,2$. In particular, as discussed in Appendix C.1 of
GG22, AQR estimators are smooth w.p.a. $1$, which ensures, 
\begin{equation*}
\left\vert Q_{L}\left( \theta ,\widehat{\psi }\right) -Q\left( \theta ,%
\widehat{\psi }\right) \right\vert \leq \sup_{\theta \in \Theta ,\psi \in 
\mathcal{D}_{1}\times \mathcal{D}_{2}}\left\vert Q_{L}\left( \theta ,\psi
\right) -Q\left( \theta ,\psi \right) \right\vert =o_{p}\left( 1\right) .
\end{equation*}%
The uniform convergence of $Q_{L}$ to $Q$ in probability is a consequence of Lemma C.3.

Next, take $q\left( \theta ,\widehat{\psi }\right) $ and $q\left( \theta
,\psi _{0}\right) $ in $\mathcal{Q}$ for any $\theta $. We have,%
\begin{equation*}
\left\vert q\left( \theta ,\widehat{\psi }\right) ^{2}-q\left( \theta ,\psi
_{0}\right) ^{2}\right\vert \leq 2c_{\mathcal{Q}^{0}}\left\vert q\left(
\theta ,\widehat{\psi }\right) -q\left( \theta ,\psi _{0}\right) \right\vert
,
\end{equation*}%
due to the triangle inequality and bounding by an envelope of $\mathcal{Q}$%
.\ It will be instructive here to show that $q$ is (pointwise) Lipschitz
continuous with respect to $\psi $ for any $\theta $\ in the following sense:%
\begin{equation*}
\left\vert q\left( \theta ,\psi \right) -q\left( \theta ,\psi ^{\prime
}\right) \right\vert \leq c\left[ \left\Vert \psi _{1}-\psi _{1}^{\prime
}\right\Vert _{\infty }+\left\Vert \psi _{2}-\psi _{2}^{\prime }\right\Vert
_{\infty }\right] ,
\end{equation*}%
for some function $c$\ such that $E\left[ c^{2}\right] <\infty $. The main
calculation for establishing Lipschitz
continuity involves showing that $T\left( \psi \right) =\left( \psi
_{1}\circ \psi _{2}\right) \left( 1-\psi _{2}\right) $\ is Lipschitz
continuous, which will also be useful in other parts of the appendix, so we
provide some details here.

Take any $\psi ,\psi ^{\prime }$ and consider,%
\begin{eqnarray*}
T\left( \psi \right) -T\left( \psi ^{\prime }\right) &=&\left( \psi
_{1}\circ \psi _{2}-\psi _{1}^{\prime }\circ \psi _{2}\right) \left( 1-\psi
_{2}\right) \\
&&+\left( \psi _{1}^{\prime }\circ \psi _{2}-\psi _{1}^{\prime }\circ \psi
_{2}^{\prime }\right) \left( 1-\psi _{2}\right) \\
&&+\left( \psi _{1}^{\prime }\circ \psi _{2}^{\prime }\right) \left( \left(
1-\psi _{2}\right) -\left( 1-\psi _{2}^{\prime }\right) \right) .
\end{eqnarray*}%
Note that the image of functions in $\mathcal{D}_{2}$\ is in $A_{\delta }$
so that $\max \{\sup_{\psi _{2}\in \mathcal{D}_{2}}\left\Vert 1-\psi
_{2}\right\Vert _{\infty },\sup_{\psi _{2}\in \mathcal{D}_{2}}\left\Vert
\psi _{2}\right\Vert _{\infty }\}<1$. The first term of the sum above then
satisfies $\left\Vert \left( \psi _{1}\circ \psi _{2}-\psi _{1}^{\prime
}\circ \psi _{2}\right) \left( 1-\psi _{2}\right) \right\Vert _{_{\infty
}}\leq \left\Vert \psi _{1}-\psi _{1}^{\prime }\right\Vert _{\infty }$. For
the second term, we can apply a mean value expansion and impose the upper
bound on $\psi _{1}^{\left( 1\right) }$\ to obtain $\left\Vert \psi
_{1}^{\prime }\circ \psi _{2}-\psi _{1}^{\prime }\circ \psi _{2}^{\prime
}\right\Vert _{_{\infty }}\leq C_{\mathcal{D}_{1}^{1}}\left\Vert \psi
_{2}-\psi _{2}^{\prime }\right\Vert _{\infty }$. The last term is bounded
above by $C_{\mathcal{D}_{1}^{1}}\left\Vert \psi _{2}-\psi _{2}^{\prime
}\right\Vert _{\infty }$, since $\psi _{1}^{\prime }$\ is bounded above by $%
C_{\mathcal{D}_{1}^{0}}$. Together, they imply%
\begin{equation}
\left\Vert T\left( \psi \right) -T\left( \psi ^{\prime }\right) \right\Vert
_{\infty }\leq \left( 1+C_{\mathcal{D}_{1}^{1}}+C_{\mathcal{D}%
_{1}^{0}}\right) \left( \left\Vert \psi _{1}-\psi _{1}^{\prime }\right\Vert
_{\infty }+\left\Vert \psi _{2}-\psi _{2}^{\prime }\right\Vert _{\infty
}\right) .  \label{T Lipschitz}
\end{equation}%
By S1(iii), $\left\vert U_{\theta }^{\left( 1\right) }\right\vert \leq
c_{\Theta ^{1}}$. We then have,%
\begin{equation}
\left\vert q\left( \theta ,\psi \right) -q\left( \theta ,\psi ^{\prime
}\right) \right\vert \leq \left( 1+C_{\mathcal{D}_{1}^{1}}+C_{\mathcal{D}%
_{1}^{0}}\right) c_{\Theta ^{1}}\left[ \left\Vert \psi _{1}-\psi
_{1}^{\prime }\right\Vert _{\infty }+\left\Vert \psi _{2}-\psi _{2}^{\prime
}\right\Vert _{\infty }\right] ,  \label{q-psi Lipschitz}
\end{equation}%
as desired.

Since $\widehat{\psi }_{2}\in \mathcal{D}_{2}$ w.p.a. $1$, it follows that, 
\begin{eqnarray*}
\sup_{\theta \in \Theta }\left\vert Q\left( \theta ,\widehat{\psi }\right)
-Q\left( \theta ,\psi _{0}\right) \right\vert &\leq &2\left( 1+C_{\mathcal{D}%
_{1}^{1}}+C_{\mathcal{D}_{1}^{0}}\right) E\left[ c_{\mathcal{Q}^{0}}^{2}%
\right] ^{1/2}E\left[ c_{\Theta ^{1}}^{2}\right] ^{1/2} \\
&&\times \left( E\left[ \left\Vert \widehat{\psi }_{1}-\psi _{10}\right\Vert
_{\infty }\right] +E\left[ \left\Vert \widehat{\psi }_{2}-\psi
_{20}\right\Vert _{\infty }\right] \right) \\
&=&o\left( 1\right) ,
\end{eqnarray*}%
where we have used Cauchy Schwarz inequality to bound $E\left[ \left\vert c_{%
\mathcal{Q}^{0}}c_{\Theta ^{1}}\right\vert \right] $, and $E\left[
\left\Vert \widehat{\psi }_{i}-\psi _{i0}\right\Vert _{\infty }\right]
=o\left( 1\right) $ for $i=1,2$ is a result of Vitali convergence theorem,
as $\left\Vert \widehat{\psi }_{i}-\psi _{i0}\right\Vert _{\infty
}=o_{p}\left( 1\right) $ and functions in $\mathcal{D}_{i}$ are uniformly
bounded.$\blacksquare $

\bigskip

\textbf{Proof of Theorem 1.}

This is an immediate consequence of Lemma 4 and Lemma 5. For example, see
Theorem 2.1 in Newey and McFadden (1994).$\blacksquare $

\bigskip

\textbf{Proof of Lemma 6.}

Let us define $Y_{L}\left( \theta ,\psi \right) =\frac{1}{L}%
\sum\limits_{l=1}^{L}y\left( Z_{l},\theta ,\psi \right) $, where $y\left(
Z_{l},\theta ,\psi \right) =q\left( Z_{l},\theta ,\psi \right) \frac{%
\partial }{\partial \theta }q\left( Z_{l},\theta ,\psi \right) $, and let $%
Y\left( \theta ,\psi \right) =\int q\left( Z_{l},\theta ,\psi \right) \frac{%
\partial }{\partial \theta }q\left( Z_{l},\theta ,\psi \right) dP_{Z}$.\ We
want to show $Y\left( \theta _{0},\widehat{\psi }\right) =\dot{Y}\left(
\theta _{0},\psi _{0}\right) \left[ \widehat{\psi }-\psi _{0}\right]
+o_{p}\left( L^{-1/2}\right) $, where $\dot{Y}\left( \theta _{0},\psi
_{0}\right) \left[ \widehat{\psi }-\psi _{0}\right] $ denotes the pathwise
derivative of $Y$ at $\left( \theta _{0},\psi _{0}\right) $ in direction $%
\left[ \widehat{\psi }-\psi _{0}\right] $, and $\sqrt{L}\dot{Y}\left( \theta
_{0},\psi _{0}\right) \left[ \widehat{\psi }-\psi _{0}\right] $ satisfies a
suitable CLT.

To do this, we need to derive $\dot{Y}\left( \theta _{0},\psi _{0}\right) %
\left[ \widehat{\psi }-\psi _{0}\right] $. The main functional components in 
$Y$\ can be traced to $T\left( \psi \right) =\left( \psi _{1}\circ \psi
_{2}\right) \left( 1-\psi _{2}\right) $, which appears in both $q$ and $%
\frac{\partial }{\partial \theta }q$. Let us first calculate the pathwise
derivative of $T$ at $\psi $\ in direction $\omega $. It follows from simple
algebra that 
\begin{equation*}
T\left( \psi +\epsilon \omega \right) -T\left( \psi \right) =\epsilon \omega
_{1}\circ \psi _{2}\left( 1-\psi _{2}\right) +\epsilon \left( \psi
_{1}^{\left( 1\right) }\circ \psi _{2}\left( 1-\psi _{2}\right) -\psi
_{1}\circ \psi _{2}\right) \omega _{2}+O\left( \epsilon ^{2}\right) ,
\end{equation*}%
so that $\lim_{\epsilon \rightarrow 0}\frac{T\left( \psi +\epsilon \omega
\right) -T\left( \psi \right) }{\epsilon }$ exists and is equal to 
\begin{equation*}
\dot{T}\left( \psi \right) \left[ \omega \right] =\omega _{1}\circ \psi
_{2}\left( 1-\psi _{2}\right) +\left( \psi _{1}^{\left( 1\right) }\circ \psi
_{2}\left( 1-\psi _{2}\right) -\psi _{1}\circ \psi _{2}\right) \omega _{2}.
\end{equation*}%
The perturbation $T\left( \psi +\epsilon \omega \right) -T\left( \psi
\right) $\ can also be refined to obtain the second derivative. It can be
checked through some more algebra that the $O\left( \epsilon ^{2}\right) $
term in the display above consists of $\epsilon ^{2}\Xi \left( \omega
\right) +O\left( \epsilon ^{3}\right) $, where%
\begin{equation*}
\Xi \left( \omega \right) =\left( \omega _{1}^{\left( 1\right) }\circ \psi
_{2}\left( 1-\psi _{2}\right) -\omega _{1}\circ \psi _{2}\right) \omega
_{2}+\left( \frac{1}{2}\psi _{1}^{\left( 2\right) }\circ \psi _{2}\left(
1-\psi _{2}\right) -\psi _{1}^{\left( 1\right) }\circ \psi _{2}\right)
\omega _{2}^{2}.
\end{equation*}%
Focusing on the derivative at $\psi _{0}$, when $\widehat{\psi }-\psi
_{0}=\epsilon \omega $, we have%
\begin{eqnarray*}
\Xi \left( \epsilon \omega \right) &=&O_{p}\left( \left\Vert \epsilon \omega
_{1}\times \epsilon \omega _{2}\right\Vert _{\infty }+\left\Vert \epsilon
\omega _{1}^{\left( 1\right) }\times \epsilon \omega _{2}\right\Vert
_{\infty }+\left\Vert \left( \epsilon \omega _{2}\right) ^{2}\right\Vert
_{\infty }\right) \\
&=&O_{p}\left( \left\Vert \left( \widehat{\psi }_{1}^{\left( 1\right) }-\psi
_{10}^{\left( 1\right) }\right) \times \left( \widehat{\psi }_{2}-\psi
_{20}\right) \right\Vert _{\infty }\right) ,
\end{eqnarray*}%
where $\left\Vert \epsilon \omega _{1}^{\left( 1\right) }\times \epsilon
\omega _{2}\right\Vert _{\infty }$\ is the leading term, as it can be shown
that $\left\Vert \widehat{\psi }_{1}^{\left( 1\right) }-\psi _{10}^{\left(
1\right) }\right\Vert _{\infty }=O_{p}\left( \sqrt{\frac{\log L}{Lh^{3}}}%
+h^{s-1}\right) $ and $\left\Vert \widehat{\psi }_{2}-\psi _{20}\right\Vert
_{\infty }=O_{p}\left( \sqrt{\frac{\log L}{L}}+h^{s+1}\right) $. These rates
correspond to the convergence rates in Proposition 4 with $j=2$ and
Proposition 5 respectively. We then have $\Xi \left( \epsilon \omega \right)
=O_{p}\left( \frac{\log L}{Lh^{3/2}}+h^{2s}\right) =o_{p}\left(
L^{-1/2}\right) $, where the latter holds due to our assumption on $h$. This
ensures that $T\left( \widehat{\psi }\right) -T\left( \psi _{0}\right) =\dot{%
T}\left( \psi _{0}\right) \left[ \widehat{\psi }-\psi _{0}\right]
+o_{p}\left( L^{-1/2}\right) $.

With $\dot{T}\left( \psi \right) \left[ \omega \right] $\ on hand, it
follows immediately that the pathwise derivatives of $q\left( \theta ,\cdot
\right) $ and $\frac{\partial }{\partial \theta }q\left( \theta ,\cdot
\right) $\ at $\psi $\ in direction $\omega $ are $U_{\theta }^{\left(
1\right) }\dot{T}\left( \psi \right) \left[ \omega \right] $ and $U_{\theta
}^{\left( 2\right) }\dot{T}\left( \psi \right) \left[ \omega \right] $
respectively for any $\theta $. The pathwise derivatives of $Y\left( \theta
,\cdot \right) $ can then be readily obtained from applying the product rule
of differentiation to $q\left( \theta ,\cdot \right) \frac{\partial }{%
\partial \theta }q\left( \theta ,\cdot \right) $ and integrating this over $%
P_{Z}$. Specifically,%
\begin{equation*}
\lim_{\epsilon \rightarrow 0}\frac{Y\left( \theta ,\psi +\epsilon \omega
\right) -Y\left( \theta ,\psi \right) }{\epsilon }=\dot{Y}\left( \theta
,\psi \right) \left[ \omega \right] =\int \frac{\partial }{\partial \theta }%
q\left( \theta ,\psi \right) U_{\theta }^{\left( 1\right) }\dot{T}\left(
\psi \right) \left[ \omega \right] +q\left( \theta ,\psi \right) U_{\theta
}^{\left( 2\right) }\dot{T}\left( \psi \right) \left[ \omega \right] dP_{Z}.
\end{equation*}%
Focusing on the derivative of $Y$\ at $\psi _{0}$\ in direction $\left[ 
\widehat{\psi }-\psi _{0}\right] $ when $\theta =\theta _{0}$, this yields:%
\begin{eqnarray}
\dot{Y}\left( \theta _{0},\psi _{0}\right) \left[ \widehat{\psi }-\psi _{0}%
\right] &=&\int c_{Y_{1}}\left( \widehat{\psi }_{1}-\psi _{10}\right) \circ
\psi _{20}+c_{Y_{2}}\left( \widehat{\psi }_{2}-\psi _{20}\right) dP_{Z},%
\text{ where}  \notag \\
c_{Y_{1}} &=&\left( 1-\psi _{20}\right) \left( \frac{\partial }{\partial
\theta }q\left( \theta _{0},\psi _{0}\right) U_{\theta _{0}}^{\left(
1\right) }+q\left( \theta _{0},\psi _{0}\right) U_{\theta _{0}}^{\left(
2\right) }\right) ,  \label{cY1} \\
c_{Y_{2}} &=&\left( \psi _{10}^{\left( 1\right) }\circ \psi _{20}\left(
1-\psi _{20}\right) -\psi _{10}\circ \psi _{20}\right) \left( \frac{\partial 
}{\partial \theta }q\left( \theta _{0},\psi _{0}\right) U_{\theta
_{0}}^{\left( 1\right) }+q\left( \theta _{0},\psi _{0}\right) U_{\theta
_{0}}^{\left( 2\right) }\right) .  \label{cY2}
\end{eqnarray}%
Moreover, $Y\left( \theta _{0},\widehat{\psi }\right) -Y\left( \theta
_{0},\psi _{0}\right) =\dot{T}\left( \psi _{0}\right) \left[ \widehat{\psi }%
-\psi _{0}\right] +o_{p}\left( L^{-1/2}\right) $.

To show asymptotic normality, we apply Theorem 4 of GG22 where they prove a
general CLT\ holds for an integral functional of the bids quantile and its
derivative. To make clear why their result applies, it shall be useful to
input the pointwise arguments of our pathwise derivative,%
\begin{equation*}
\dot{Y}\left( \theta _{0},\psi _{0}\right) \left[ \widehat{\psi }-\psi _{0}%
\right] =\int \left. 
\begin{array}{c}
c_{Y_{1}}\left( w,r,x\right) \left( \widehat{\psi }_{1}\left( \psi
_{20}\left( r,x\right) ,x\right) -\psi _{10}\left( \psi _{20}\left(
r,x\right) ,x\right) \right) \\ 
+c_{Y_{2}}\left( w,r,x\right) \left( \widehat{\psi }_{2}\left( r,x\right)
-\psi _{20}\left( r,x\right) \right)%
\end{array}%
\right. P_{R,W,X}\left( dr,dw,dx\right) .
\end{equation*}%
Next, we re-write the integrand above as a functional of the observed bid's
quantile and its derivative,$\ B$ and $B^{\left( 1\right) }$ respectively. Let us define
\begin{equation*}
\Gamma \left( w,r,x,B,B^{\left( 1\right) }\right) =c_{Y_{1}}\left(
w,r,x\right) \phi ^{\prime }\left( \psi _{20}\left( r,x\right) \right)
B^{\left( 1\right) }\left( \psi _{20}\left( r,x\right) |x\right)
+c_{Y_{2}}\left( w,r,x\right) \phi ^{-1}(B^{-1}\left( r,x\right) ).
\end{equation*}%
Our $\Gamma $\ is analogous to the integrand that defines the functional in
equation (3.1) of GG22. It is easy to see that $\Gamma $\ is twice
continuously differentiable with respect to $B,B^{\left( 1\right) }$, so
that $\Gamma $\ satisfies Assumption F in their article. To apply their
result, we define the following functionals: 
\begin{eqnarray*}
\widehat{\xi }\left( w,x\right) &=&\int \Gamma \left( w,r,x,\widehat{B},%
\widehat{B}^{\left( 1\right) }\right) P_{R|W,X}\left( dr|w,x\right) , \\
\xi _{0}\left( w,x\right) &=&\int \Gamma \left( w,r,x,B,B^{\left( 1\right)
}\right) P_{R|W,X}\left( dr|w,x\right) , \\
\widehat{\xi } &=&\int \widehat{\xi }\left( w,x\right) P_{W,X}\left(
dw,dx\right) , \\
\xi _{0} &=&\int \xi _{0}\left( w,x\right) P_{W,X}\left( dw,dx\right) .\text{
\ }
\end{eqnarray*}%
Since $\widehat{B}$ and $\widehat{B}^{\left( 1\right) }$ are AQR estimators
and all conditions of their Theorem 4 are satisfied, it follows that $\sqrt{L%
}\left( \widehat{\xi }-\xi _{0}\right) \overset{d}{\rightarrow }N\left(
0,\sigma _{0}^{2}\right) $ for some $\sigma _{0}^{2}>0$. The fact that our
integral functionals are defined with different measures to GG22 does not
affect the conclusion of their proof as a change of variables can be
applied. We refer the reader to Appendix E.3 in GG-SM for further details.$%
\blacksquare $

\bigskip

\textbf{Proof of Theorem 2.}

We use $Y_{L},Y,y$ as defined in the proof of Lemma 6. We begin with a
statement that $\mathcal{Q}_{1}=\left\{ \frac{\partial }{\partial \theta }q%
\text{ for }q\in \mathcal{Q}\right\} $ is a Donsker class, which
subsequently implies $\mathcal{Q}_{2}=\left\{ q\frac{\partial }{\partial
\theta }q\text{ for }q\in \mathcal{Q}\right\} $ is Donsker due to
preservation of the Donsker properties of product of two Donsker classes of
functions (VW23, Theorem 2.10.8). We omit details on proving $\mathcal{Q}%
_{1} $\ is Donsker, as the argument used in proving Lemma C.2 directly
applies once $U_{\theta }^{\left( j\right) }$ is replaced by $U_{\theta
}^{\left( j+1\right) }$.

Let us now consider the first-order condition: $0=Y_{L}\left( \widehat{%
\theta },\widehat{\psi }\right) $. Since $\mathcal{Q}_{2}$\ is Donsker, we
want to exploit the stochastic equicontinuity property of the empirical
process to obtain 
\begin{equation}
0=Y_{L}\left( \widehat{\theta },\widehat{\psi }\right) =Y\left( \widehat{%
\theta },\widehat{\psi }\right) +o_{p}\left( 1/\sqrt{L}\right) ,
\label{SE for FOC}
\end{equation}%
since $\left\vert \widehat{\theta }-\theta _{0}\right\vert =o_{p}\left(
1\right) $ and $\left\Vert \widehat{\psi }_{i}-\psi _{i0}\right\Vert
_{\infty }=o_{p}\left( 1\right) $ for $i=1,2$. This is valid as long as we
can show $\left\Vert y-y^{\prime }\right\Vert _{L_{2}\left( P_{Z}\right)
}\lesssim \rho \left( \left( \theta ,\psi \right) ,\left( \theta ^{\prime
},\psi ^{\prime }\right) \right) $ in a neighborhood of $\left( \theta
_{0},\psi _{0}\right) $\ for some semi-metric $\rho $. We use $\rho \left(
\left( \theta ,\psi \right) ,\left( \theta ^{\prime },\psi ^{\prime }\right)
\right) =\left\vert \theta -\theta ^{\prime }\right\vert +\left\Vert \psi
_{1}-\psi _{1}^{\prime }\right\Vert _{\infty }+\left\Vert \psi _{2}-\psi
_{2}^{\prime }\right\Vert _{\infty }$, hence it suffices to show $\left(
\theta ,\psi \right) \mapsto q\left( \theta ,\psi \right) \frac{\partial }{%
\partial \theta }q\left( \theta ,\psi \right) $ is Lipschitz with respect to 
$\rho $. To do this, it is enough to verify that $q$ and $\frac{\partial }{%
\partial \theta }q$\ are Lipschitz, because it readily follows that a
product of bounded Lipschitz functions is Lipschitz.

In the proof of Lemma 5, using (\ref{T Lipschitz}), we have shown in (\ref%
{q-psi Lipschitz}) that $q$ is Lipschitz in $\psi $ for fixed $\theta $. It
is easy to extend this over $\Theta $, as $U_{\theta }^{\left( j\right) }$
is Lipschitz due to continuity of $U_{\theta }^{\left( j+1\right) }$\ on $%
\Theta $\ for $j=0,1$. It can then be verified that,%
\begin{eqnarray*}
\left\vert q\left( \theta ,\psi \right) -q\left( \theta ^{\prime },\psi
^{\prime }\right) \right\vert &\leq &\left( 2c_{\Theta ^{1}}+C_{\mathcal{D}%
_{1}^{0}}c_{\Theta ^{2}}\right) \left\vert \theta -\theta ^{\prime
}\right\vert \\
&&+\left( 1+C_{\mathcal{D}_{1}^{1}}+C_{\mathcal{D}_{1}^{0}}\right) c_{\Theta
^{1}}\left[ \left\Vert \psi _{1}-\psi _{1}^{\prime }\right\Vert _{\infty
}+\left\Vert \psi _{2}-\psi _{2}^{\prime }\right\Vert _{\infty }\right] .
\end{eqnarray*}%
Since $c_{\Theta ^{j}}$\ are uniformly bounded functions under S1(iii), we
have%
\begin{equation*}
\left\vert q\left( \theta ,\psi \right) -q\left( \theta ^{\prime },\psi
^{\prime }\right) \right\vert \lesssim \rho \left( \left( \theta ,\psi
\right) ,\left( \theta ^{\prime },\psi ^{\prime }\right) \right) ,
\end{equation*}%
as desired. Proving that $\frac{\partial }{\partial \theta }q$ is Lipschitz\
can be done analogously. The only difference is $U_{\theta }^{\left(
j\right) }$\ in $q$\ becomes $U_{\theta }^{\left( j+1\right) }$ in $\frac{%
\partial }{\partial \theta }q$, which gives%
\begin{eqnarray*}
\left\vert \frac{\partial }{\partial \theta }q\left( \theta ,\psi \right) -%
\frac{\partial }{\partial \theta }q\left( \theta ^{\prime },\psi ^{\prime
}\right) \right\vert &\leq &\left( 2c_{\Theta ^{2}}+C_{\mathcal{D}%
_{1}^{0}}c_{\Theta ^{3}}\right) \left\vert \theta -\theta ^{\prime
}\right\vert \\
&&+\left( 1+C_{\mathcal{D}_{1}^{1}}+C_{\mathcal{D}_{1}^{0}}\right) c_{\Theta
^{2}}\left[ \left\Vert \psi _{1}-\psi _{1}^{\prime }\right\Vert _{\infty
}+\left\Vert \psi _{2}-\psi _{2}^{\prime }\right\Vert _{\infty }\right] ,
\end{eqnarray*}%
where the extra smoothness condition on $U_{\theta }$\ is assumed in S2(i).
We can then conclude that $\left\vert y\left( \theta ,\psi \right) -y\left(
\theta ^{\prime },\psi ^{\prime }\right) \right\vert \lesssim \rho \left(
\left( \theta ,\psi \right) ,\left( \theta ^{\prime },\psi ^{\prime }\right)
\right) $ and (\ref{SE for FOC}) holds.

Next, we linearize $Y\left( \widehat{\theta },\widehat{\psi }\right) $\
around $\theta _{0}$, where there is an intermediate value $\widetilde{%
\theta }$, between $\widehat{\theta }$\ and $\theta _{0}$, and re-arrange to
obtain%
\begin{eqnarray*}
0 &=&Y\left( \theta _{0},\widehat{\psi }\right) +\frac{\partial }{\partial
\theta }Y\left( \widetilde{\theta },\widehat{\psi }\right) \left( \widehat{%
\theta }-\theta _{0}\right) +o_{p}\left( 1/\sqrt{L}\right) , \\
\widehat{\theta }-\theta _{0} &=&\left[ \frac{\partial }{\partial \theta }%
Y\left( \widetilde{\theta },\widehat{\psi }\right) \right] ^{-1}Y\left(
\theta _{0},\widehat{\psi }\right) +o_{p}\left( 1/\sqrt{L}\right) .
\end{eqnarray*}

To complete the proof, we will show that,%
\begin{equation*}
\sqrt{L}\left( \widehat{\theta }-\theta _{0}\right) =\left[ \frac{\partial }{%
\partial \theta }Y\left( \theta _{0},\psi _{0}\right) \right] ^{-1}\sqrt{L}%
\dot{Y}\left( \theta _{0},\psi _{0}\right) \left[ \widehat{\psi }-\psi _{0}%
\right] +o_{p}\left( 1/\sqrt{L}\right) ,
\end{equation*}%
where $\sqrt{L}\dot{Y}\left( \theta _{0},\psi _{0}\right) \left[ \widehat{%
\psi }-\psi _{0}\right] $ converges in distribution to a normal random
variable as established by Lemma 6. I.e., we are left to prove that $%
\left\vert \frac{\partial }{\partial \theta }Y\left( \widetilde{\theta },%
\widehat{\psi }\right) -\frac{\partial }{\partial \theta }Y\left( \theta
_{0},\psi _{0}\right) \right\vert =o_{p}\left( 1\right) $\ where $\frac{%
\partial }{\partial \theta }Y\left( \theta _{0},\psi _{0}\right) $\ is
invertible and the linearization error is $o_{p}\left( L^{-1/2}\right) $.

Adding null to $\frac{\partial }{\partial \theta }Y\left( \widetilde{\theta }%
,\widehat{\psi }\right) -\frac{\partial }{\partial \theta }Y\left( \theta
_{0},\psi _{0}\right) $ and apply the triangle inequality gives,%
\begin{equation*}
\left\vert \frac{\partial }{\partial \theta }Y\left( \widetilde{\theta },%
\widehat{\psi }\right) -\frac{\partial }{\partial \theta }Y\left( \theta
_{0},\psi _{0}\right) \right\vert \leq \left\vert \frac{\partial }{\partial
\theta }Y\left( \widetilde{\theta },\widehat{\psi }\right) -\frac{\partial }{%
\partial \theta }Y\left( \widetilde{\theta },\psi _{0}\right) \right\vert
+\left\vert \frac{\partial }{\partial \theta }Y\left( \widetilde{\theta }%
,\psi _{0}\right) -\frac{\partial }{\partial \theta }Y\left( \theta
_{0},\psi _{0}\right) \right\vert .
\end{equation*}%
We will prove that $\sup_{\theta \in \Theta }\left\vert \frac{\partial }{%
\partial \theta }Y\left( \theta ,\widehat{\psi }\right) -\frac{\partial }{%
\partial \theta }Y\left( \theta ,\psi _{0}\right) \right\vert =o_{p}\left(
1\right) $ and $\left\vert \frac{\partial }{\partial \theta }Y\left( 
\widetilde{\theta },\psi _{0}\right) -\frac{\partial }{\partial \theta }%
Y\left( \theta _{0},\psi _{0}\right) \right\vert =o_{p}\left( 1\right) $,
which implies $\left\vert \frac{\partial }{\partial \theta }Y\left( 
\widetilde{\theta },\widehat{\psi }\right) -\frac{\partial }{\partial \theta 
}Y\left( \theta _{0},\psi _{0}\right) \right\vert =o_{p}\left( 1\right) $.

First, let us write $\frac{\partial }{\partial \theta }y$\ in terms of $q$
and its derivatives:%
\begin{equation*}
\frac{\partial }{\partial \theta }y\left( \theta ,\psi \right) =q\left(
\theta ,\psi \right) \frac{\partial ^{2}}{\partial \theta ^{2}}q\left(
\theta ,\psi \right) +\left( \frac{\partial }{\partial \theta }q\left(
\theta ,\psi \right) \right) ^{2}.
\end{equation*}%
Note that $\frac{\partial }{\partial \theta }y\left( \theta _{0},\psi
_{0}\right) =\left( \frac{\partial }{\partial \theta }q\left( \theta
_{0},\psi _{0}\right) \right) ^{2}$, since $q\left( \theta _{0},\psi
_{0}\right) =0$ a.s., and $\int \left( \frac{\partial }{\partial \theta }%
q\left( \theta _{0},\psi _{0}\right) \right) ^{2}dP_{Z}$ is invertible under
S2(iii).

Consider any $\theta $, using elementary algebra, we have%
\begin{eqnarray*}
\left\vert \frac{\partial }{\partial \theta }y\left( \theta ,\widehat{\psi }%
\right) -\frac{\partial }{\partial \theta }y\left( \theta ,\psi _{0}\right)
\right\vert &\leq &c_{\mathcal{Q}^{0}}\left\vert \frac{\partial ^{2}}{%
\partial \theta ^{2}}q\left( \theta ,\widehat{\psi }\right) -\frac{\partial
^{2}}{\partial \theta ^{2}}q\left( \theta ,\psi _{0}\right) \right\vert +c_{%
\mathcal{Q}^{2}}\left\vert q\left( \theta ,\widehat{\psi }\right) -q\left(
\theta _{0},\psi _{0}\right) \right\vert \\
&&+2c_{\mathcal{Q}^{1}}\left\vert \frac{\partial }{\partial \theta }q\left(
\theta ,\widehat{\psi }\right) -\frac{\partial }{\partial \theta }q\left(
\theta _{0},\psi _{0}\right) \right\vert ,\text{ where\ }
\end{eqnarray*}%
$\sup_{\left( \theta ,\psi _{1},\psi _{2}\right) \in \Theta \times \mathcal{D%
}_{1}\times \mathcal{D}_{2}}\left\vert \frac{\partial ^{j}}{\partial \theta
^{j}}q\left( \theta ,\psi \right) \right\vert \leq c_{\mathcal{Q}^{j}}$ for $%
j=1,2$\ such that for all $\left( w,r,x\right) $, 
\begin{equation*}
c_{\mathcal{Q}^{j}}\left( w,r,x\right) =\left\vert c_{\Theta ^{j}}\left(
w\right) \right\vert +\left\vert c_{\Theta ^{j}}\left( r\right) \right\vert
+C_{\mathcal{D}_{1}^{0}}\left\vert c_{\Theta ^{j+1}}\left( r\right)
\right\vert .
\end{equation*}%
$c_{\mathcal{Q}^{j}}$ extends (\ref{Q - envelope}) where $c_{\mathcal{Q}%
^{0}} $\ is defined. The same argument for proving $q\left( \theta ,\psi
\right) $\ is Lipschitz in $\psi $\ for all $\theta $\ also applies to $%
\frac{\partial ^{j}}{\partial \theta ^{j}}q\left( \theta ,\psi \right) $,
giving us%
\begin{equation*}
\left\vert \frac{\partial ^{j}}{\partial \theta ^{j}}q\left( \theta ,\psi
\right) -\frac{\partial ^{j}}{\partial \theta ^{j}}q\left( \theta ,\psi
^{\prime }\right) \right\vert \leq \left( 1+C_{\mathcal{D}_{1}^{1}}+C_{%
\mathcal{D}_{1}^{0}}\right) c_{\Theta ^{j+1}}\left[ \left\Vert \psi
_{1}-\psi _{1}^{\prime }\right\Vert _{\infty }+\left\Vert \psi _{2}-\psi
_{2}^{\prime }\right\Vert _{\infty }\right] ,\text{ for }j=1,2,
\end{equation*}%
which extends (\ref{q-psi Lipschitz}) in Lemma 5 when $j=0$. Since $\int
\left\vert c_{\Theta ^{j+1}}\right\vert ^{2}dP_{Z_{0}}<\infty $, we have $%
\int \left\vert c_{\mathcal{Q}^{j}}\right\vert ^{2}dP_{Z_{0}}<\infty $ and $%
\int \left\vert c_{\mathcal{Q}^{j}}c_{\Theta ^{l+1}}\right\vert
dP_{Z_{0}}<\infty $ by repeated applications of Cauchy Schwarz inequality
for $j,l=0,1,2$. Then it follows from Jensen's inequality that $\sup_{\theta
\in \Theta }\left\vert \frac{\partial }{\partial \theta }Y\left( \theta ,%
\widehat{\psi }\right) -\frac{\partial }{\partial \theta }Y\left( \theta
,\psi _{0}\right) \right\vert =O_{p}(\left\Vert \widehat{\psi }_{1}-\psi
_{10}\right\Vert _{\infty }+\left\Vert \widehat{\psi }_{2}-\psi
_{20}\right\Vert _{\infty })=o_{p}\left( 1\right) $.

For $\left\vert \frac{\partial }{\partial \theta }Y\left( \widetilde{\theta }%
,\psi _{0}\right) -\frac{\partial }{\partial \theta }Y\left( \theta
_{0},\psi _{0}\right) \right\vert $, using $y\left( \theta _{0},\psi
_{0}\right) =0$ a.s. and\ a similar argument to previously, we have: 
\begin{eqnarray*}
\left\vert \frac{\partial }{\partial \theta }y\left( \widetilde{\theta }%
,\psi _{0}\right) -\frac{\partial }{\partial \theta }y\left( \theta
_{0},\psi _{0}\right) \right\vert &\leq &c_{\mathcal{Q}^{2}}\left\vert
q\left( \widetilde{\theta },\psi _{0}\right) -q\left( \theta _{0},\psi
_{0}\right) \right\vert +2c_{\mathcal{Q}^{1}}\left\vert \frac{\partial }{%
\partial \theta }q\left( \widetilde{\theta },\psi _{0}\right) -\frac{%
\partial }{\partial \theta }q\left( \theta _{0},\psi _{0}\right) \right\vert
\\
&\leq &3c_{\mathcal{Q}^{1}}c_{\mathcal{Q}^{2}}\left\vert \widetilde{\theta }%
-\theta _{0}\right\vert .
\end{eqnarray*}%
Integrating the above with respect to $P_{Z}$ and apply Jensen's inequality
gives,%
\begin{equation*}
\left\vert \frac{\partial }{\partial \theta }Y\left( \widetilde{\theta }%
,\psi _{0}\right) -\frac{\partial }{\partial \theta }Y\left( \theta
_{0},\psi _{0}\right) \right\vert =O_{p}\left( \left\vert \widetilde{\theta }%
-\theta _{0}\right\vert \right) ,
\end{equation*}%
where Cauchy Schwarz inequality ensures $\int \left\vert c_{\mathcal{Q}%
^{1}}c_{\mathcal{Q}^{2}}\right\vert dP_{Z}<\infty $. The proof is completed
as $\left\vert \widetilde{\theta }-\theta _{0}\right\vert =o_{p}\left(
1\right) $.$\blacksquare $

\newpage

\end{document}